\documentclass[apj,twocolumn]{emulateapj}
\usepackage{natbib}

\newcommand{\msun}{M$_\odot$}

\newcommand{\qpah}{$q_{\rm PAH}$}

\shorttitle{SMC PAH Fraction}
\shortauthors{Sandstrom et al.}

\begin{document}


\title{The \emph{Spitzer} Survey of the Small Magellanic Cloud (S$^3$MC):
Insights into the Life-Cycle of Polycyclic Aromatic Hydrocarbons}

\author{Karin M. Sandstrom\altaffilmark{1,2}, 
	Alberto D. Bolatto\altaffilmark{3},
	Bruce Draine\altaffilmark{4},
	Caroline Bot\altaffilmark{5},
	Sne\v{z}ana Stanimirovi\'{c}\altaffilmark{6}}

\affil{$^1$Astronomy Department, 601 Campbell Hall, University of
  California, Berkeley, CA 94720, USA} 
\affil{$^2$Max Planck Institut f\"{u}r Astronomie, D-69117 Heidelberg,
  Germany}
\affil{$^3$Department of Astronomy and Laboratory for Millimeter-wave
  Astronomy, University of Maryland, College Park, MD 20742, USA}
\affil{$^4$Department of Astrophysical Sciences, Princeton University, Princeton NJ
  08544, USA}
\affil{$^5$UMR 7550, Observatoire Astronomiques de Strasbourg,
  Universite Louis Pasteur, F-67000 Strasbourg, France}
\affil{$^6$Astronomy Department, University of Wisconsin, Madison, 475
  North Charter Street, Madison, WI 53711, USA}

\email{sandstrom@mpia.de}


\begin{abstract}

We present the results of modeling dust spectral energy distributions
(SEDs) across the Small Magellanic Cloud (SMC) with the aim of mapping
the distribution of polycyclic aromatic hydrocarbons (PAHs) in a
low-metallicity environment.  Using \emph{Spitzer} Survey of the SMC
(S$^3$MC) photometry from 3.6 to 160 \micron\ over the main
star-forming regions of the Wing and Bar of the SMC along with
spectral mapping observations from 5 to 38 \micron\ from the
\emph{Spitzer} Spectroscopic Survey of the Small Magellanic Cloud
(S$^4$MC) in selected regions, we model the dust spectral energy
distribution and emission spectrum to determine the fraction of dust
in PAHs across the SMC.  We use the regions of overlaping photometry
and spectroscopy to test the reliability of the PAH fraction as
determined from SED fits alone.  The PAH fraction in the SMC is low
compared to the Milky Way and variable--with relatively high fractions
(\qpah$\sim 1-2$\%) in molecular clouds and low fractions in the
diffuse ISM (average $\langle$\qpah$\rangle = 0.6$\%). We use the map
of PAH fraction across the SMC to test a number of ideas regarding the
production, destruction and processing of PAHs in the ISM.  We find
weak or no correlation between the PAH fraction and the distribution
of carbon AGB stars, the location of supergiant H I shells and young
supernova remnants, and the turbulent Mach number.  We find that the
PAH fraction is correlated with CO intensity, peaks in the dust
surface density and the molecular gas surface density as determined
from 160 \micron\ emission.  The PAH fraction is high in regions of
active star-formation, as predicted by its correlation with molecular
gas, but is supressed in H II regions.  Because the PAH fraction in
the diffuse ISM is generally very low--in accordance with previous
work on modeling the integrated SED of the SMC--and the PAH fraction
is relatively high in molecular regions, we suggest that PAHs are
destroyed in the diffuse ISM of the SMC and/or PAHs are forming in
molecular clouds.  We discuss the implications of these observations
for our understanding of the PAH life cycle, particularly in
low-metallicity and/or primordial galaxies.

\end{abstract}



\section{Introduction}\label{intro}

Polycyclic Aromatic Hydrocarbons (PAHs) are thought to be the carrier
of the ubiquitously observed mid-IR emission bands \citep[][among
others]{allamandola89}.  The bands are the result of vibrational
de-excitation of the PAH skeleton through bending and stretching modes
of C-H and C-C bonds after the absorption of a UV photon.  The
emission in these bands can be very bright and can comprise a
significant fraction, up to 10$-$20\% \citep{smith07}, of the total
infrared emission from a galaxy.  For this reason, PAH emission has
been suggested to be a useful tracer of the star-formation rate, even
out to high redshifts \citep{calzetti07}.  Making use of PAHs as a
tracer, however, requires understanding how the abundance and emission
from PAHs depends on galaxy properties such as metallicity and
star-formation history.

PAHs also play a number of important roles in the interstellar medium
(ISM).  In particular, these small dust grains can dominate
photoelectric heating rates \citep{bakes94}.  In dense clouds, PAHs
can alter chemical reaction networks by providing a neutralization
route for ionized species \citep{bakes98,weingartner01} and contribute
large amounts of surface area for chemical reactions that occur on
grain surfaces.  PAHs are a crucial component of interstellar dust so
we would like to understand the processes that govern their abundance
and physical state. 

The life-cycle of PAHs, however, is not yet well understood.  PAHs are
thought to form in the carbon-rich atmospheres of some evolved stars
\citep{latter91,cherchneff92}. Emission from PAHs has been observed
from carbon-rich asymptotic giant branch stars \citep{sloan07} and
more frequently in carbon-rich post-AGB stars where the radiation
field is more effective at exciting the mid-IR bands \citep{buss93}.
A ``stardust'' origin (i.e. formation in the atmospheres of evolved
stars) for the majority of PAH material is controversial, however,
because it has yet to be demonstrated that PAHs can be produced in AGB
stars faster than they are destroyed in the ISM \citep[for a recent
review, see][]{draine09}, i.e. the timescale for destruction of dust
by SNe shocks is shorter than the timescale over which the ISM is
enriched with dust from AGB stars \citep[for example,][]{jones94}.  In
addition to destruction by supernova shocks, PAH material may be
destroyed by UV fields, a process that can dominate near a hot star or
in the ISM of low metallicity and/or primordial galaxies.  If PAHs are
mostly not ``stardust'', they must have formed in the ISM itself, by
some mechanism which is not yet characterized.  A variety of
mechanisms have been suggested
\citep{tielens87,puget89,herbst91,greenberg00} however there is little
observational support for any one model as of yet.

In recent years, observations with ISO and \emph{Spitzer} have allowed
us to study the abundance and physical state of PAHs in a variety of
ISM conditions beyond those we observe in the Milky Way.  One of the
most striking results is the abrupt change in the fraction of dust in
PAHs as a function of metallicity.  A deficit of PAH emission from low
metallicity galaxies has been widely observed
\citep{madden00,engelbracht05,madden06,wu06,jackson06,engelbracht08}.
\citet{engelbracht05} found that the ratio of the 8 to 24 \micron\
surface brightness undergoes a transition from a SED with typical PAH
emission to an SED essentially devoid of PAH emission at a metallicity
of 12 + log(O/H) $\sim 8$.  The weakness of PAH emission in
low-metallicity galaxies has been confirmed spectroscopically
\citep{wu06,engelbracht08}.  Using the SINGS galaxy sample
\citet{draine07b} modeled the integrated SEDs and determined that the
deficit of PAH emission corresponds to a decrease in the PAH fraction
rather than a change in excitation of the PAHs.  They found that
\qpah\ (defined as the fraction of the total dust mass that is
contributed by PAHs containing less than $10^3$ carbon atoms) changes
from a median of $\sim 4$\% \citep[comparable to the Milky Way PAH
fraction of 4.6\%;][]{li01} in galaxies with 12 + log(O/H) $> 8.1$ to
a median of $\sim 1$\% in more metal poor galaxies.
\citet{munoz-mateos09} have investigated the radial variation of
\qpah\ and metallicity in the SINGS sample and find results consistent
with \citet{draine07b}.

There have been a number of suggestions as to what in the PAH
life-cycle changes at low metallicity leading to the observed
deficiency.  \citet{galliano08} suggested that the delay between
enrichment of the ISM by supernova-produced dust relative to that from
AGB stars could lead to a lower PAH fraction at low metallicity.  This
model relies on the assumption that supernovae contribute a
significant amount of dust to the ISM, an assumption  which is
controversial
\citep{moseley89,dunne03,krause04,sugerman06,meikle07,draine09}, as
well as long timescales for dust production in carbon-rich AGB stars,
which may be shorter than previously thought \citep{sloan09}.
Fundamentally, the \citet{galliano08} model assumes a ``stardust''
origin for PAHs, which may not be the case.  Other models explaining
the low metallicity deficiency rely on enhanced destruction of PAHs.
This can be accomplished through more efficient destruction via
supernova shocks \citep{ohalloran06} or via the harder and more
intense UV fields in these galaxies \citep[e.g.][]{madden06,gordon08}.  

In order to investigate the PAH life cycle at low metallicity, we
performed two surveys of the Small Magellanic Cloud (SMC) with the
\emph{Spitzer} Space Telescope.  The SMC is a nearby dwarf irregular
galaxy that is currently interacting with the MW and the Large
Magellanic Cloud.  Its proximity \citep[61 kpc;][]{hilditch05}, low
metallicity \citep[12 + log(O/H) $\sim 8$, $Z \sim 0.2
Z_{\odot}$;][]{kurt98} and tidally disrupted ISM make it an ideal
location in which to study the life cycle of PAHs in an environment
very different from the Milky Way.  The SMC has a low dust-to-gas
ratio, $\sim 10$ times smaller than in the Milky Way
\citep{bot04,leroy07}, leading to more pervasive UV fields.  Because
of its proximity we can observe the ISM at high spatial resolution and
sensitivity in order to characterize the processes driving the PAH
fraction.

The PAH fraction in the SMC has been controversial.  \citet{li02},
using IRAS and COBE data, found that the PAHs in the SMC Bar contained
only 0.4\% of the interstellar carbon, corresponding to \qpah $\approx
$ 0.2\%, whereas \citet{bot04} concluded that PAHs accounted for 4.8\%
of the total dust mass in the diffuse ISM of the SMC, similar to the
Milky Way.  In the following, we present results on the fraction of
PAHs in the SMC using observations from the \emph{Spitzer} Survey of
the Small Magellanic Cloud (S$^3$MC). We use spectroscopy in the
regions covered by the \emph{Spitzer} Spectroscopic Survey of the
Small Magellanic Cloud (S$^4$MC) to verify that our models for the
photometry are correctly gauging the PAH fraction.  We defer a
detailed analysis of the spectroscopy to an upcoming paper (Sandstrom
et al. 2010, in prep)  In Section~\ref{sec:obs} we describe the
observations and data reduction, particularly focusing on the
foreground subtraction and cross-calibration of the IRAC, MIPS and IRS
observations.  In Section~\ref{sec:models} we describe the SED fitting
procedure using the models of \citet{draine07a} and the modifications
necessary to incorporate the S$^4$MC spectroscopy into the fit.  In
Sections~\ref{sec:results} and~\ref{sec:discussion} we present the
results of the SED modeling and discuss their implications for our
understanding of the PAH life cycle both in low metallicity galaxies
and in the Milky Way.

\section{Observations and Data Reduction}\label{sec:obs}

\subsection{\emph{Spitzer} Survey of the Small Magellanic Cloud (S$^3$MC)
Observations}\label{sec:s3mc}

\begin{figure*}
\centering
\epsscale{1.0}
\includegraphics[width=7in]{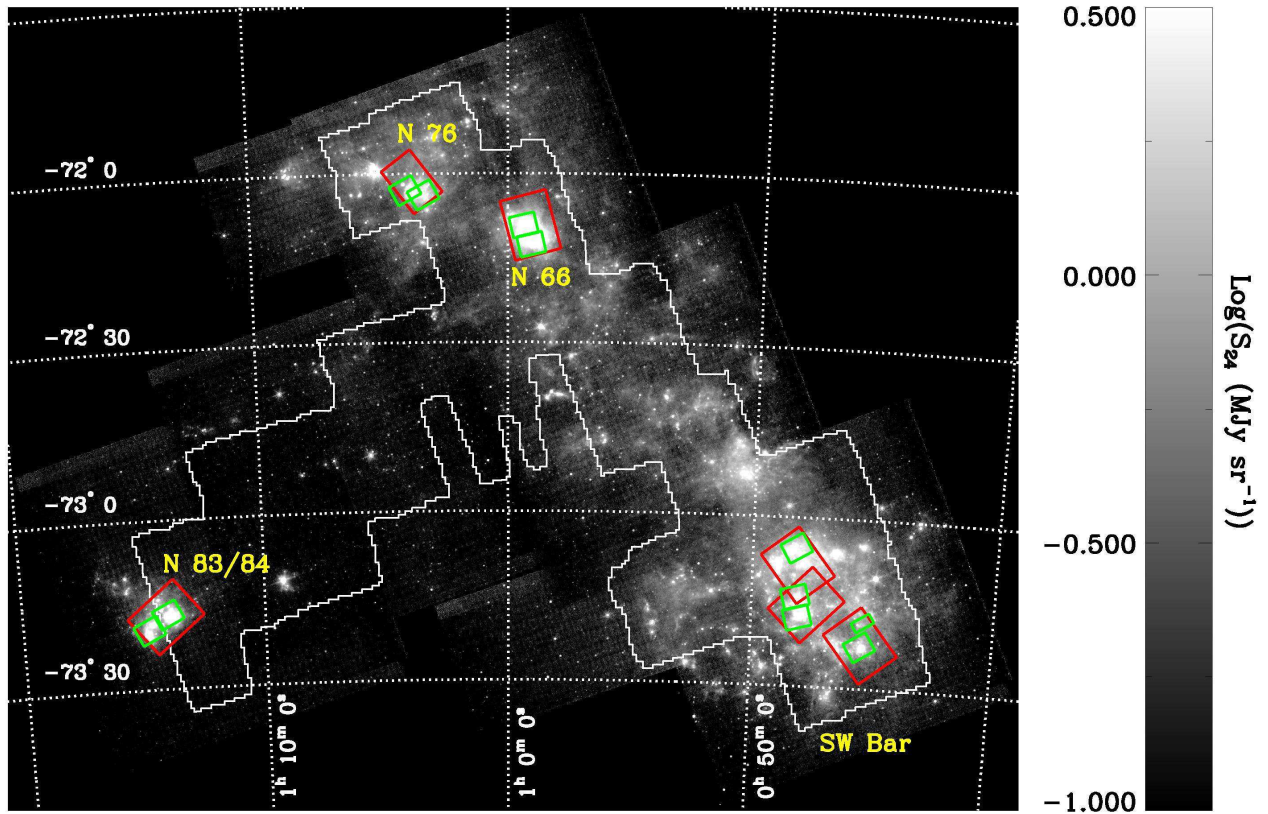}
\caption{The coverage of the S$^3$MC and S$^4$MC surveys overlayed on
the MIPS 24 \micron\ map.  The color scale is logarithmic, with the stretch
illustrated in the colorbar.  The red boxes show the coverage of the LL1
order maps (the LL2 maps are shifted by $\sim 3\arcmin$) and the green
boxes show the coverage of the SL1 order maps (the SL2 maps are
shifted by $\sim 1\arcmin$).  We also identify the various regions of
the galaxy by the names we will refer to in the remainder of this
paper.}
\label{fig:coverage}
\end{figure*}

We mapped the main star-forming areas of the Bar and Wing of the SMC
using the IRAC and MIPS instruments as part of the S$^3$MC project (GO
3316).  A more comprehensive description of the observations and data
reduction can be found in \citet{bolatto07} and  \citet{leroy07}.  The
region where the coverage of the IRAC and MIPS observations overlap is
shown in Figure~\ref{fig:coverage} overlayed on the 24 \micron\ image.
The mosaics were constructed using the MOPEX software provided by the
SSC\footnote{\url{http://ssc.spitzer.caltech.edu/postbcd/mopex.html}}.
The IRAC and MIPS mosaics were corrected for a number of artifacts as
described in \citet{bolatto07}.  The most important of these for the
purposes of this work is the large additive gradients at IRAC
wavelengths (primarily 5.8 and 8.0 \micron) caused by varying offsets
in the detectors and the mosaicing algorithm implemented in MOPEX.  We
will discuss these gradients further in Section~\ref{sec:fg} since
they become important for determining the foregrounds present in our
maps.

Since the initial processing of the MIPS mosaics, the calibration
factors recommended by the SSC have been revised.  We correct the
mosaics to use the recommended factors of 0.0454, 702 and 41.7 MJy
sr$^{-1}$ per instrumental data unit, which are 3\%, 11\% and 0.7\%
different from the earlier values used by \citet{bolatto07} and
\citet{leroy07}.  In addition, the 70 \micron\ observations suffer
from non-linearities at high surface brightness as noted by
\citet{dale07}.  Although relatively few pixels in our map are
affected by the correction, these regions in particular are most
likely to overlap with our spectroscopic observations.  We use the
most recent non-linearity correction described in Gordon et al.
(2010), in prep.  No calibration correction for extended emission is
necessary for the MIPS mosaics \citep{cohen09}. The 1$\sigma$
sensitivities of the observations are listed in Table~\ref{tab:s3mc}.
The noise level at 70 \micron\ is higher than the predicted detector
noise  due to pattern noise, visible as striping in the map \citep[for
further discussion of 70 \micron\ noise properties, see][]{bolatto07}.
Aside from the non-linearity corrections, we use the MIPS calibration
factors listed on the SSC website, but we note that \citet{leroy07}
found an offset between the MIPS 160 \micron\ and the DIRBE 140
\micron\ photometry of the SMC which may be the results of a
calibration difference.  They adjusted the calibration by a factor of
1.25 to match DIRBE.  We do not apply this correction, and we briefly
discuss the implications of that choice in Section~\ref{sec:unc}.

\begin{deluxetable}{lccc}
\tablewidth{0pt}
\tabletypesize{\scriptsize}
\tablecolumns{4}
\tablecaption{S$^3$MC Observation Details}
\tablehead{ \multicolumn{1}{l}{Band} &
\multicolumn{1}{c}{Map Noise Level} &
\multicolumn{1}{c}{Cal. Uncertainty} &
\multicolumn{1}{c}{Resolution} \\
\multicolumn{1}{c}{} & 
\multicolumn{1}{c}{(MJy sr$^{-1}$)} & 
\multicolumn{1}{c}{(\%)} &
\multicolumn{1}{c}{(\arcsec)}}
\startdata
3.6 & 0.015 & 10\tablenotemark{a} & 1.66 \\
4.5 & 0.017 & 10\tablenotemark{a} & 1.72 \\ 
5.8 & 0.055 & 10\tablenotemark{a} & 1.88 \\ 
8.0 & 0.042 & 10\tablenotemark{a} & 1.98 \\
24 & 0.047 & 4 & 6.0 \\ 
70 & 0.664 & 7 & 18  \\   
160 & 0.695 & 12 & 40 
\enddata
\label{tab:s3mc}
\tablecomments{The noise levels for the IRAC maps shown here have not
been multiplied by the extended source calibration.}
\tablenotetext{a}{Calibration uncertainties in the IRAC bands have
been increased to 10\% because of the extended source corrections.}
\end{deluxetable}

Because the calibration of the IRAC bands is based on stellar point
sources and we are dealing with extended objects, we also apply the
extended source calibration factors of 0.955, 0.937, 0.772 and 0.737
to the 3.6, 4.5, 5.8 and 8.0 bands as recommended by \citet{reach05}.
Recent work by \citet{cohen07} verified the 36\% correction at 8.0
\micron.  The correction factors depend on the structure of the
emission for each region of the map, which ranges from diffuse to
point-like.  Thus, applying on uniform correction factor across the
entire map introduces some systematic uncertainty in our IRAC flux
densities.  To account for these uncertainties we assume a 10\%
calibration uncertainty for the IRAC bands.  The details of the
calibration are discussed further in an Appendix and the convolution
and alignment to a common resolution will be discussed in
Section~\ref{sec:conv}.

\subsection{\emph{Spitzer} Spectroscopic Survey of the Small Magellanic Cloud
(S$^4$MC)}\label{sec:s4mc}

In order to directly probe the physical state of PAHs in the SMC, we
performed spectral mapping of six star-forming regions using the low
spectral resolution orders of IRS on \emph{Spitzer} (GO 30491).  These
observations are primarily intended to investigate spectral variations
in the PAH emission, to be described in Sandstrom et al. (2010), in
prep.  The coverage of the maps are shown overlayed on the MIPS 24
\micron\ image from our S$^3$MC observations in
Figure~\ref{fig:coverage}.

The spectral coverage of the low-resolution orders of IRS extends from
5.2 to 38.0 \micron, covering the major PAH emission bands in the
mid-infrared except the 3.3 \micron\ feature.  The maps are fully
sampled by stepping perpendicular to the slit one-half slit width
between each slit position (1.85\arcsec and 5.08\arcsec, for SL and LL
respectively). The LL maps are made of 98 pointings perpendcular to
the slit and 7 pointings parallel for a coverage of $493\arcsec \times
474\arcsec$, except for the map of N 76 which covers 75 by 6 pointings
and an area of $376\arcsec \times 395\arcsec$.  The SL maps are made
of 120 pointings perpendcular and 5 pointings parallel covering an
area of $220\arcsec \times 208\arcsec$ in each map, except for the region
around SMC B1 where we use 60 by 4 pointings covering an area of
$109\arcsec \times 156\arcsec$.  

The spectra were processed with either version 15.3.0 or 16.1.0 of the
IRS Pipeline.  The only difference of note between these versions is a
change in the processing of radiation hits that does not affect the
results we present.  The details of the data reduction are discussed
in \citet{sandstrom09}.  In brief, the maps were assembled using
Cubism\footnote{\url{http://ssc.spitzer.caltech.edu/archanaly/contributed/cubism/}},
wherein a ``slit loss correction function'' was applied, analogous to
the extended source correction for the IRAC bands, to adjust the
calibration from point sources to extended objects \citep{smith07}.
Each mapping observation was followed or preceded by an observation of
a designated ``off'' position at R.A.  1$^{\rm h}$9$^{\rm
m}$40$^\mathrm{s}$ and Dec $-$73$^\circ$31$\arcmin$30$\arcsec$.  This
position was seen to have minimal SMC emission in our MIPS
observations.  The ``off'' spectra, in addition to subtracting the
zeroth-order foregrounds, help to mitigate the effects of rogue
pixels.  Additional bad pixel removal was done within Cubism.  Beyond
$\sim 35$ \micron, there are increasing numbers of ``hot'' pixels
which degrade the sensitivity of our spectra, we trim the LL orders to
35 \micron\ to avoid issues with the long wavelength data.

Prior to further reduction steps we determine corrections to match the
SL2 (5.2$-$7.6 \micron), SL1 (7.5$-$14.5 \micron), LL2 (14.5$-$20.75
\micron) and LL1 (20.5$-$38.5 \micron) orders in their overlap
regions.  For the SL2/SL1 and LL2/LL1 overlap, we find that the
offsets are best explained by small additive effects, which may be due
to a temporally and spatially varying dark current analogous to the
``dark settle'' effect seen in high-resolution IRS spectroscopy.
Further discussion of these offsets will be presented in Sandstrom et
al (2010), in prep.  In brief, we apply correction factors determined
by examining the unilluminated parts of the IRS detector, very similar
to what is done in the ``darksettle'' software for LH available
through the SSC.  After applying these correction factors the orders
generally match-up to within their respective errors.  In regions of
low surface brightness small residual effects have the appearance of a
bump around 20 \micron\ in the stitched spectra.  These offsets only
affect a small portion of the spectrum and do not measurably alter the
results of the fit.

\subsection{Foreground Subtraction}\label{sec:fg}

\subsubsection{IRAC and MIPS Foreground Subtraction}\label{sec:fgfit}

In the mid- and far-infrared there are three major
foreground/background contributions that contaminate our observations
of the SMC: the zodiacal emission, Milky Way cirrus emission and the
cosmic infrared background (CIB).  In addition, for the IRAC mosaics
there is a planar offset introduced by varying detector offsets and
the mosaicing algorithm in MOPEX \citep[see ][for more
details]{bolatto07}.  These need to be removed from the maps to
isolate the emission from the SMC.  In the following we will briefly
discuss the foreground subtraction that has been carried out on our
data and the major uncertainties in this process.  The approach we use
to subtract these foregrounds is motivated by the following
limitations of our observations: (1) at longer wavelengths the mosaics
do not extend to areas with no SMC emission, (2) the observations of
neutral hydrogen which we use to subtract the Milky Way cirrus
emission have a lower angular resolution than our \emph{Spitzer} maps
and (3) the residual mosaicing gradients in the 5.8, 8.0 and to a
lesser degree 4.5 \micron\ IRAC bands interfere with directly fitting
and subtracting the zodiacal light. 

Our foreground subtraction has three steps: (1) we determine the
coefficients of a planar surface that describe the combination of
zodiacal light and the residual mosaicing offsets, (2) we subtract
this plane from each map at its native resolution, (3) we convolve the
maps to the MIPS 160 \micron\ resolution of 40\arcsec\ and subtract
the MW cirrus foreground and CIB.  Throughout this procedure we fix
the CIB level at 70 and 160 \micron\ to be 0.23 and 1.28 MJy sr$^{-1}$
as determined from the \emph{Spitzer} Observation Planning Tool
(SPOT)\footnote{\url{http://ssc.spitzer.caltech.edu/documents/background/bgdoc\_release.html}}.
We also assume a fixed proportionality between the infrared cirrus
emission and the column density of MW H I \citep{boulanger96}, using
coefficients derived from the model of \citet{draine07a} for MW dust
heated by the local interstellar radiation field.  The
\citet{draine07a} model reproduces the DIRBE observations of the
cirrus from \citet{arendt98} to within their quoted 20\%
uncertainties. We convolve the \citet{draine07a} model emissivity
spectrum with the IRAC and MIPS spectral response curves to obtain the
coefficients, which are listed in Table~\ref{tab:mwcoeff}.  We use a
map of Galactic hydrogen obtained by combining data from ATCA and
Parkes \citep[see ][for further information on the technique and the
original observations]{stanimirovic99} re-reduced and provided to us
by E.  Muller (private communication) to subtract the MW cirrus
contribution.

\begin{deluxetable}{lc}
\tablewidth{0pt}
\tabletypesize{\scriptsize}
\tablecolumns{2}
\tablecaption{Milky Way Foreground Coefficients}
\tablehead{ \multicolumn{1}{l}{Band} &
\multicolumn{1}{c}{Coefficient} \\
\multicolumn{1}{c}{} & 
\multicolumn{1}{c}{(MJy sr$^{-1}$ (10$^{21}$ H)$^{-1}$)}}
\startdata
3.6 & 0.018 \\
4.5 & 0.006 \\ 
5.8 & 0.070 \\ 
8.0 & 0.215 \\
24 & 0.162 \\ 
70 & 2.669 \\   
160 & 11.097  
\enddata
\label{tab:mwcoeff}
\end{deluxetable}

The first step in our foreground subtraction is to determine the
coefficients describing the planar contributions from the zodiacal
light and the mosaicing offsets.  Over the area of the SMC, the
zodiacal light is well-described by a plane and the gradients
introduced by the mosaicing algorithm also have a planar dependence
\citep{bolatto07}.  Unfortunately, there are no ``off'' locations
where we could simply fit a plane to the foreground level, since no
regions are totally free of SMC emission in the S$^3$MC maps.  At each
position in the map the surface brightness is a combination of: SMC
emission, MW cirrus emission, zodiacal light, mosaicing offsets and
CIB.  To isolate the planar component, we first extract photometry for
a number of $200\arcsec \times 200\arcsec$ regions from the IRAC and
MIPS maps. We choose the regions to be outside of star-forming areas,
where the SMC emission is dominantly from dust heated by the general
interstellar radiation field, where we can assume a proportionality
between the dust emission and the SMC H I column.  We choose the
region size of $200\arcsec \times 200\arcsec$ to be larger than the
98\arcsec\ beam of the H I observations and large enough to robustly
determine the mean level in each box even when contaminated by point
sources at the shorter wavelengths.  We then subtract off the Milky
Way cirrus contribution for those regions determined with the
coefficients in Table~\ref{tab:mwcoeff} and the MW H I map,  and, for
the 70 and 160 \micron\ maps, the CIB level.  We call this MW cirrus
and CIB subtracted value S$_{\nu,{\rm resid}}$.   S$_{\nu,{\rm
resid}}$ is a combination of the zodiacal light, emission from the
diffuse ISM of the SMC and whatever residual gradients there remain
from the mosaicing for the IRAC maps.  Next, using the SMC neutral H I
observations of \citet{stanimirovic99}, we perform a least-squares fit
to the foreground values with the following function:
\begin{equation}
S_{\nu, {\rm resid}} = A(1 + B\Delta_{\alpha} + C\Delta_{\delta}) +
D\times H I_{SMC} .
\end{equation}
Here $A$, $B$ and $C$ are the coefficients describing a plane;
$\Delta_{\alpha}$ and $\Delta_{\delta}$ are the gradients in Right
Ascension and Declination across the SMC;  $D$ is dust emission per H
in the given waveband; and $S_{\nu,{\rm resid}}$ is the residual
emission in each box after subtracting the MW foreground and CIB
components. With this fit we determine the coefficients of the 
best fit planar foreground while excluding emission in the map
coming from the SMC itself.

We take two additional steps to improve the determination of the
planar foreground coefficients: (1) we use the 2MASS point source
catalog to avoid regions of high stellar density in the IRAC bands so
we do not subtract unresolved starlight, which can masquerade as a
foreground, and (2) we fix the gradient of the zodiacal light ($B$ and
$C$) at 70 and 160 \micron\  to the results of the fit at 24 \micron.
Since there are no mosiacing offsets for the MIPS bands, the zodiacal
light is by far the dominant foreground at 24 \micron\ and the
zodiacal light should have the same spatial dependence at all of the
MIPS wavelengths, fixing the zodiacal light gradient to what we
measure at 24 \micron\ is more effective than trying to fit for those
coefficients at the longer wavelengths. The fixed pattern noise at 70
\micron\ and the increasingly dominant SMC emission at 70 and 160
\micron\ make the fitting procedure less robust compared to the 24
\micron\ results.  

The results of the fits are listed in Table~\ref{tab:fg}.  For the
IRAC bands at 3.6 and 4.5 \micron\ we do not detect any emission
correlated with the SMC neutral hydrogen.  At 4.5 \micron, there is a
quite significant residual gradient from the mosaicing procedure.  At
3.6 \micron\ the zodiacal foreground and its gradient are not detected
at the sensitivity of the map.  At both 3.6 and 4.5 unresolved
starlight can play a role in the foreground determination, so we
choose regions to avoid high stellar densities.  In Table~\ref{tab:fg}
we also list the standard deviations of the fit residuals to show the
quality of the foreground determination for the mosaics.  Finally, we
subtract the planar fit listed in Table~\ref{tab:fg} from the mosaics
at their full resolution.  

\begin{deluxetable*}{lccccc}
\tablewidth{0pt}
\tabletypesize{\scriptsize}
\tablecolumns{6}
\tablecaption{Foreground Properties}
\tablehead{ \multicolumn{1}{l}{Band} &
\multicolumn{1}{c}{A} &
\multicolumn{1}{c}{B} &
\multicolumn{1}{c}{C} &
\multicolumn{1}{c}{D} &
\multicolumn{1}{c}{Std. Dev. of Fit} \\
\multicolumn{1}{c}{} &
\multicolumn{1}{c}{(MJy sr$^{-1}$)} &
\multicolumn{2}{c}{(deg$^{-1}$)} &
\multicolumn{1}{c}{(MJy sr$^{-1}$ (10$^{20}$ H)$^{-1}$)} &
\multicolumn{1}{c}{(MJy sr$^{-1}$)}}
\startdata
3.6 & $<$0.004 & \nodata & \nodata & \nodata & 0.011 \\
4.5 & $-$0.11 & 0.019   & 4.074\tablenotemark{a}   & \nodata & 0.014 \\
5.8 & 4.99    & $-$0.002 & 0.002 & $9.14\times 10^{-4}$ & 0.023  \\
8.0 & 3.69    &  0.0014  & 0.0014 & $7.43\times 10^{-5}$ & 0.023 \\
24  & 22.34 &  $-$0.0026 & 0.0010 & $1.63\times 10^{-3}$ & 0.039  \\
70  & 6.43 &  $-$0.0026\tablenotemark{b} & 0.0010\tablenotemark{b} & $3.97\times 10^{-2}$ & 1.23  \\
160 & 1.83 &  $-$0.0026\tablenotemark{b} & 0.0010\tablenotemark{b} & $1.81\times 10^{-1}$ & 3.30
\enddata
\label{tab:fg}
\tablecomments{See Section~\ref{sec:fgfit} for a description of the
coefficients.}
\tablenotetext{a}{Large gradient at 4.5 \micron\ due to detector
offsets and mosaicing algorithm.}
\tablenotetext{b}{Fixed from fit at 24 \micron.}
\end{deluxetable*}
 
Because it is at lower resolution (98\arcsec), we subtract the Milky
Way foreground after convolving to the MIPS 160 \micron\ resolution
($\sim 40$\arcsec).  At the signal-to-noise ratio of the map, the
Galactic H I in front of the SMC does not show any high contrast
features, although there is a gradient ranging between 2 and 4$\times
10^{20}$ cm$^{-2}$ across the region.  We note that the resolution
difference between our mosaics and the H I observations means that
there could be features with spatial scales less than 98\arcsec\ that
are not adequately subtracted from our maps.  However, assuming a
distance of 1 kpc for the H I foreground of the Galaxy, features left
in our map would have to have spatial scales of $\sim 0.5$ pc.  The
typical foreground of Milky Way gas in this direction is $\sim 3
\times 10^{20}$ cm$^{-3}$, an 0.5 pc cloud of cold neutral medium
having a typical volume density of $\sim 40$ cm$^{-3}$
\citep{heiles03} would contribute a column of $6\times 10^{19}$
cm$^{-3}$, a factor of 5 less than the average foreground we see.
Thus, we expect that inadequate subtraction of small scale structure
will not greatly affect the quality of the foreground subtraction.

\subsubsection{IRS Foreground Subtraction}

The ``off'' observations for each spectral cube will remove foreground
emission to first order.  However, there are gradients between the
``off'' position and the cubes that are still present in the data.  To
remove the remaining foregrounds we add back to the spectra the
difference between the ``off'' position and the map position at each
wavelength calculated using the zodiacal light predictions from SPOT
and the MW cirrus emission spectrum from \citet{draine07a} multiplied
by the MW H I column density at those locations.

\subsection{Further Processing}

\subsubsection{Point Source Removal}

The SED models which we use to determine the PAH fraction assume a
stellar component in the Rayleigh-Jeans tail with $F_{\nu}
\propto \nu^2$.  In cases where this condition is not met, the
presence of a stellar point source can corrupt the results of our
fitting.  In the vast majority of cases we see no problems relating to
stellar point sources in the SMC, and thus we do not perform a
comprehensive point source extraction on the observations.  In
addition, we are not able to remove any unresolved stellar
contribution, so obtaining a map with the stellar component completely
removed would require more detailed modeling.  Instead, we mask out
bright point sources which do not have a well behaved SED in the
mid-IR.  These sources typically fall into one of two categories: very
bright stars which are saturated at one or more of the IRAC bands or
stars with non-typical infrared SEDs, such as YSOs or carbon stars.
We mask these objects out with a circular aperture which we fill in
with the local background value.

\subsubsection{Convolution and Alignment}\label{sec:conv}

The IRAC and MIPS mosaics are all convolved to match the resolution of
the 160 \micron\ observations using the kernels derived by
\citet{gordon08} and available from the SSC.  After convolution we
regrid the maps to match the astrometry and pixel scale of the 160
\micron\ mosaic.

For the IRS cubes, the convolution and alignment procedure involves a
few additional steps.  We convolve directly with the 160 \micron\ PSF
at each wavelength, which is much larger than the PSF of IRS even at
its long wavelength end, so wavelength dependence of the PSF makes
little difference to the final map.  After convolution we align the
cubes using the polygon clipping technique described in
\citet{sandstrom09}.  Throughout this process, we appropriately
propagate the uncertainty cubes produced by Cubism.  Although the AORs
are the largest possible size given the observation length limitations
on \emph{Spitzer}, the maps are only at most a few resolution elements
across at 160 \micron.  Thus, we take some care to make sure that the
spectra we use are only those where the PSF is not sampling regions
outside the observed cube.  For the following analysis we use only
spectra where 90\% of the area of the PSF or more is within the
observed cube, which cuts down the number of viable spectra we can
extract from the cubes to 63.  The small number of resolution elements
across the cube results in the introduction of scatter into the
comparison between the MIPS and IRAC mosaics and the spectral maps.
This is the result of information from outside the cube not being
``convolved in''.  This scatter represents a fundamental limitation of
our observations.  We estimate the magnitude of this scatter to be
$\sim 10$\% by comparing convolved and aligned IRAC 8.0 and MIPS 24
\micron\ maps that have been cropped to the coverage of the spectral
cubes to those that have not.  For the SMC B1 cube we do not convolve
or align the map because of the small number of resolution elements
and the loss of signal due to the subsequent regridding to match at
160 \micron.  Instead we extract the spectrum for the whole SMC B1
spectral map and extract matching photometry from the IRAC and MIPS
observations.  After convolution and alignment we apply the previously
mentioned correction factors to match the orders and stitch the cubes
together in the overlaping spectral regions.

\section{SED Model Fitting}\label{sec:models}

We use the SED models of \citet{draine07a} to determine the dust mass,
radiation field properties and PAH fraction using our MIPS and IRAC
mosaics in every independent pixel of the map where all of the MIPS
and IRAC measurements are detected above $3\sigma$. We also perform
simultaneous fits to photometry and spectroscopy in the regions
covered by S$^4$MC as described in Section~\ref{sec:photospectrofit}.
To distinguish between these two types of fits we introduce the
following terminology: ``photofit'' refers to the best-fit model using
only the photometry and ``photospectrofit'' refers to the best-fit
model to the combined photometry and spectroscopy.

The model fit involves searching through a pre-made grid of models and
finding the model which minimizes the following pseudo-$\chi^2$: 
\begin{equation}
\chi^2 = \sum_{b} \frac{\left(F_{\mathrm{obs},b} -
\langle F_{\nu,\mathrm{model}}\rangle_b\right)^2}{\sigma^2_{\mathrm{obs},b} +
\sigma^2_{\mathrm{model},b}} 
\end{equation} 
Here $F_{\mathrm{obs},b}$ is the observed surface brightness in band
$b$, $\langle F_{\nu,\mathrm{model}}\rangle_b$ is the model spectrum
convolved with the spectral response curve of band $b$,
$\sigma_{\mathrm{obs},b}$ is the uncertainty in the observed surface
brightness and $\sigma_{\mathrm{model},b}$ is a factor which allows us
to account for the systematics associated with the modeling.
Following \citet{draine07b} we use $\sigma_{\mathrm{model},b} = 0.1
F_{\nu,\mathrm{model},b}$.  

The radiation field heating the dust is described by a power-law plus
a delta function at the lowest radiation field:
\begin{equation}
\frac{dM_{D}}{dU} = (1-\gamma) M_{D} \delta (U - U_{min}) +
\gamma M_{D} \frac{(\alpha - 1)}{U_{min}^{1-\alpha} -
U_{max}^{1-\alpha}} U^{-\alpha}
\end{equation}
Here $U$ is the radiation field in units of the interstellar radiation
field in the Solar neighborhood from \citet{mathis83}, $\gamma$ is the
fraction of the dust heated by the minimum radiation field, $\alpha$
is the power-law index of the radiation field distribution and M$_{D}$
is the dust mass surface density.  This parametrization allows us to
approximately account for both the dust heated by the general
interstellar radiation field and the dust heated by nearby massive
stars.   

The adjustable parameters in the model are: the stellar luminosity per
unit area ($\Omega_{*}$), the dust mass surface density (M$_{D}$), the
PAH fraction (\qpah), the minimum and maximum radiation field
(U$_{min}$ and U$_{max}$), the fraction of dust heated by the minimum
radiation field ($1-\gamma$) and the exponent of the radiation field
power-law ($\alpha$).  $\Omega_{*}$, M$_{D}$ and $\gamma$ are
continuous variables which normalize the grid of models to match the
observations. The other variables are adjusted in discrete steps and
can have the range of values listed in Table~\ref{tab:models}.  Using
the results of the fit we compute a number of useful parameters
describing the outcome. These are $f_{PDR}$, the fraction of the total
infrared power radiated by dust grains illuminated by radiation fields
$U > 10^2$, and $\bar{U}$, the average radiation field.

\begin{deluxetable}{lcc}
\tablewidth{0pt}
\tabletypesize{\scriptsize}
\tablecolumns{3}
\tablecaption{\citet{draine07a} Model Parameters}
\tablehead{ \multicolumn{1}{l}{Parameter} &
\multicolumn{1}{c}{Range} }
\startdata
$\Omega_{*}$  & $> 0$  \\
M$_{D}$    & $> 0$  \\ 
$q_{\rm PAH}$ & 0.4$-$4.6\% \\ 
U$_{min}$     & 0.6$-$30 \\
U$_{max}$     & $10^3-10^7$ \\ 
$\gamma$      & $> 0$  \\   
$\alpha$      & 1.5$-$2.5     
\enddata
\label{tab:models}
\end{deluxetable}

In the analysis we present here, we use the \citet{draine07a} Milky
Way (MW) dust model, so we take a moment to justify and explain this
decision and its implications. In order to model the dust emission
spectrum, it is necessary to choose a physical model for the dust
which prescribes the abundances of different grains and their size
distributions.  At present, a dust grain size distribution model
specific to the SMC including a \emph{variable} PAH fraction and
conforming to constraints on the dust mass and raw materials available
has not been established, and producing such a model is complex and
beyond the scope of this paper \citep[it would require extending the
model by][]{weingartner01}.  Note that because we are primarily
interested in the PAH fraction, we must employ a model that includes
a size distribution that extends into the PAH regime.  In addition, we
aim to compare our results with other studies of the PAH fraction 
(particularly that on the SINGS sample) which use the Milky Way dust
model developed by \citet{draine07a} and \citet{draine07b}. 

There is evidence that the dust grain size distributions in the SMC
and the Milky Way are different
\citep{rodrigues97,weingartner01,gordon03}, and that low metallicity
galaxies in general may have an excess of small dust grains relative
to the Milky Way. These differences, however, should not affect our
measurements of the PAH fraction. \citet{draine07b} found that the
dust mass and PAH fraction from the best fit models for the galaxies
in the SINGS sample that fell in the range of \qpah\ covered by the
LMC/SMC dust models were not significantly different if the Milky Way
model was used instead, hence they employed Milky Way models.  Given
our desire to find the range of PAH fractions in the SMC and the
benefit of being able to compare our results directly with those of
\citet{draine07b}, we will fit MW dust models to the photometry. 

We expect the choice of the MW dust models will introduce some
systematic effects into the results of the model fitting.  In
particular, there is evidence that the SMC, like other low-metallicity
galaxies, has a larger contribution from small grains compared to the
Milky Way which results in ``excess'' emission at 24 and 70 \micron\
\citep{gordon03,bot04,galliano05,bernard08}.  The effect of such a
population on our modeling will be to increase the best-fit radiation
field in order to match the 24 and 70 \micron\ brightness.  Since we
use a power-law distribution of radiation fields, and the equilibrium
emission from large grains determines the minimum radiation field, the
general effect is to decrease the power-law exponent and increase
U$_{max}$, making a small fraction of the dust be heated by a more 
intense radiation field.  Because the PAH emission is produced by
single-photon heating, \qpah\ is simply proportional to the ratio of
the PAH emission in the 8 \micron\ band to the total far-IR emission,
which is robustly constrained by the 70 and 160 \micron\ photometry.
Hence, the derived \qpah\ is relatively insensitive to variations in
the U$_{min}$, U$_{max}$ and $\alpha$, provided the observed total
infrared luminosity is reproduced by the models. The regions with
overlapping spectroscopy will provide a good test for this reasoning,
since the $5-38$ \micron\ continuum in conjunction with the 24, 70 and
160 \micron\ SED provides more stringent constraints for the models.
We will revisit this subject in Section~\ref{sec:results}.

\subsection{The IRAC 4.5 \micron\ Br$\alpha$ Contribution}

The models we employ only calculate contributions to the IRAC and MIPS
photometry from starlight, dust continuum and dust emission features.
However, the emission spectrum from the ISM of the SMC is likely to
contain a number of emission lines, particularly from H II regions,
that contaminate the photometry.  Most of these lines are very weak
with respect to the dust continuum except in the case of the Brackett
$\alpha$ hydrogen recombination line at 4.05 \micron.  Recent work by
\citet{smith09} has shown that low-metallicity dwarf galaxies with
recent star formation can have a Br$\alpha$ contribution that is a
significant fraction of their integrated 4.5 \micron\ flux.  This
problem is exacerbated within the H II regions themselves, where the
4.5 \micron\ emission can be almost entirely from Br$\alpha$
\citep[see for instance work on M 17 by][]{povich07}.  

The SED models contain starlight continuum with a fixed 3.6 \micron\
to 4.5 \micron\ ratio. If there is excess 4.5 \micron\ emission due to
the hydrogen line, the starlight continuum will be too high, altering
the ratio of stellar to non-stellar emission at 5.8 and 8.0 \micron.
To correct for the Br$\alpha$ emission, we use the Magellanic Cloud
Emission Line Survey image of H$\alpha$ \citep[kindly provided by C.
Smith and F.  Winkler;][]{smith99} along with the Case B factors at
10,000 K from \citet{osterbrock06} to convert H$\alpha$ to Br$\alpha$.
This correction will underestimate the Br$\alpha$ emission where there
is extinction, but since the intrinsic extinction in the SMC is low to
begin with we expect this estimate of Br$\alpha$ to be adequate.  This
Br$\alpha$ correction is highest in H II regions, where we find
Br$\alpha$ emission can account for 20-30\% of the total 4.5 \micron\
emission.  Outside of H II regions, the correction is negligible.

\subsection{Additional Systematic Uncertainties}\label{sec:unc}

\citet{leroy07} investigated the integrated far-IR SED of the SMC,
using observations from IRAS, ISO, DIRBE and TOPHAT.  They found that
the MIPS 160 \micron\ photometry is high by approximately 25\%
compared to the values predicted by interpolating from the DIRBE
observations at 140 and 240 \micron, a significant offset given how
close in wavelength the DIRBE 140 and MIPS 160 bands are.  More recent
MIPS observations of the SMC obtained through the SAGE-SMC legacy
survey (Gordon et al 2010, in prep.) agree very well with the S$^3$MC
photometry, suggesting that this offset is real.  The source of this
offset is not known, it may be partly due to a calibration difference
between the instruments or to [C II] 158 \micron\ emission
contributing in the MIPS bandpass.  In their analysis, \citet{leroy07}
divided the S$^3$MC map by a factor of 1.25.  We do not use this
factor in our analysis.  Dividing the 160 \micron\ photometry by 1.25
would produce a lower PAH fraction from our analysis because the
70/160 \micron\ ratio would increase, the radiation field needed for
the large grains to acheive the necessary temperature would be higher
and, given the emission at 8 \micron, the PAH fraction necessary will
be lower.

If the contribution of the [C II] line is significant, there may be a
systematic offset in the PAH fraction we determine.  \citet{rubin09}
find that [C II] and 8 \micron\ emission are correlated in the LMC.
If regions of the SMC with 8 \micron\ emission have significant [C II]
emission in the 160 \micron\ band, we will systematically overestimate
the PAH fraction, since it will seem that the large grains are colder
than they really are.   Future Herschel observations of the [C II]
line in the SMC will help to understand what the level of
contamination of the 160 \micron\ observation.

\subsection{Simultaneous SED and Spectral
Fitting}\label{sec:photospectrofit}

In the following, we use the S$^4$MC data is to investigate the
reliability of the determination of \qpah\ from SEDs alone (i.e.
photofit).  The photospectrofit model fitting procedure is very
similar to the SED fits, and involves searching the same grid of
models for the best fit.  

In using SED fits to measure the PAH fraction, we must assume that the
IRAC 8.0 \micron\ band, which samples the 7.7 \micron\ PAH feature,
traces the total PAH emission.  This has been seen to be a good
assumption when looking at the integrated spectra of galaxies
\citep{smith07}, but variations in the relative band strengths can be
averaged out on galaxy-scales and the 7.7 \micron\ feature may not
trace the total PAH emission as effectively within individual
star-forming regions.  The simultaneous SED and spectral coverage
provided by S$^3$MC and S$^4$MC allow us to test the effectiveness of
SED fits to determine \qpah, since the photofit results will be solely
determined by the 7.7 feature, while the photospectrofit results will,
to first-order, match the total PAH emission in the mid-IR.  

The photospectrofit models will only reproduce to total PAH emission
to first order because the models use fixed spectral profiles for the
various PAH bands, and a fixed PAH ionization fraction.  There are
variations in the PAH bands observed in the SMC which are not
reproduced in detail by the models.  However, fitting the full mid-IR
spectrum in the S$^4$MC regions will approximately reproduce the total
PAH emission in all of the major mid-IR bands, and therefore be a
better tracer than the 7.7 feature alone.  Although the comparison
between the photofit and photospectrofit results is not the ideal way
to quantify the dependence of \qpah\ on the band ratios, it is the
best that can currently be done without further modeling which is
outside the scope of this paper.

Prior to fitting the IRS spectroscopy, we remove the emission lines
from the spectrum using the PAHFIT spectral fitting package
\citep{smith07}.  We also use the PAHFIT results to estimate
the uncertainties in the spectra employing the following procedure.
The IRS pipeline produces uncertainties based on the slope fits to
individual ramps and propagates these through the various reduction
steps.  These errors do not include any systematic effects, and
underestimate the scatter in our spectra significantly.  In addition,
the assumption that the uncertainties are random, uniform and
uncorrelated for propagation through our analysis does not apply:
striping in the spectral cube does not average out spatially or
spectrally.  To get a better idea of the uncertainties, we estimate
the average deviation of the spectrum from the best PAHFIT result in a
5 pixel-wide sliding window.  Although this technique may artificially
increase the uncertainties in regions that are poorly fit by PAHFIT,
we find that uncertainties estimated in this manner are much more
reasonable than those propagated from the pipeline.  Finally, because
the number of points associated with the spectrum is much larger than
the SEDs it was necessary to artificially increase the weight of the
SED points to achieve a decent fit.  We find that applying weighting
factors of 40 to all the photometric points yield reasonable joint
fits.

\section{Results}\label{sec:results}

\subsection{Photofit and Photospectrofit \qpah\ Consistency}

\begin{figure*}
\centering
\epsscale{1.0}
\includegraphics[width=4in]{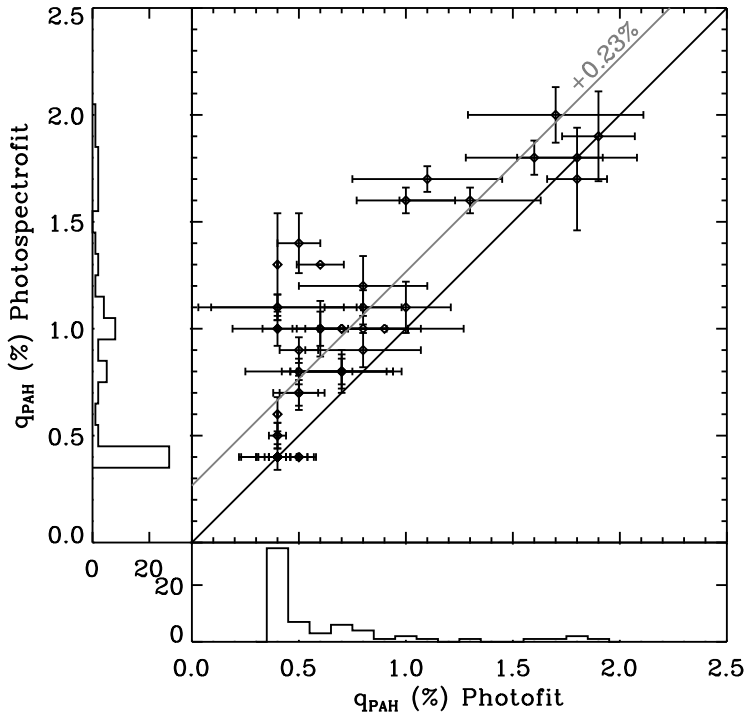}
\caption{\qpah\ from photofit and photospectrofit models.  
Because of the gridding of the models, a number of
points can overlap on this plot.  The gray line shows the average
offset between the two values excluding the points which have
\qpah= $0.4$\% while the black line shows a one-to-one relationship. 
Most of the points have a best fit value for
\qpah\ at the lower limit of the model range.}
\label{fig:photvspec}
\end{figure*}

Figure~\ref{fig:photvspec} shows the values of \qpah\ for the best fit
models in the regions with overlapping photometry and spectroscopy.
Because of the gridded nature of the models, there are typically a
number of points overlapping in the plot, so we additionally show a
histogram of the values on either axis.  The majority of points in our
spectral map regions have \qpah\ at the lower limit of the range and
the majority of those points yield the same value of \qpah\ from the
photofit and photospectrofit results.  At the high end of the
range of \qpah, we also see good consistency between
the photofit and photospectrofit values.  In the intermediate
regions, the photofit models tend to underestimate \qpah\ by a small
amount.  Excluding all of the points having the minimum PAH fraction,
the photospectrofit \qpah\ value is, on average, larger than the
photofit value by only $\Delta$\qpah$\approx 0.23$\%.

\begin{figure*}
\centering
\epsscale{1.1}
\plotone{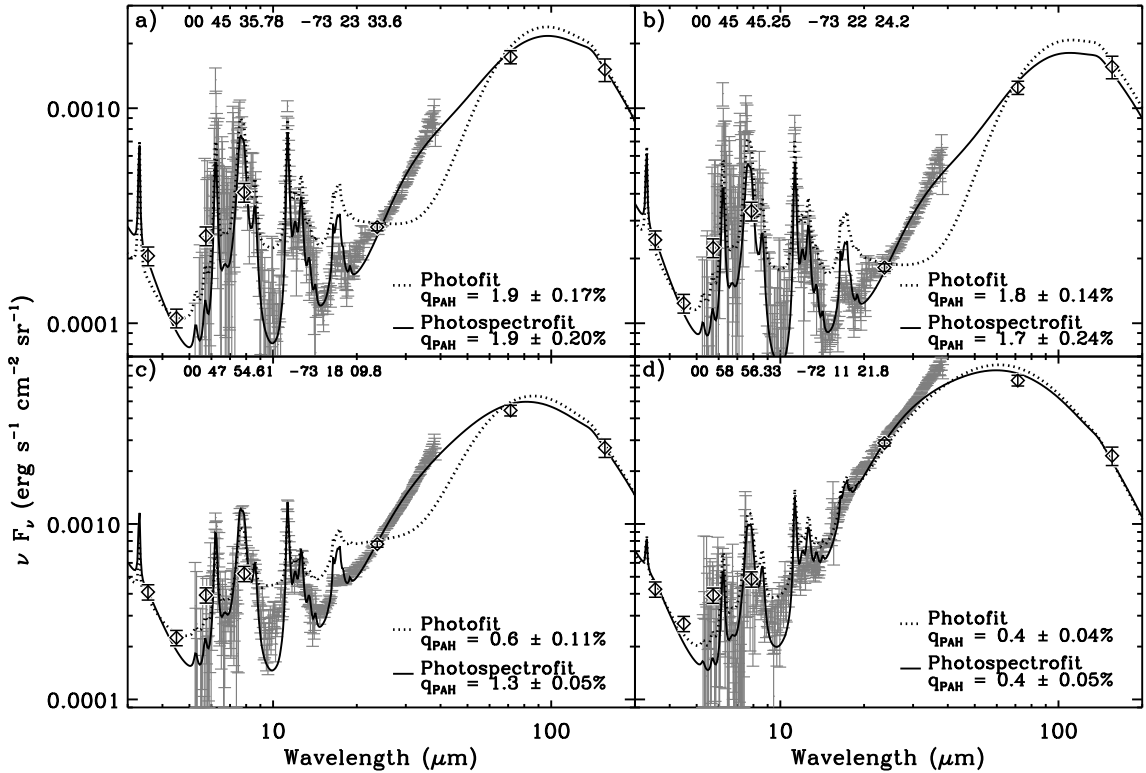}
\caption{Four examples of the photofit and photospectrofit models for
regions with overlapping photometry and spectroscopy. The IRS spectrum
is shown in gray and the MIPS and IRAC photometry are shown with
diamond-shaped symbols.  For clarity we have not overplotted the model
photometry.  The photofit model is shown as a dashed black line and the
photospectrofit model is shown as a solid black line.  The
differences between the models represent a redistribution of the
radiation field, the parameters of which are listed in
Table~\ref{tab:plotparams}.  The R.A. and Dec position of each of
these regions is listed in the upper left corner of the plot. Panel d)
shows a representative spectrum from the N 66 region.}
\label{fig:photvspecmod}
\end{figure*}

\begin{figure*}
\centering
\epsscale{1.1}
\plottwo{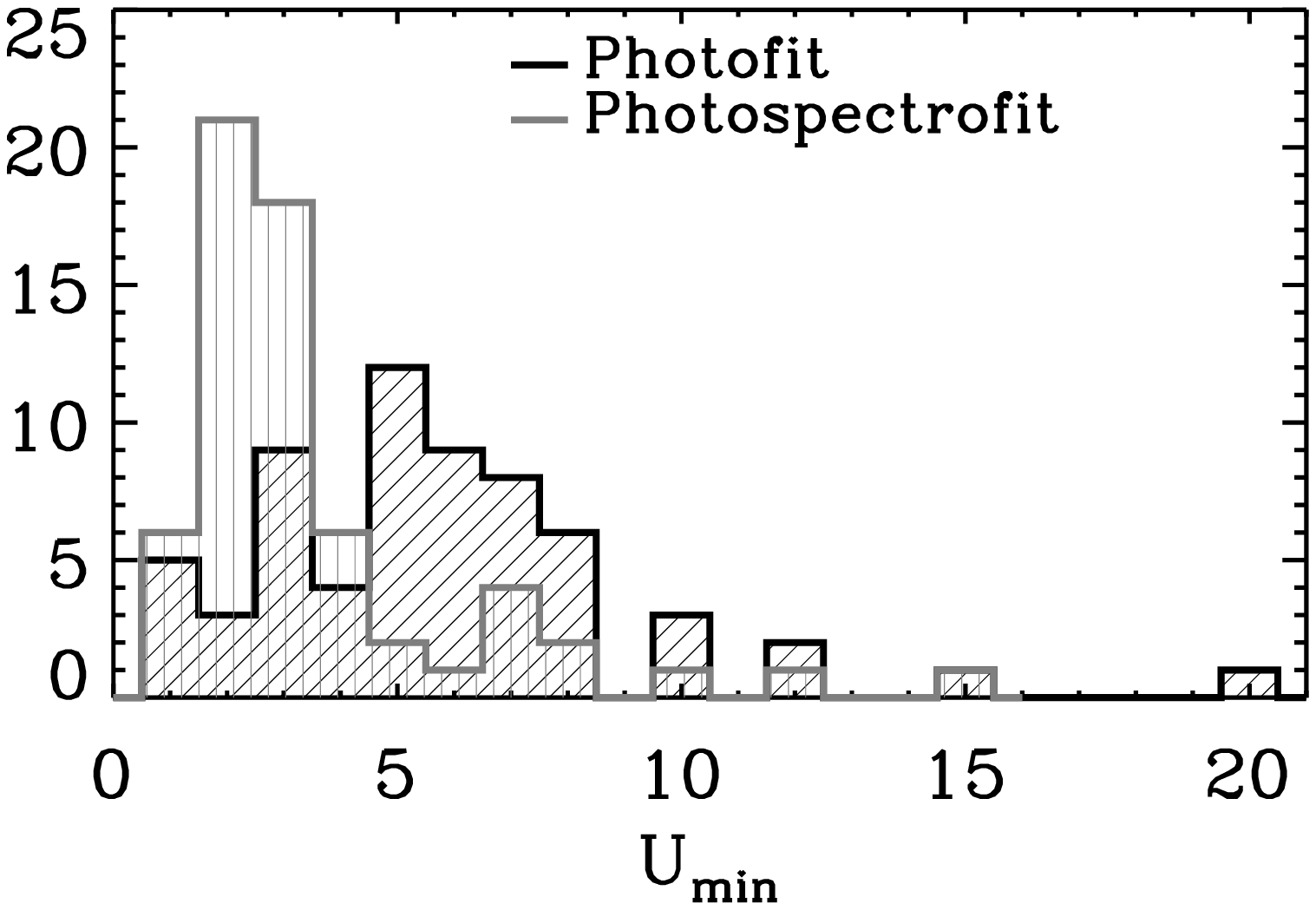}{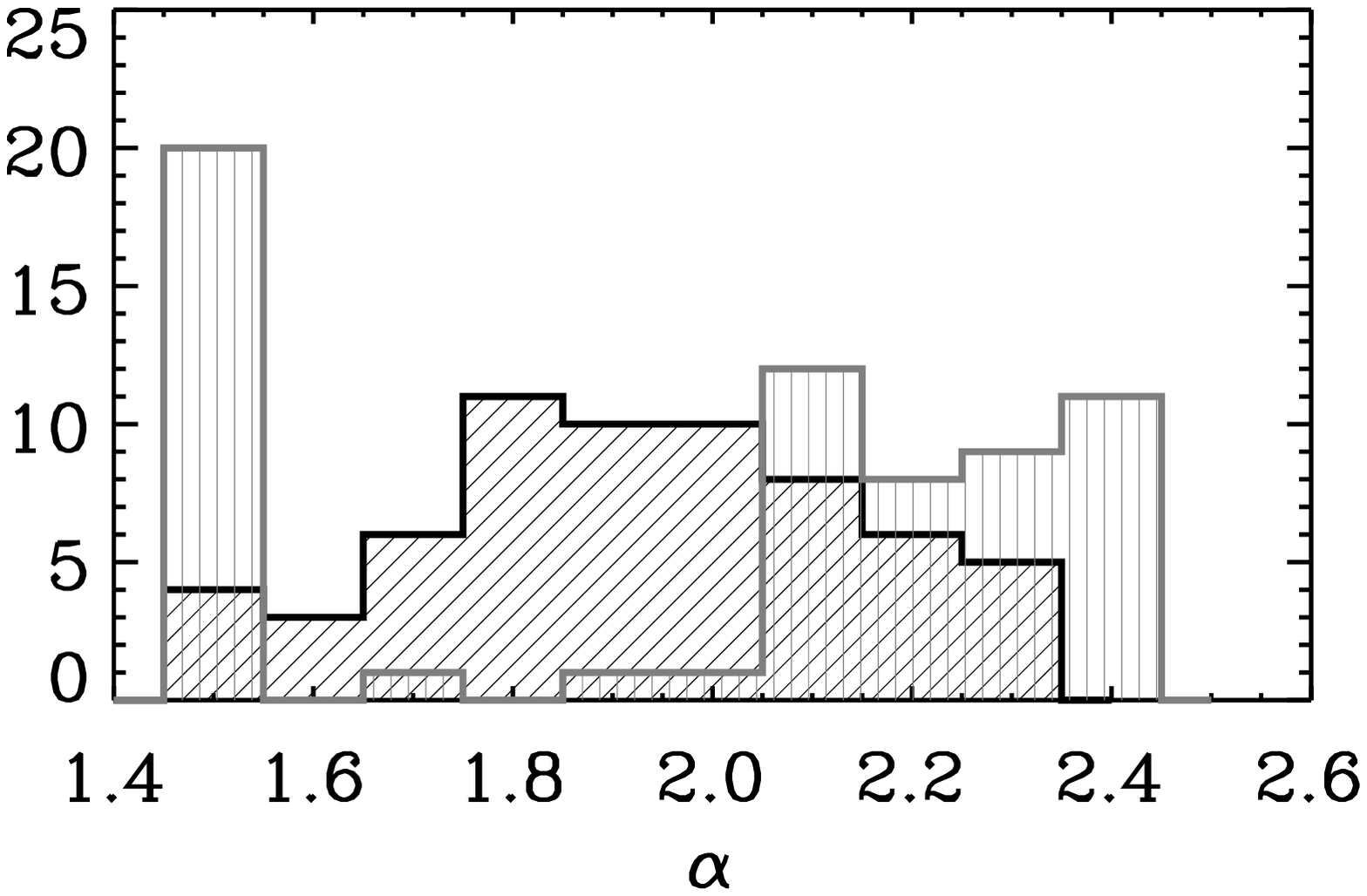}
\plottwo{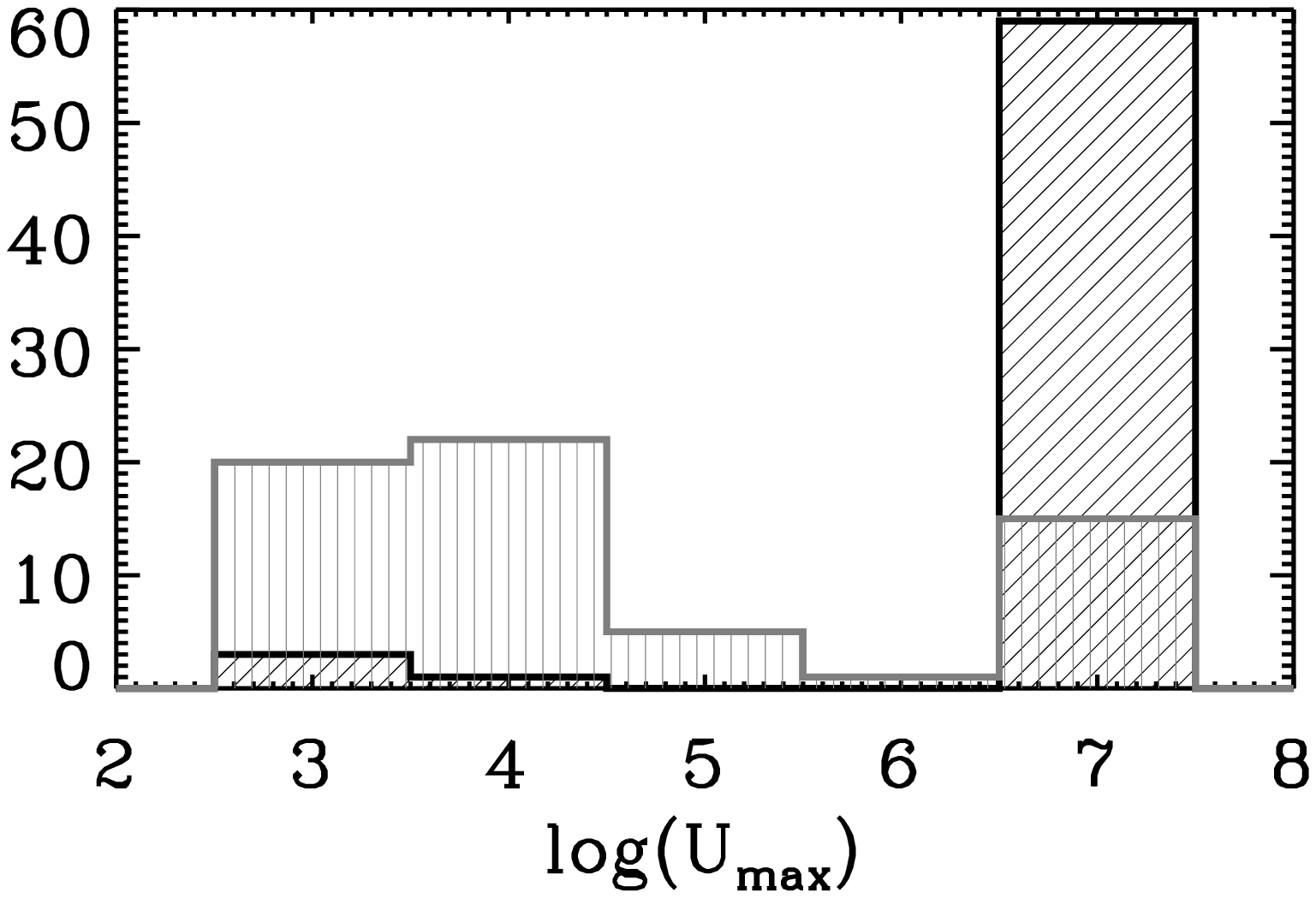}{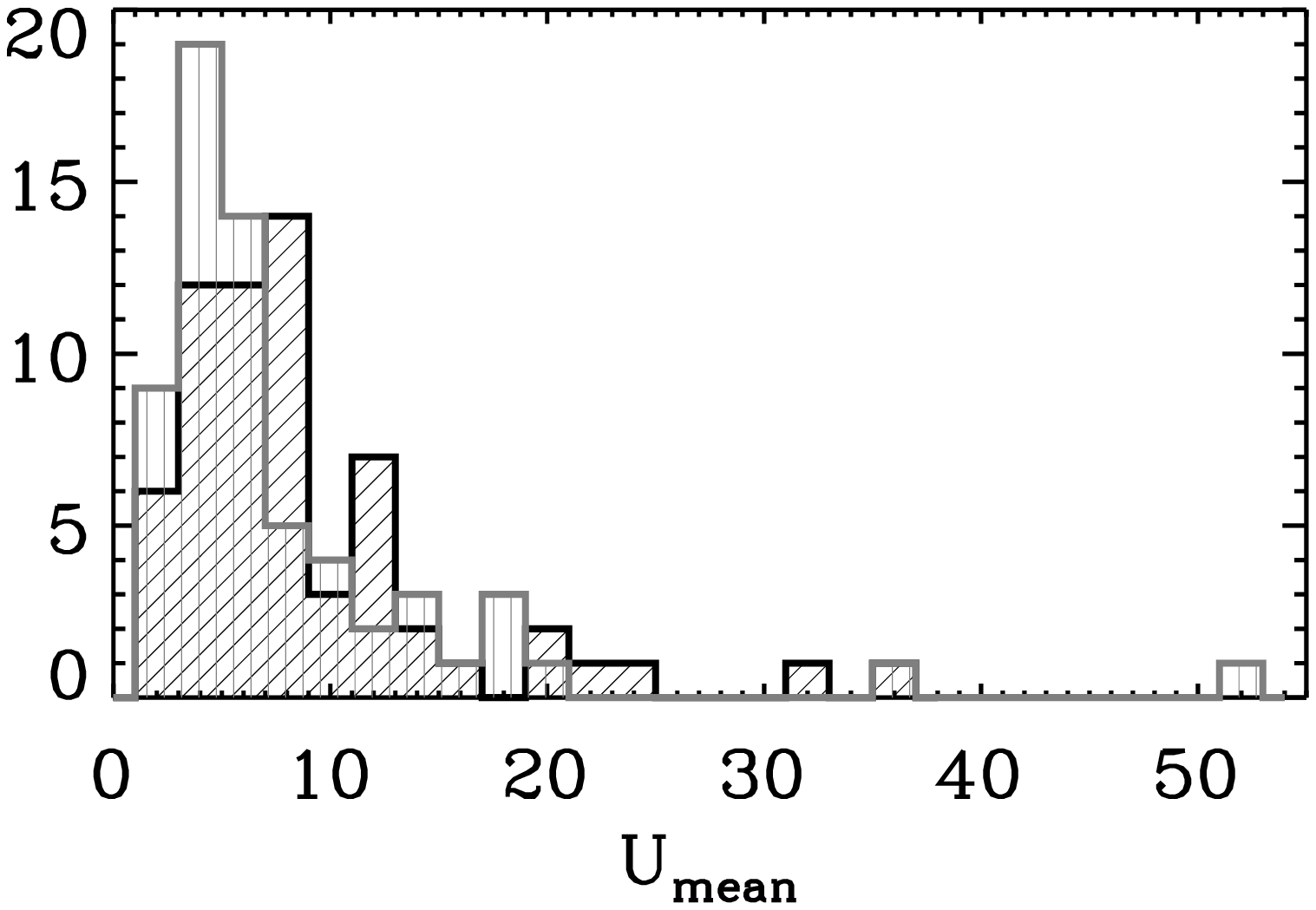}
\caption{Histograms illustrating the variation of the radiation field
paramters for the photofit models (in black) and photospectrofit
models (in gray). The mean radition field ($\bar{U}$) is shown in the
lower right panel.  Despite the redistribution of the radiation field,
the mean field is essentially unchanged in the two best fit models.}
\label{fig:photvspecall}
\end{figure*}

\begin{figure*}
\centering
\epsscale{1.0}
\includegraphics[width=4in]{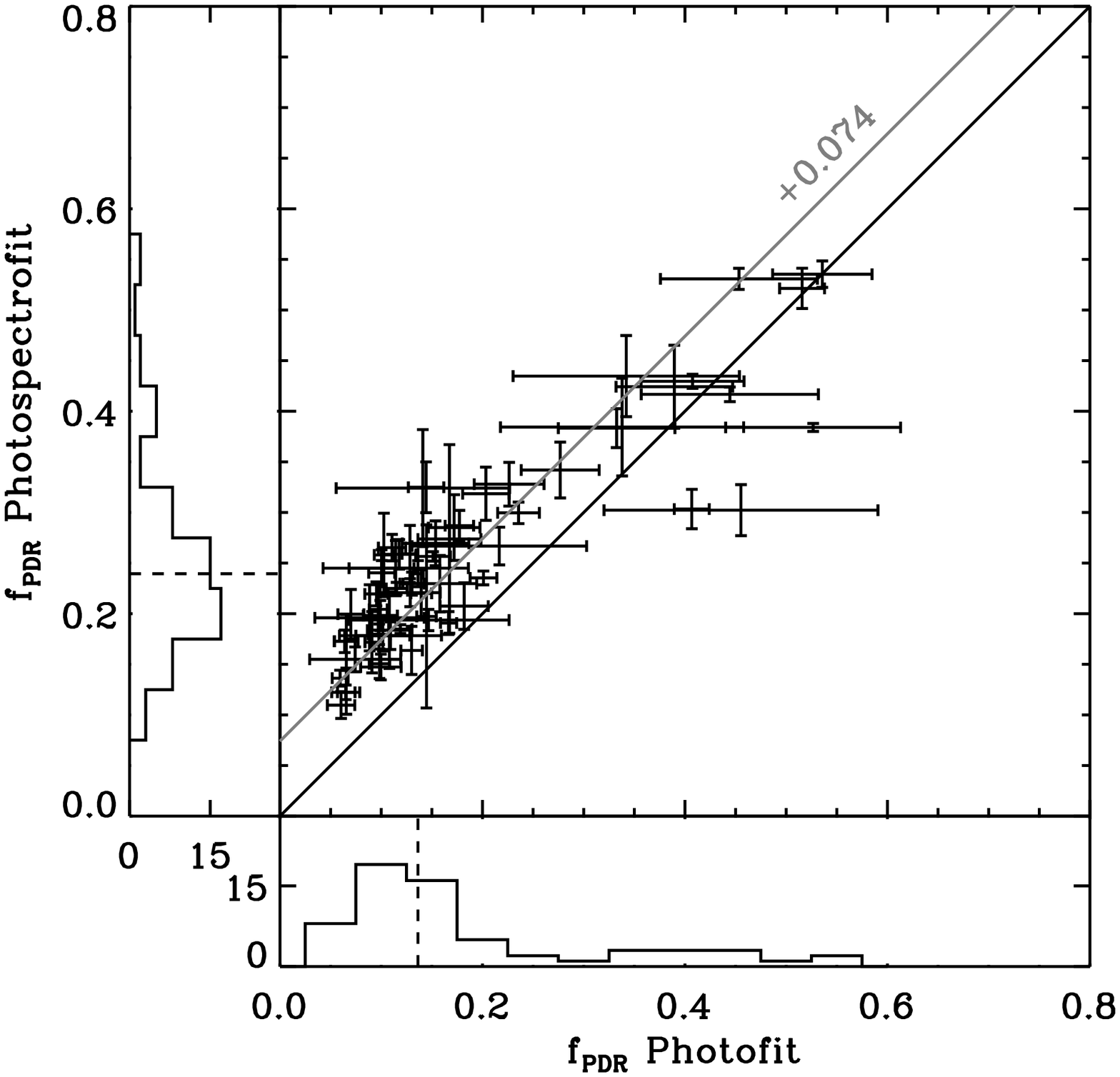}
\caption{A comparison of $f_{PDR}$ for the photofit and
photospectrofit models.  The dashed line on
the histograms illustrates the mean of the $f_{PDR}$ values.  On
average $f_{PDR}$ increases by 0.074 when the spectroscopic
information is included in the fit.}
\label{fig:photvspecfpdr}
\end{figure*}

Despite the good agreement between the photofit and photospectrofit
\qpah, the best fit models in the two cases show striking systematic
differences.  Figure~\ref{fig:photvspecmod} shows a comparison of a
few of the photofit and photospectrofit models, chosen to highlight
the range of \qpah\ we measure from the spectra.
Table~\ref{tab:plotparams} shows the best fit radiation field
parameters for the plotted SEDs and spectra.  In all cases, the
spectroscopic information shows that the mid-IR continuum below the
PAH features is lower than the photofit model predicts.  
The differences arise because the photofit models have the observed
SED unconstrained over the factor of $\sim 3$ gaps in wavelength
between 8 and 24 \micron\ and between 24 and 70 \micron, whereas the
photospectrofit models add continuous constraints on the SED from 5 to
38 \micron.  The photofit models tend to overpredict the continuum
between 8 and 24 \micron, and to underpredict the continuum between 24
and 70 \micron, using $\alpha$ values that are too small, and leading
to overestimation of $f_{PDR}$.  The overprediction of the $8-24$
\micron\ continuum leads to underprediction of \qpah, but as seen in
Figure~\ref{fig:photvspec}, the bias is not large, amounting to
$\Delta$\qpah$\approx 0.23$\%, although in some cases the errors are
larger. In general the dust surface density and stellar luminosity are
not changed in a systematic way.  What this amounts to is a
redistribution of the radiation field to increase the fraction of dust
that is being heated by very high radiation fields and to decrease the
radiation field necessary in the diffuse ISM.  In fact, the average
radiation field in the photofit and photospectrofit models is similar.
We show a series of plots illustrating these changes in
Figures~\ref{fig:photvspecall} and~\ref{fig:photvspecfpdr}.

\begin{deluxetable*}{lccccccccc}
\tablewidth{0pt}
\tabletypesize{\scriptsize}
\tablecolumns{10}
\tablecaption{Parameters of Selected Photofit and Photospectrofit Models Shown in Figure~\ref{fig:photvspecmod}}
\tablehead{ 
\multicolumn{1}{c}{} &
\multicolumn{4}{c}{Photofit} &
\multicolumn{1}{c}{} &
\multicolumn{4}{c}{Photospectrofit} \\
\cline{2-5} \cline{7-10} \\
\multicolumn{1}{c}{Panel} &
\multicolumn{1}{c}{$\alpha$} & \multicolumn{1}{c}{U$_{\rm min}$} & 
\multicolumn{1}{c}{U$_{\rm max}$} & \multicolumn{1}{c}{$\gamma$} &
\multicolumn{1}{c}{} &
\multicolumn{1}{c}{$\alpha$} & \multicolumn{1}{c}{U$_{\rm min}$} &
\multicolumn{1}{c}{U$_{\rm max}$} & \multicolumn{1}{c}{$\gamma$} }
\startdata
a & 1.70 $\pm$ 0.08 & 3.0 $\pm$ 1.0  & 7.0 &  4.9 $\pm 3.8 \times 10^{-4}$ & & 1.50 $\pm$ 0.70 & 2.0 $\pm$ 0.8  & 3.0 &  0.1 $\pm 4.1 \times 10^{-1}$ \\
b & 1.50 $\pm$ 0.04 & 1.5 $\pm$ 0.5  & 7.0 &  2.7 $\pm 2.2 \times 10^{-5}$ & & 1.70 $\pm$ 0.41 & 1.2 $\pm$ 0.2  & 3.0 &  0.2 $\pm 3.4 \times 10^{-1}$ \\
c & 1.80 $\pm$ 0.06 & 5.0 $\pm$ 1.1  & 7.0 &  2.1 $\pm 0.9 \times 10^{-3}$ & & 2.30 $\pm$ 0.01 & 2.0 $\pm$ 0.8  & 4.0 &  4.1 $\pm 2.9 \times 10^{-1}$ \\
d & 2.30 $\pm$ 0.19 & 5.0 $\pm$ 2.4  & 7.0 &  9.0 $\pm 3.0 \times 10^{-1}$ & & 2.20 $\pm$ 0.29 & 3.0 $\pm$ 12.  & 5.0 &  1.0 $\pm 0.7 \times 10^{0}$ 
\enddata
\label{tab:plotparams}
\end{deluxetable*}

As previously discussed, the value of \qpah\ is essentially
proportional to the ratio of the power in the 8 \micron\ PAH emission
features to the total far-IR power, and is therefore relatively
insensitive to variations in the other fitting parameters. The
agreement between the two best fit \qpah\ values is a strong
indication that the technique is robust even though our dust model is
not tailored for the SMC and does not have variable band ratios.  We
note that the spectroscopic maps are preferentially located in
star-forming regions, and most cover H II regions.  In these spots in
particular, the radiation field will deviate the most from the general
interstellar field.  Over the rest of the SMC, the increase in
$f_{PDR}$ will not be as dramatic.  We also note that the largest
differences in \qpah\ from the spectoscopic and photometric comparison
occur in N 22, which contains a point source that is highly saturated
at 24 \micron.  We have attempted to exclude the regions of N 22
affected by the saturation, but if there were excess 24 \micron\
emission from the PSF wings of this source contaminating the
photometry, it may artificially drive up the PDR fraction and change
\qpah\ more drastically.

Our conclusions from the comparison of the photofit and
photospectrofit models are that the \qpah\ values are in agreement.
The radiation field parameters from the photospectrofit models reflect
a redistribution of the radiation such that the average stays the same
while the fraction of dust heated by more intense fields increases and
the minimum field decreases.  These shifts are most likely not as
large in most regions across the SMC as they are in the star-forming
regions we probe with spectroscopy.  Finally, for regions with
intermediate values of \qpah\ we recognize the fact that we may
underestimate the PAH fraction by a few tenths of a percent on average,
however this small difference does not affect our conclusions.

\subsection{Results of the SED Models Across the
SMC}\label{sec:mapresults}

\begin{figure*}
\centering
\epsscale{1.0}
\includegraphics[width=6in]{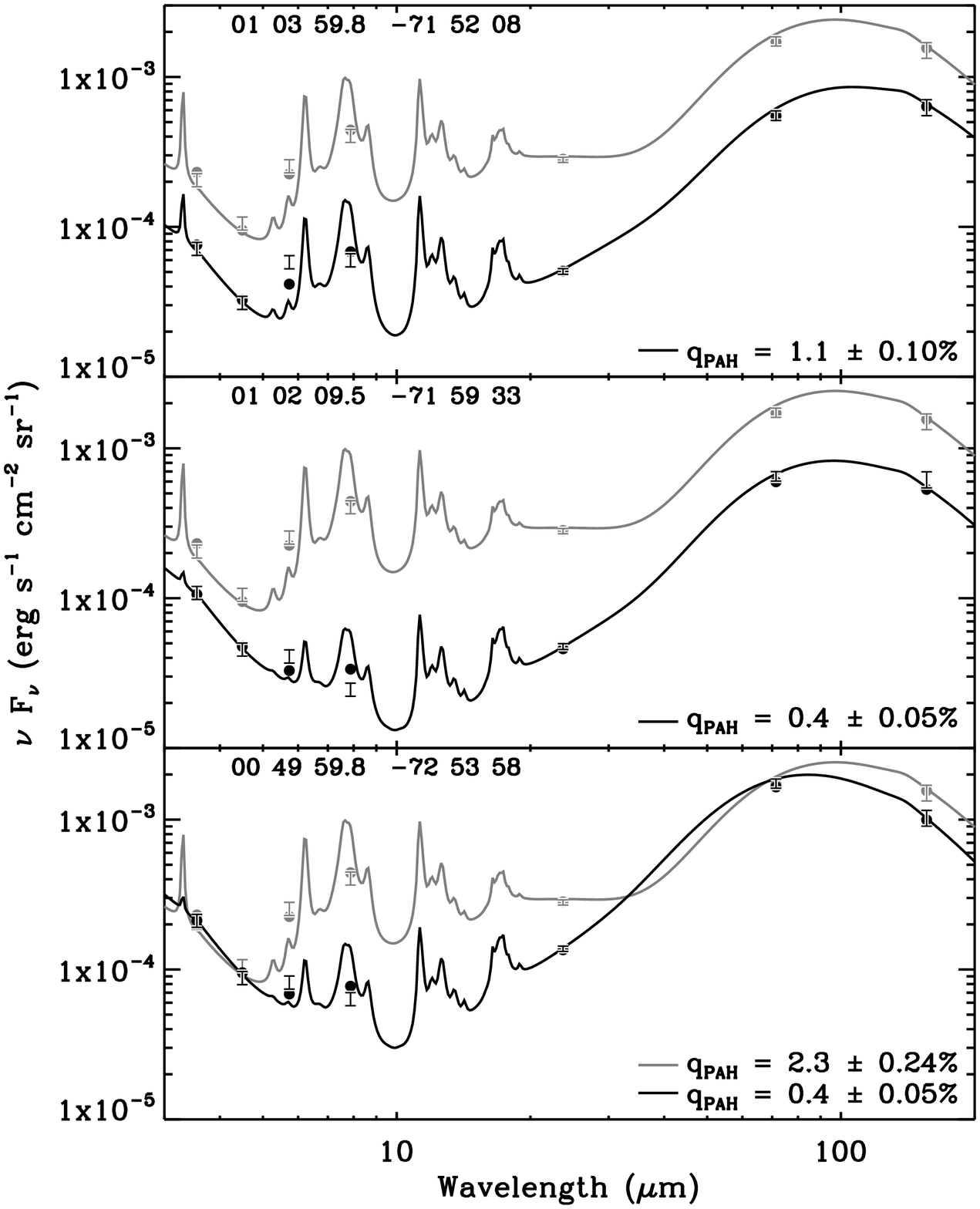}
\caption{Comparison of some representative SEDs and photofit models.
The positions of these SEDs are marked with square symbols on
Figure~\ref{fig:qpahmap}.  The gray line appears in each panel for
comparison and shows one of the highest \qpah\ values from the model
fits from the SW Bar which is marked on Figure~\ref{fig:qpahmap} with
a black square.  The location of the SEDs are listed in the top left
of the plot.  The measured photometric points are shown with error
bars and the synthetic photometry for the best fit model is shown with
a filled circle.  These panels illustrate the range of \qpah\ values
we see in the SMC.}
\label{fig:seds}
\end{figure*}

\begin{figure*}
\centering
\epsscale{1.0}
\includegraphics[width=5in]{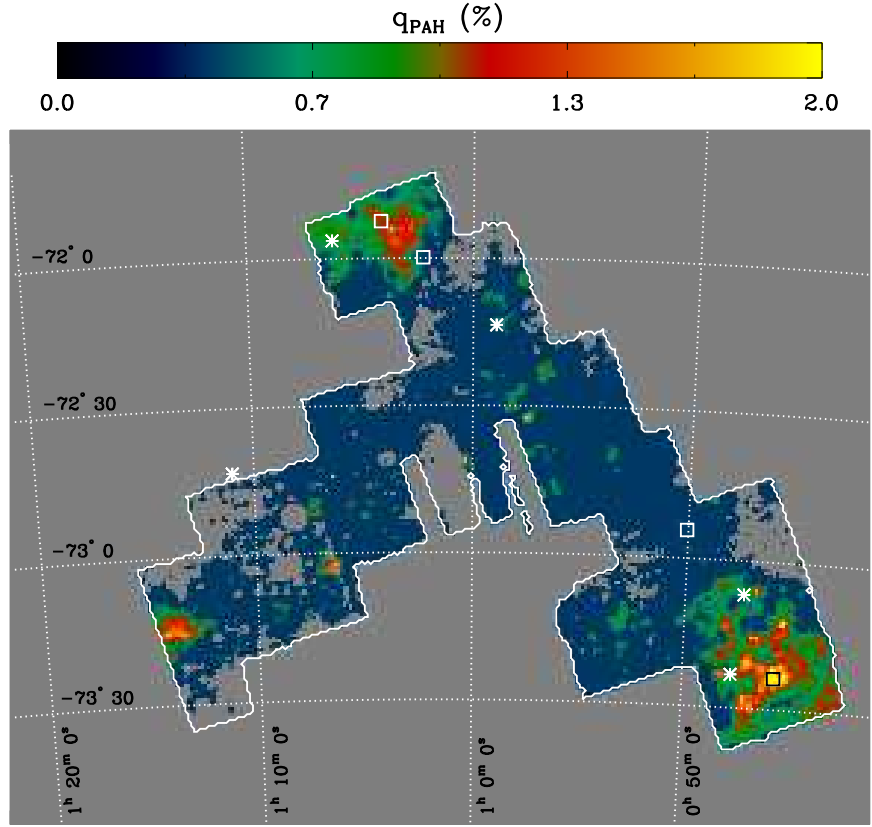}
\caption{This map shows the \qpah\ values from our fit to the
photometry at each pixel in the mapped region (40$\arcsec$
resolution).  The outer white boundary shows the overlapping coverage
of the MIPS and IRAC mosaics.  The white asterisks show the locations
of the stars with measured UV extinction curves \citep{gordon03}
discussed in Section~\ref{sec:bump}.  The
white squares show the locations of the SEDs plotted in in black in
Figure~\ref{fig:seds}. The black square in the SW Bar shows the
location of the SED plotted in gray in each panel of
Figure~\ref{fig:seds}.}
\label{fig:qpahmap}
\end{figure*}

In Figure~\ref{fig:seds} we show representative SEDs and photofit
models from four locations in the SMC and in Figure~\ref{fig:qpahmap} we
show the \qpah\ from the photofit models at every pixel in our map.  All
pixels in our map have \qpah\ less than the average Milky Way value
($q_{\rm PAH,MW}$ = 4.6\%). One of the noticable features of this map is
the large spatial variations in the PAH fraction, from essentially no
PAHs to approximately half the Milky Way PAH fraction in some
of the star-forming regions---a range that spans nearly an 
order-of-magnitude. 

The average PAH fraction in the region we mapped is $\langle$\qpah
$\rangle = 0.6$\%, determined by the following average:
\begin{equation} \langle q_{\mathrm{PAH}}\rangle = \frac{\sum_{j}
q_{\mathrm{PAH},j} M_{\mathrm{D},j}}{\sum_{j} M_{\mathrm{D},j}} .
\end{equation} Given the minimum value of \qpah\ allowed by our
models, this value is in good agreement with the SW Bar average
determined by \citet{li02} but is 8 times lower than the average from
\citet{bot04}. 

\begin{figure*}
\centering
\epsscale{1.0}
\includegraphics[width=5in]{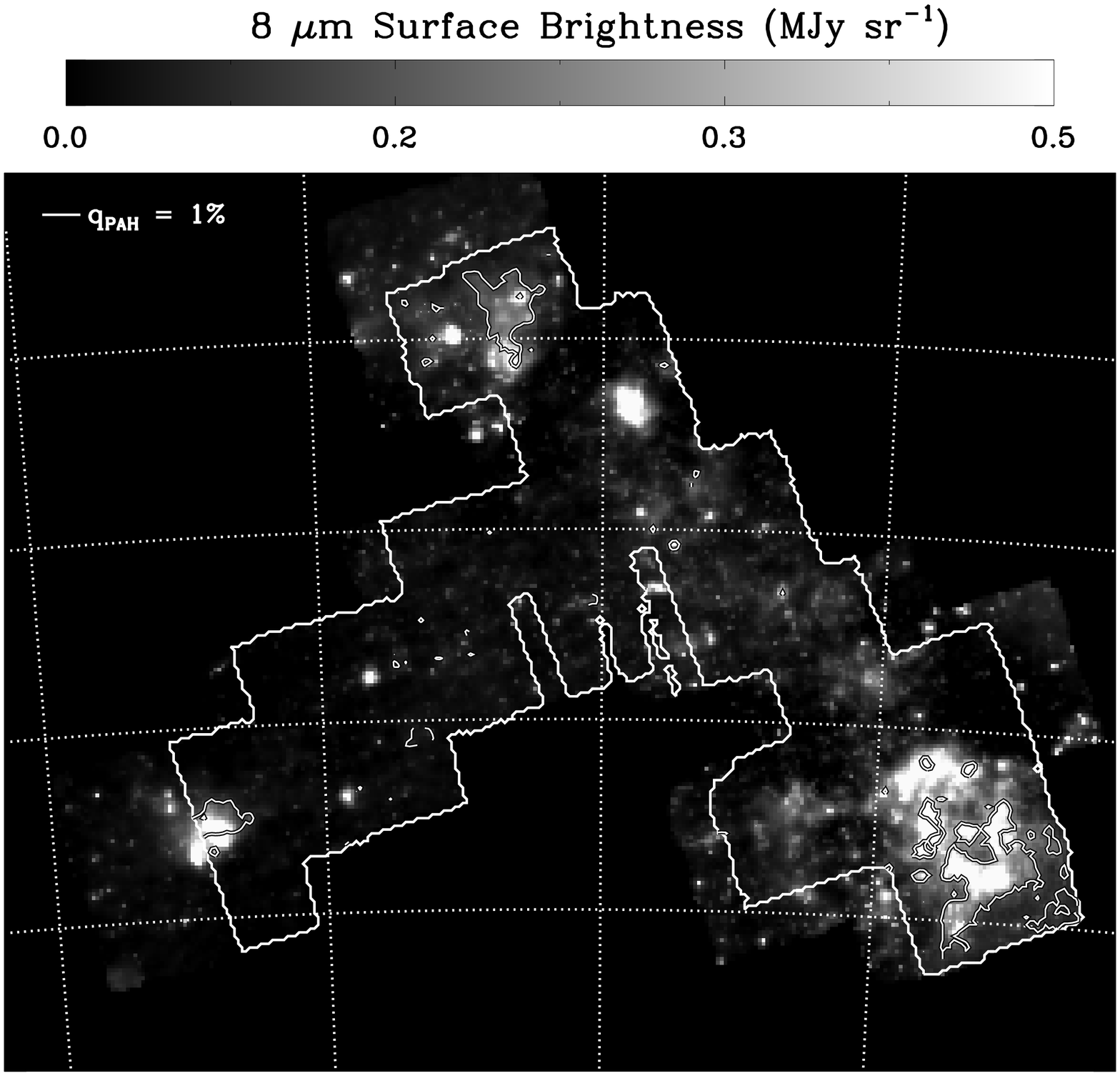}
\caption{A map of 8 \micron\ emission overlayed with the 1\% contour
of \qpah.  There are a number of regions that are very bright at 8
\micron\ that do not have high PAH fractions, particularly N 66 and
the northern part of the SW Bar.}
\label{fig:map8comp}
\end{figure*}

Previous studies of low metallicity dwarf galaxies have shown large
variations in 8 \micron\ surface brightness
\citep[e.g.][]{cannon06a,jackson06,hunter06,walter07} which we also
identify in the SMC.  Some of the regions that are brightest at 8
\micron\ have relatively low PAH fractions \citep[c.f.][]{cannon06a}.
To illustrate, we overlay the \qpah\ map on the IRAC 8 \micron\ mosaic
in Figure~\ref{fig:map8comp}.  There are a number of regions where 8
\micron\ emission is very bright while the PAH fraction is low. In
particular, N 66 and the Northern region of the SW Bar stand out as
very bright 8 \micron\ sources which have relatively low \qpah.  A
representative spectrum of N 66 is shown in the bottom panel of
Figure~\ref{fig:photvspecmod}, illustrating the low \qpah\ in this
region.  

\begin{figure*}
\centering
\epsscale{1.0}
\includegraphics[width=5in]{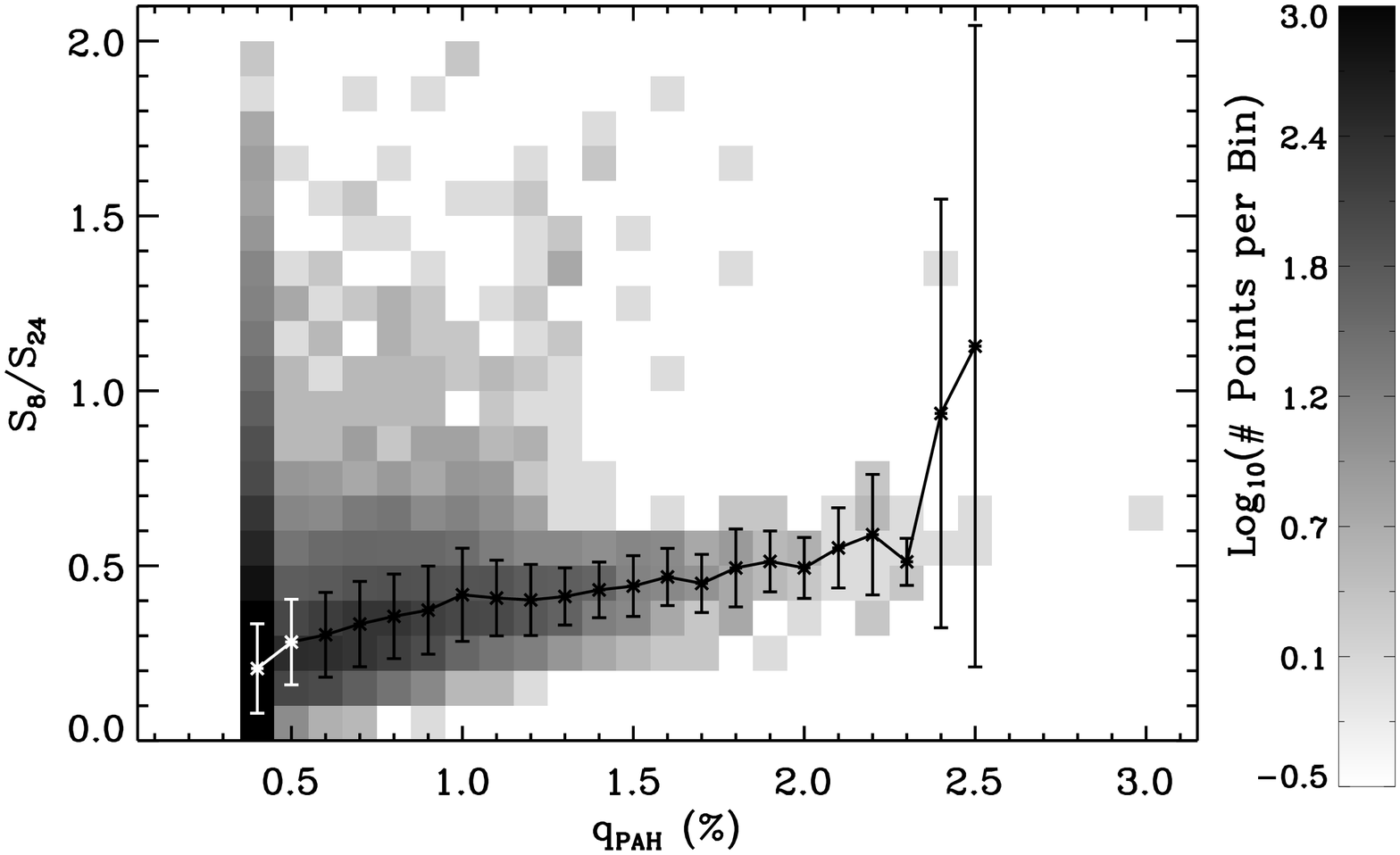}
\caption{This figure shows a two-dimensional histogram of the 8/24
ratio vs \qpah\ overlayed with the binned average of the 8/24 ratio
and error bars representing the standard deviation of the scatter at
each \qpah\ bin.  The color scale shows the number of points in each
bin.}
\label{fig:binned8to24}
\end{figure*}

To evaluate the use of the 8/24 ratio as an indicator of PAH
fraction, we show in Figure~\ref{fig:binned8to24} a plot of \qpah\ vs
8/24, with the average for each value of \qpah\ overlayed.  The PAH
fraction is correlated with the 8/24 ratio, even on small scales in
the SMC. However, it is evident that the correlation is weak, with a
large range in \qpah\ for a given value of the 8/24 ratio.  

The 8/24 ratio spans the region where \citet{engelbracht05} see a
transition from galaxies with evidence for PAH emission to galaxies
which show no PAH features, which is what one would expect given the
metallicity of the SMC.  The results of our SED modeling indicate that
the PAH fraction, though always lower than \qpah$_{\rm ,MW}$, is not
uniform across the SMC.  There are regions with \qpah\ within a factor
$\sim 2$ of \qpah$_{\rm ,MW}$\ and regions where \qpah\ is at the
lower limit of the \citet{draine07a} models ($\sim$ \qpah$_{\rm
,MW}/10$).  Since there are not comparable resolved maps of \qpah\ in
a sample of galaxies spanning this transition zone, we cannot explain
the trend in PAH fractions by looking at the SMC alone.  However, the
global PAH fraction we measure for the SMC ($\langle$\qpah$\rangle
\sim 0.6$\%) is driven by the very low \qpah\ over the majority of the
galaxy and the large regional variations in \qpah\ make it unlikely
that we are observing a uniform decrease in the SMC PAH fraction.
\emph{If the SMC is typical of galaxies at this metallicity, the
transition represents a decrease in the filling factor of the PAH-rich
regions, rather than a uniformly low global PAH fraction.}  

\begin{figure*}
\centering
\epsscale{1.0}
\plottwo{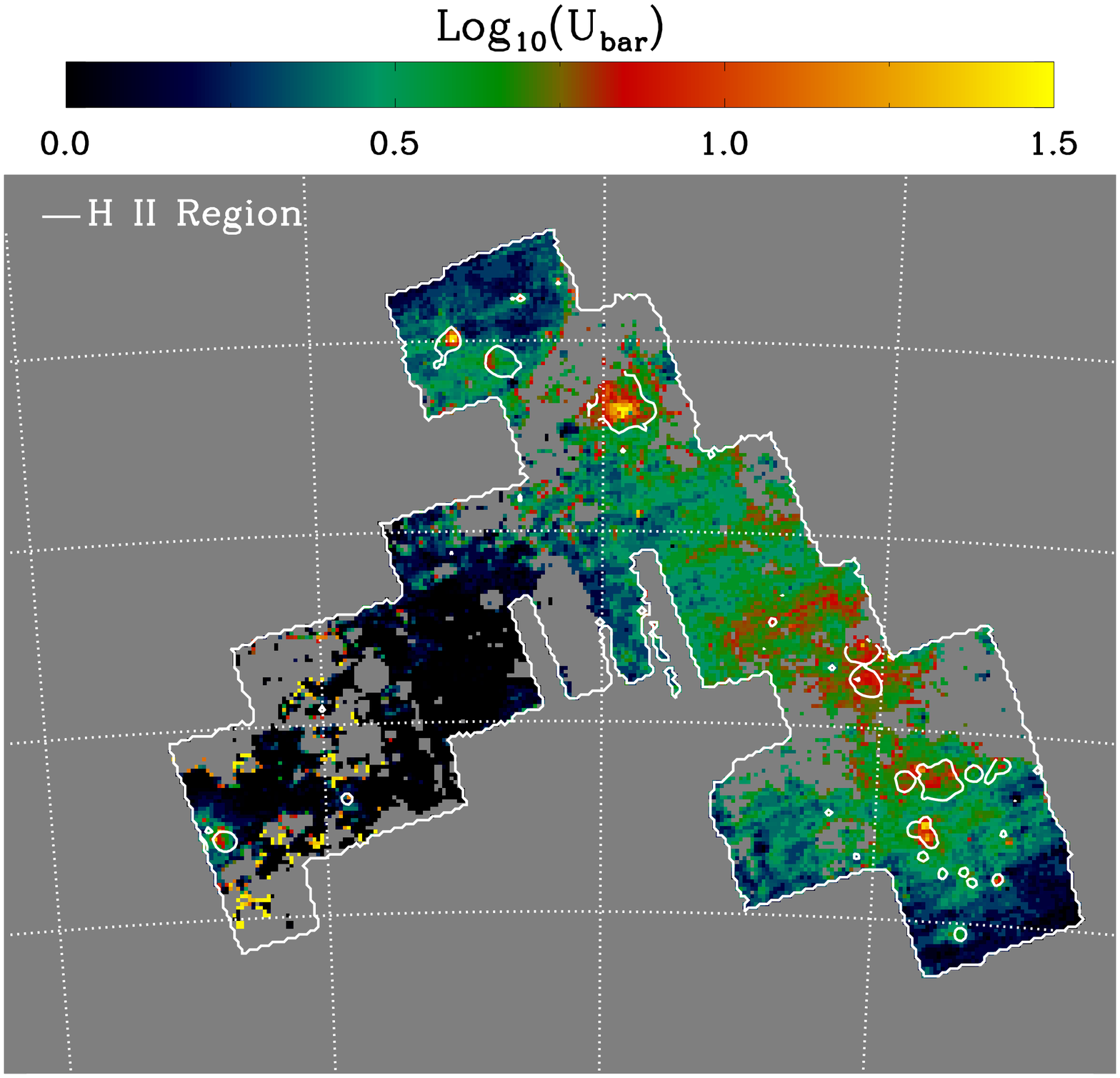}{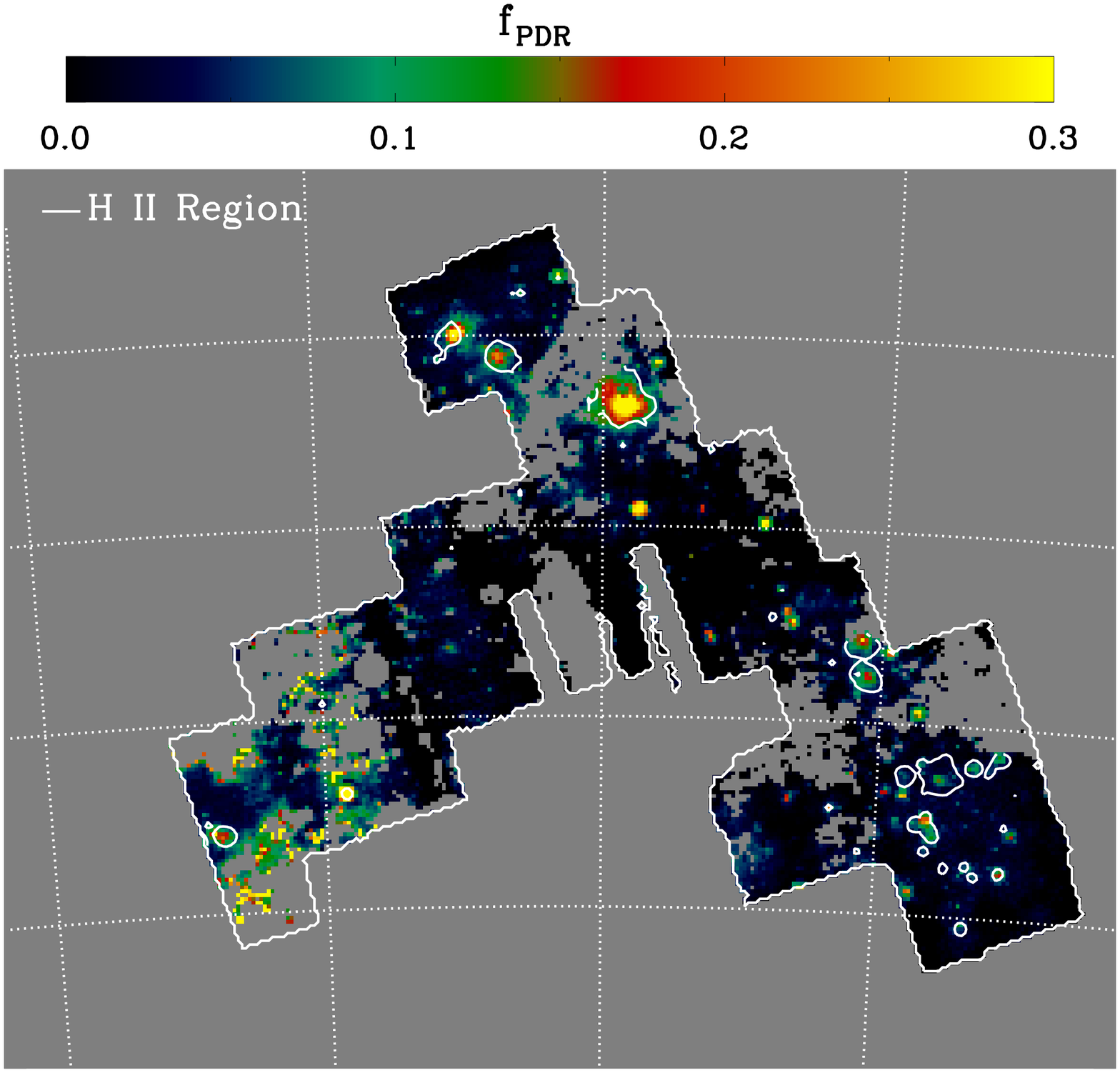}
\caption{Maps of the average radiation field ($\bar{U}$) and PDR
fraction ($f_{PDR}$) from the photofit models.  We have overlayed one
representative contour of the MCELS H$\alpha$ image to illustrate the
locations of the brightest H II regions in the SMC.  The
correspondence between $f_{PDR}$ and the location of H II regions is
very good, as expected.}
\label{fig:radiation}
\end{figure*}

\begin{figure*}
\centering
\epsscale{0.75}
\plotone{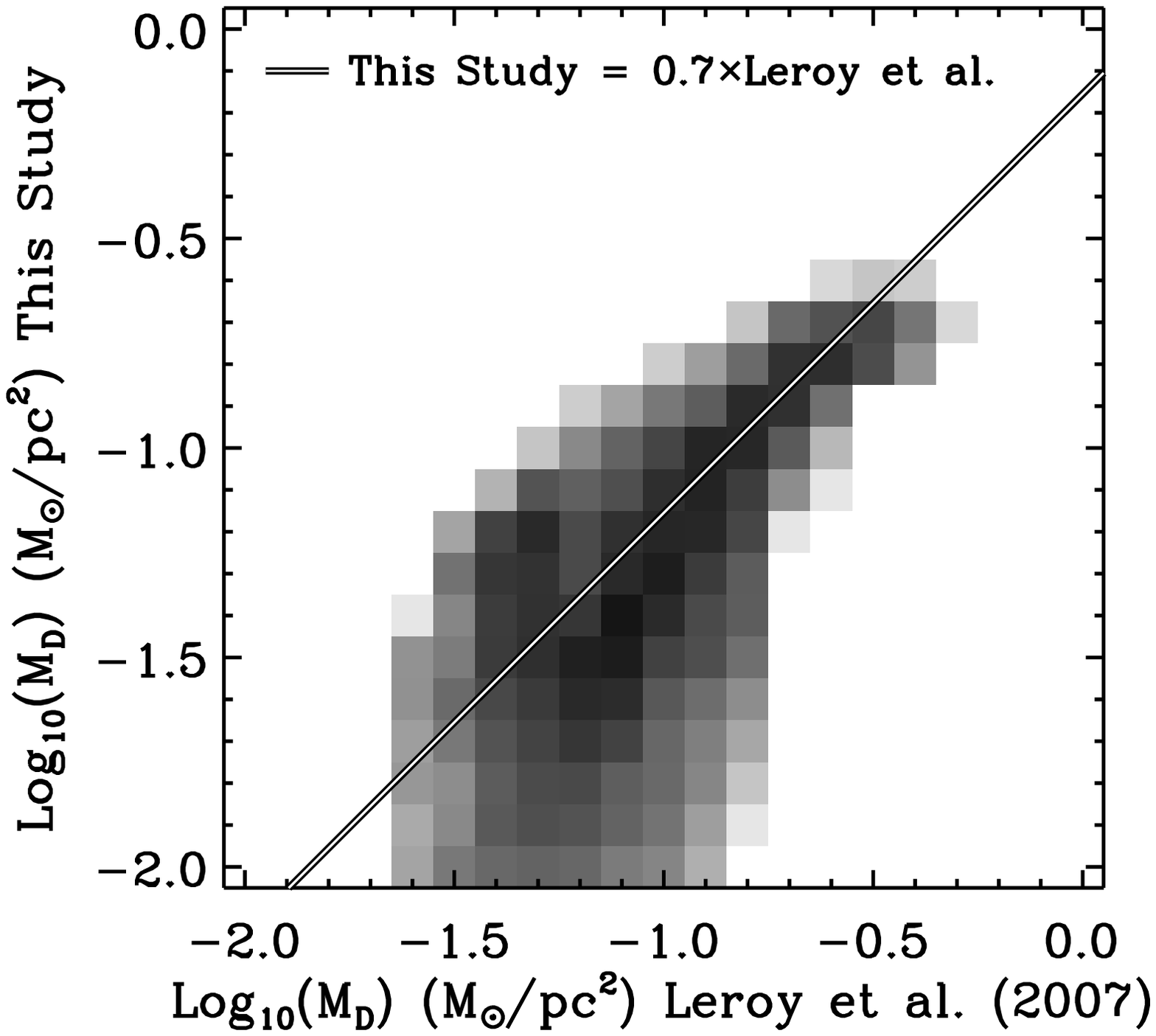}
\caption{A two-dimensional histogram showing the comparison of the
dust mass surface density M$_{D}$ from our study with that of
\citet{leroy07}.  The gray-scale, with a linear stretch, shows the
density of points in the plot.  Our M$_{D}$ is approximately 70\% of
that found by \citet{leroy07}, using the same MIPS data but different
methodolgy.  We note that the scatter at low surface densities likely
relates to the fact that we performed our analysis at 40\arcsec\
resolution but convolved to 2.6\arcmin\ for comparison with
\citet{leroy07} and we would expect higher signal-to-noise if the
analysis had been performed in the opposite order.}
\label{fig:compwadam}
\end{figure*}

The SED fits provide a number of parameters describing the radiation
field.  In Figure~\ref{fig:radiation}, we show two panels which
illustrate the average radiation field $\bar{U}$ and the PDR fraction
$f_{PDR}$. Regions where the PDR fraction is high tend to correspond
to H II regions, as expected. We overlay representative contours of
H$\alpha$ on the map of $f_{PDR}$ and $\bar{U}$ to highlight the
brightest H II regions in the Cloud.  We also measure the total dust
mass and the stellar luminosity at each pixel.  We find that the dust
mass from our fits agrees well with previous results from
\citet{leroy07} using the same MIPS observations despite different
methodologies.  Figure~\ref{fig:compwadam} shows a comparison of our
dust mass results with those of \citet{leroy07} and we find that our
masses are lower by $\sim 30$\%, well within the $\sim 50$\%
systematic uncertainties claimed by \citet{leroy07}.  

\subsection{Spatial Variations in the PAH Fraction}

Since we see clear spatial variations in the PAH fraction, we discuss
in the following section what sort of conditions foster high PAH
fractions in the SMC. We observe three trends: (1) the PAH fraction is
high in regions with high dust surface densities and/or molecular gas
as traced by CO, (2) the PAH fraction is low in the diffuse
interstellar medium and (3) the PAH fraction is depressed in H II
regions.  

\begin{figure*}
\centering
\epsscale{1.0}
\includegraphics[width=6in]{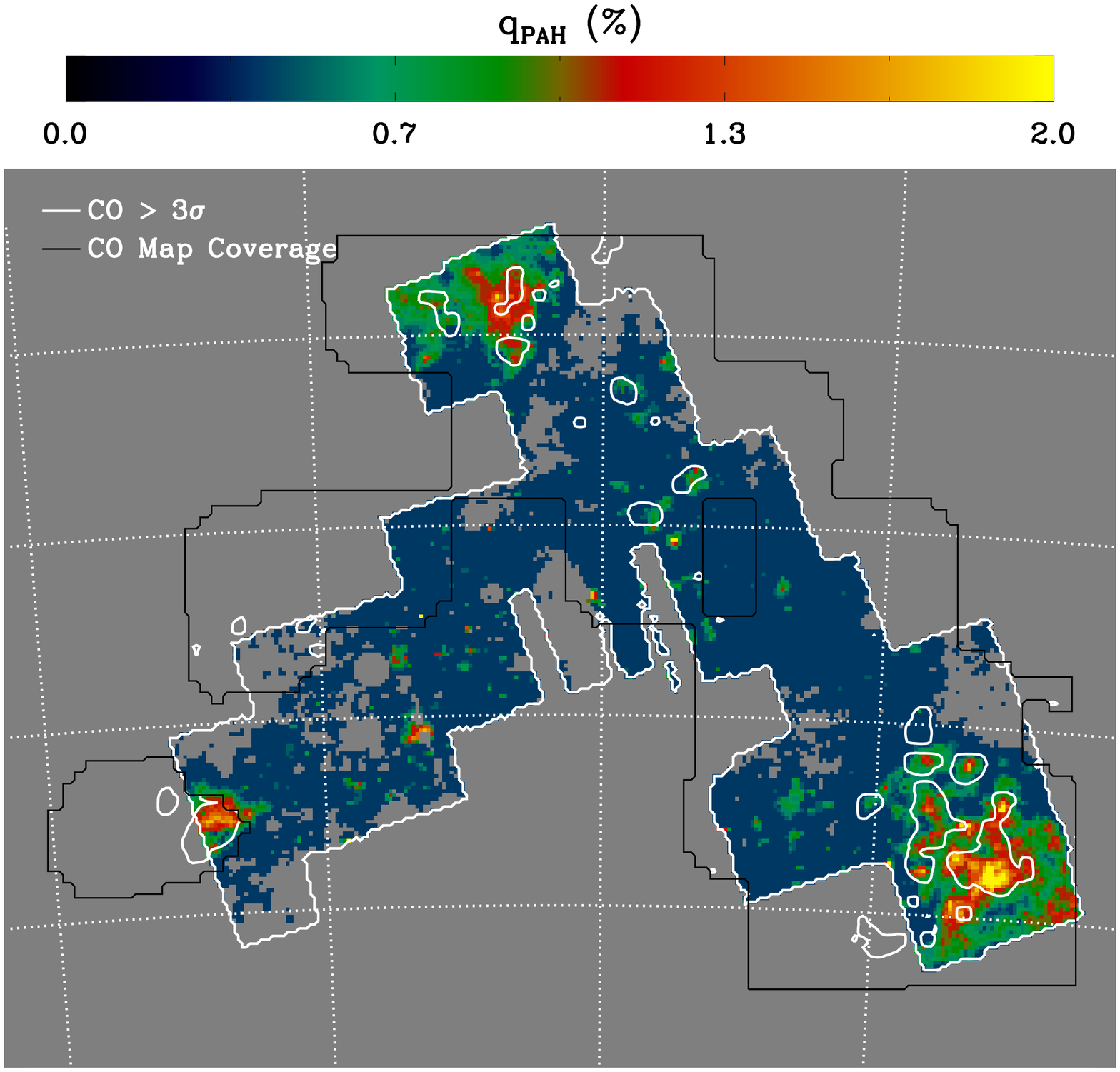}
\caption{In this figure we show the \qpah\ map from Figure~\ref{fig:qpahmap}
with contours of 3$\sigma$ CO (J=1$-$0) emission from the NANTEN survey overlayed.
The thin black line represents the coverage of the NANTEN survey. 
The NANTEN observations have a resolution of 2.6$\arcmin$.}
\label{fig:cocomp}
\end{figure*}

\begin{figure*}
\centering
\epsscale{1.0}
\includegraphics[width=6in]{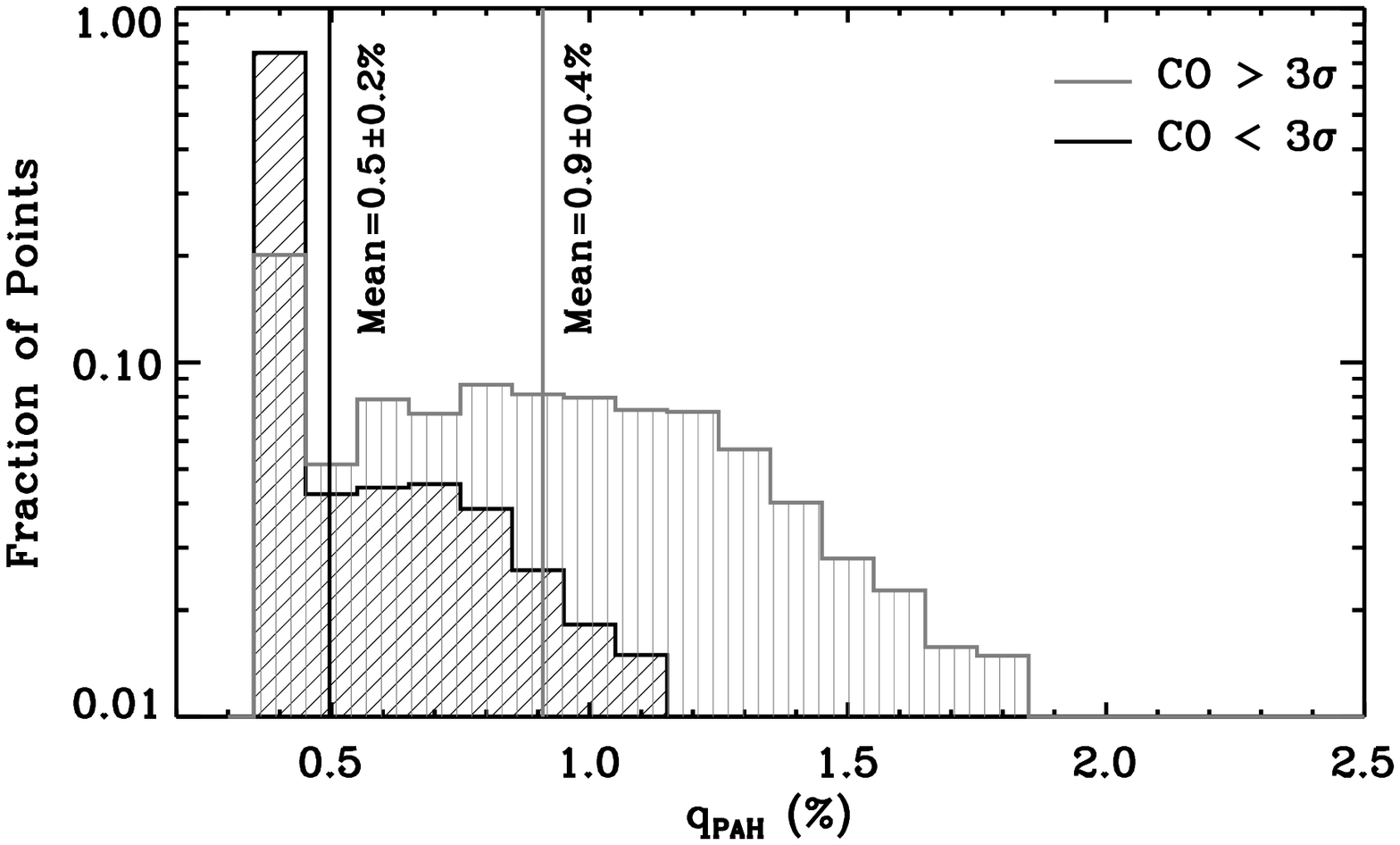}
\caption{Histogram of \qpah\ in lines-of-sight with and without
detected CO emission from the NANTEN survey.}
\label{fig:coqpah}
\end{figure*}

In Figure~\ref{fig:cocomp} we show the \qpah\ map overlayed with
contours of $^{12}$CO (J$=1-0$) emission from the NANTEN survey
\citep{mizuno01}. There is a strong correlation between the presence
of molecular gas and the regions with higher PAH fraction.  In
Figure~\ref{fig:coqpah} we show a histogram of \qpah\ for
lines-of-sight with detected CO emission in the NANTEN map of the SMC
and those without.  The mean value of \qpah\ for a line of sight with
CO is twice that for a line of sight without CO.  In addition, there
are no lines of sight through only atomic gas which have \qpah\ higher
than $\sim 1$\%. We note that the CO map has much lower resolution
than our map of \qpah, so the association of PAHs with CO is most
likely stronger than the evidence we present here.

\begin{figure*}
\centering
\epsscale{1.0}
\plottwo{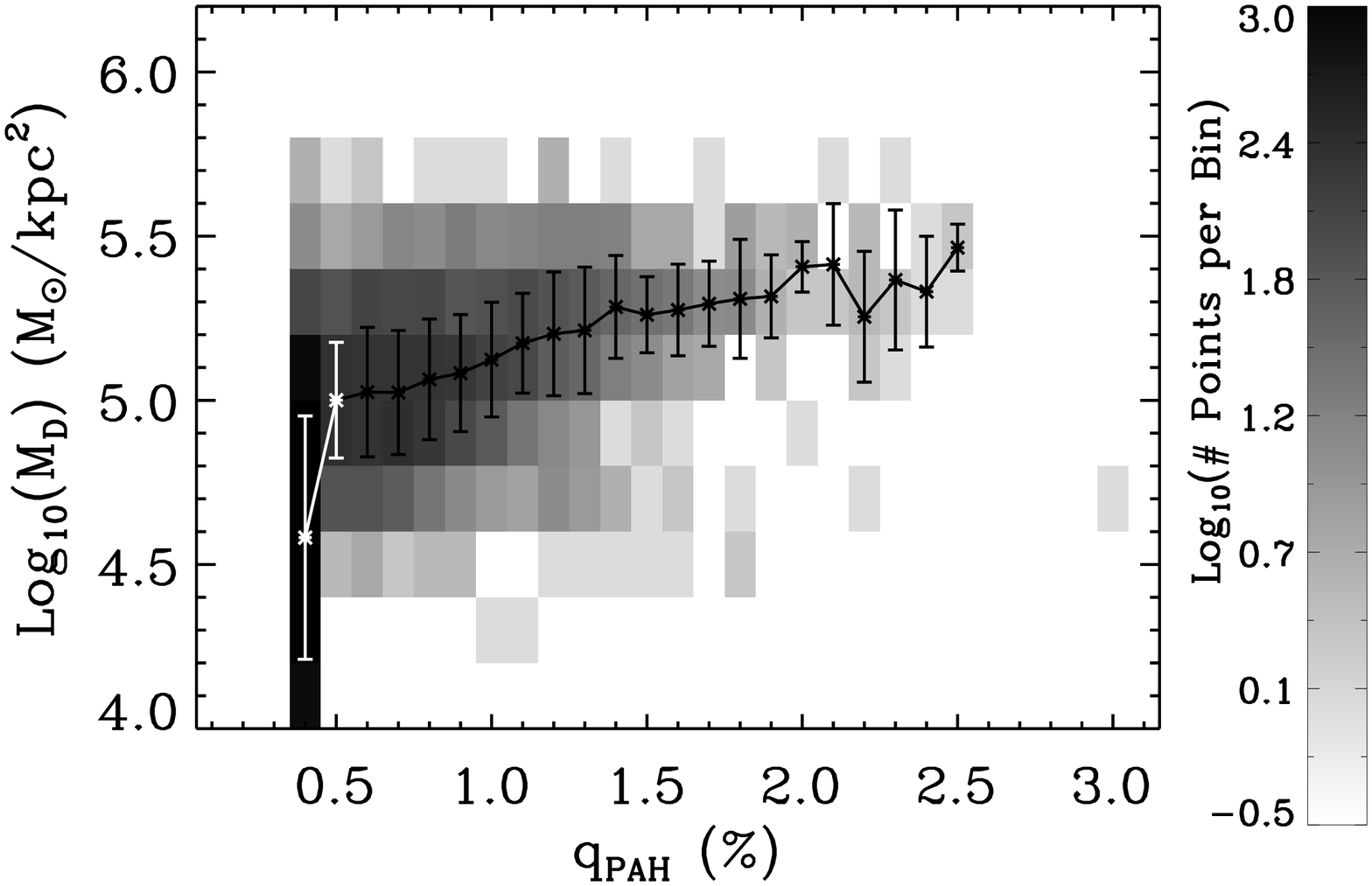}{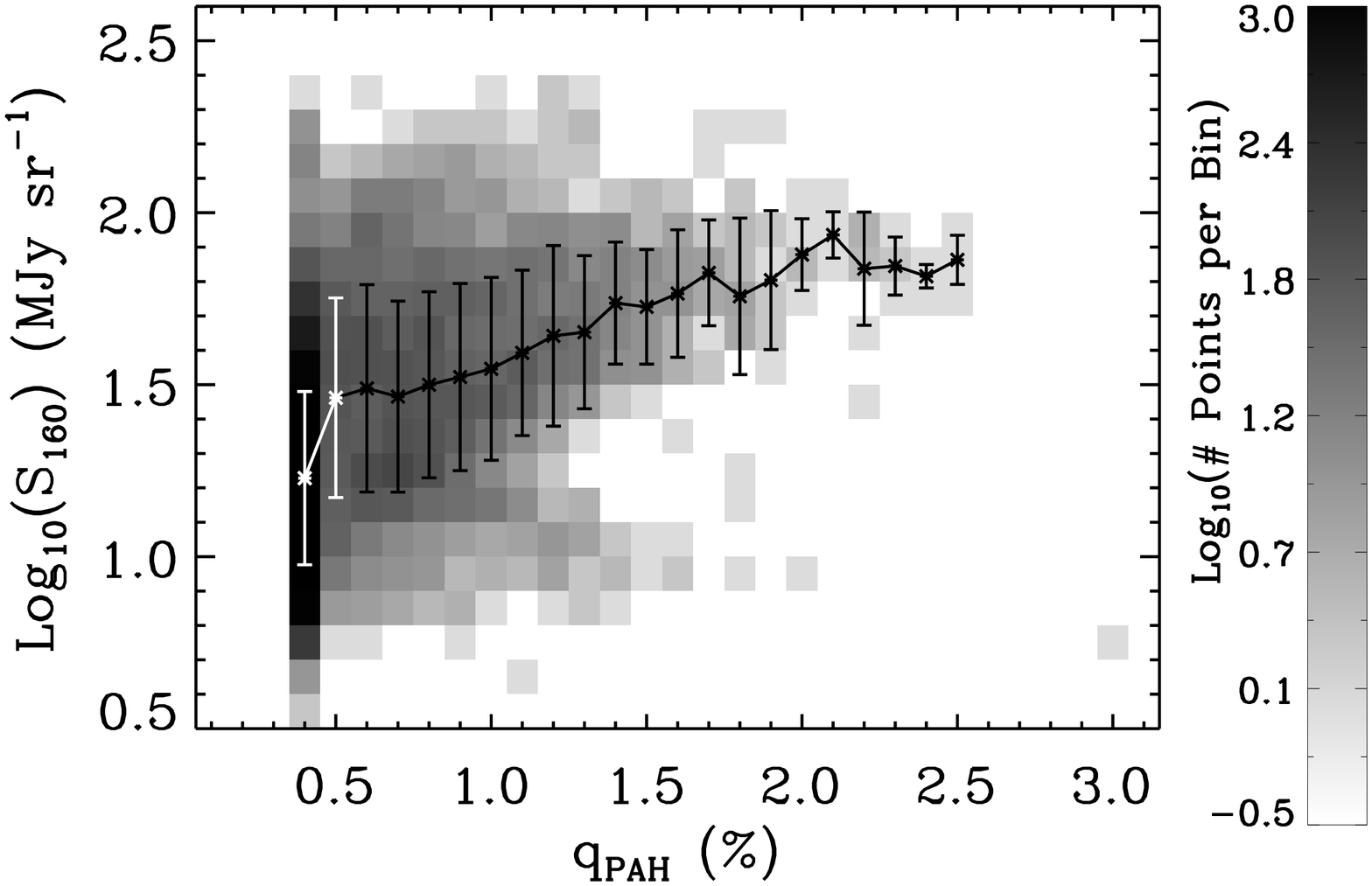}
\plottwo{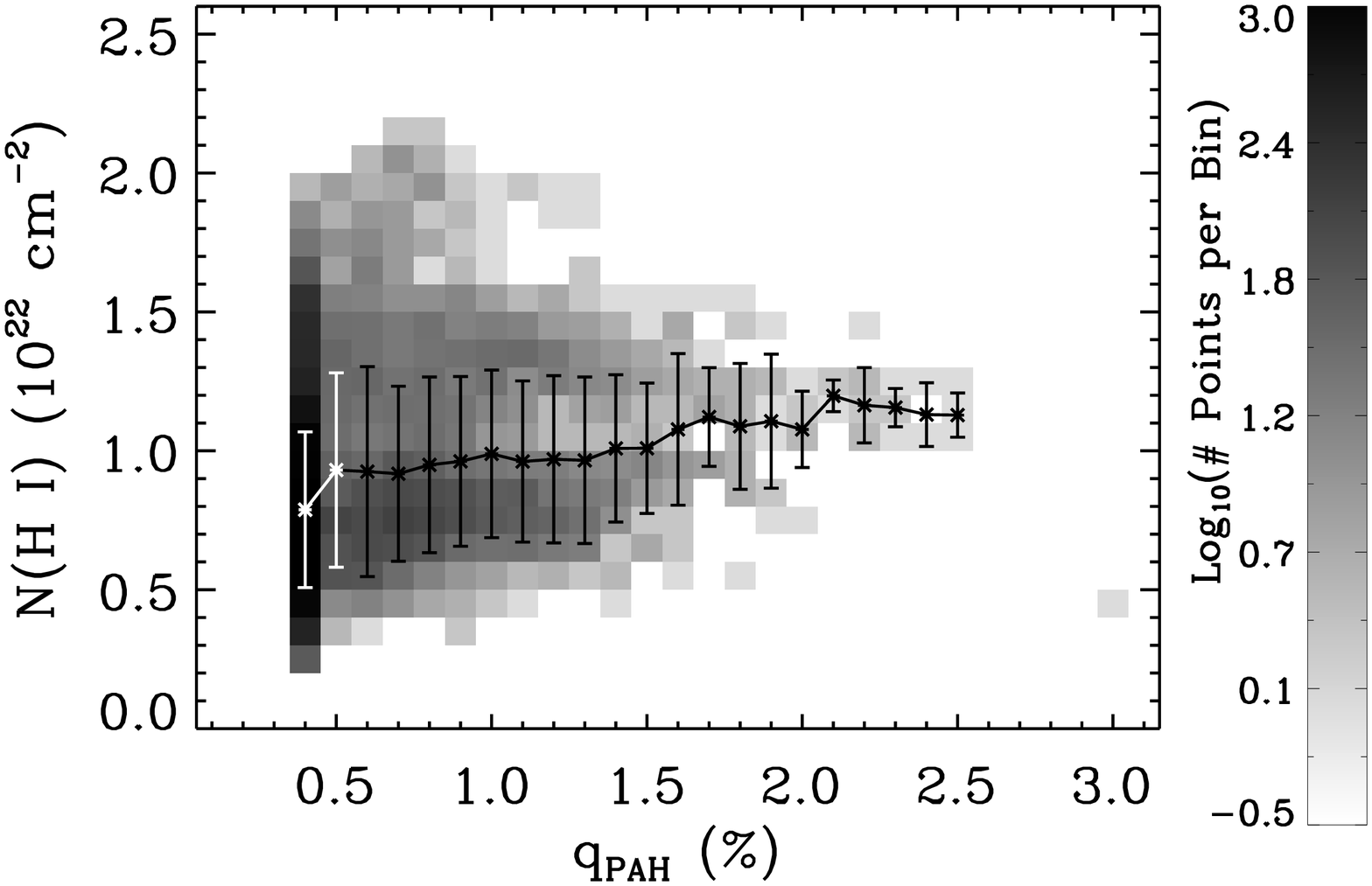}{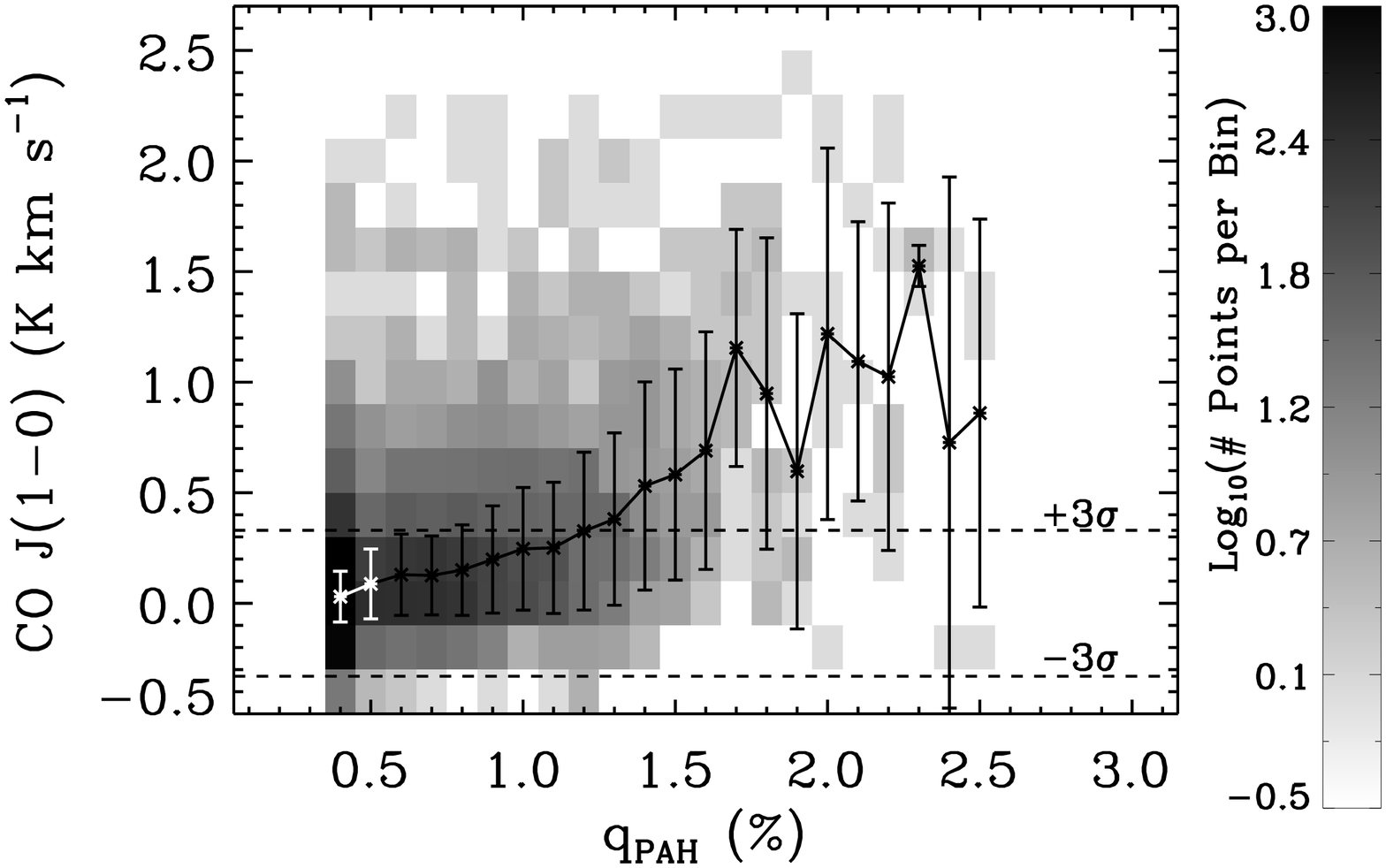}
\caption{Two-dimensional histograms of the dust mass surface density,
160 \micron\ surface brightness, neutral hydrogen column and CO
integrated intensity as a function of \qpah\ overlayed with the binned
average.  The error bars show the standard deviation of the scatter in
each bin of \qpah.  The gray scale represents the logarithm of the
number of points in each bin.}
\label{fig:binned}
\end{figure*}

The association of PAH emission with star-forming regions and
molecular clouds versus the diffuse ISM of a galaxy is a matter of
debate, and may vary depending on the galaxy type and star-formation
history.  \citet{bendo08} find that the 8/160 \micron\ ratio in 15
nearby face-on spirals suggests that the PAHs are associated with the
diffuse cold dust that produces most of the 160 \micron\ emission.  On
the other hand, \citet{haas02} find that the 8 \micron\ feature,
across a range of galaxy type and current star formation rate, is
associated with peaks of 850 \micron\ surface brightness, which
originate in molecular regions.  The distribution of PAHs in a galaxy
is one parameter that will help determine what fraction of the PAH
luminosity aries from the reprocessing of UV photons from young, hot
stars versus the general galactic distribution of B stars.  This
distinction is crucial in using PAH emission as a tracer of current
star-formation \citep{peeters04,calzetti07}.  To further understand
the distribution of PAHs in the SMC, we explore the correlation of PAH
fraction with 160 \micron\ emission and the dust mass in
Figure~\ref{fig:binned}.  In this Figure, we see that the PAH
fraction is correlated with dust surface density (M$_{D}$) and
160 \micron\ emission, but only weakly correlated with H I column
(note that the H I column is shown in a linear scale while M$_{D}$ and
160 \micron\ emission are shown on a logarithmic scale).

\begin{figure*}
\centering
\epsscale{1.0}
\includegraphics[width=6in]{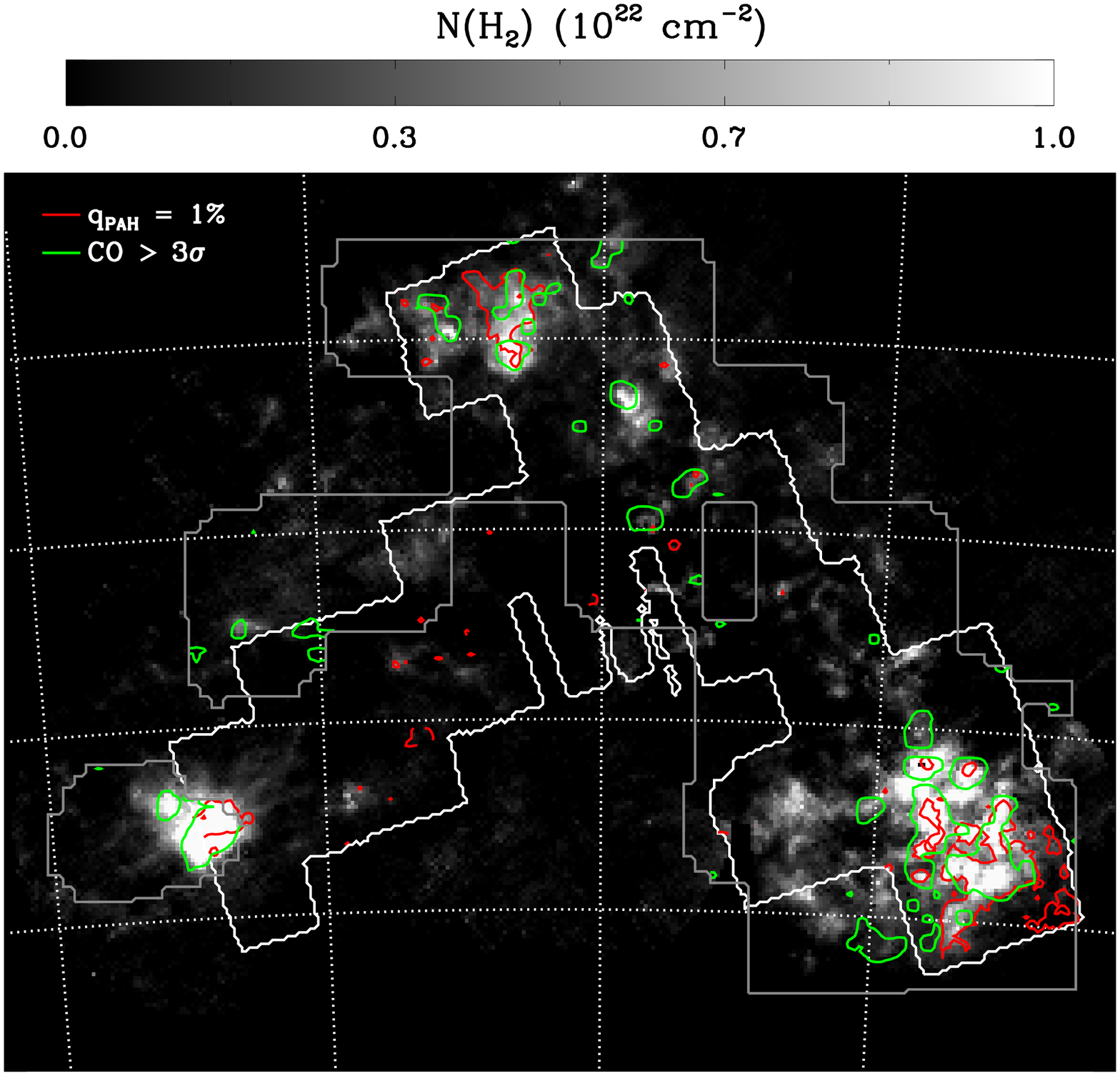}
\caption{A map of molecular gas column density inferred from excess
dust emission at 160 \micron\ from Bolatto et al 2009 (in prep)
overlayed with the 3$\sigma$ contour of CO emission in green and the
1\% contour of \qpah\ in red.  The coverage of the CO survey is
shown as a thin gray line.}
\label{fig:h2pah}
\end{figure*}

The correlation of the PAH fraction with dust surface density \emph{but
not H I} reflects the fact that PAHs are not uniformly distributed in
the SMC.  Regions with PAH fraction greater than 1\% in general have
M$_{D} > 10^5$ \msun\ kpc$^{-2}$.  However, regions with dust surface
densities above this level also tend to contain molecular gas, so the
dust surface density and H I column no longer track each other because
of the presence of H$_{2}$ \citep{leroy07}.  For this reason, we see
at best a weak correlation of \qpah\ with neutral hydrogen
column, but a stronger association with CO emission.  Bolatto et al
2010 (in prep) have used the MIPS observations of the SMC from S$^3$MC
and the SAGE-SMC survey to map the distribution of molecular gas as
inferred from regions with ``excess'' dust surface density relative to
the column of neutral hydrogen, using the same techniques as
\citet{leroy07} and \citet{leroy09}.  In Figure~\ref{fig:h2pah} we
show their map of the H$_{2}$ column density overlayed with a contour
at \qpah= $1$\% and a contour of 3$\sigma$ CO emission.  Although we
must use caution in comparing the detailed distribution of CO to
\qpah\ since the CO is at $\sim 4$ times lower resolution, it is
interesting to note that there are regions with high molecular gas
columns without CO and with low PAH fractions, for instance in the
northern region of the SW Bar.  This may indicate that the radiation
field in these regions is affecting both CO and PAHs.

\begin{figure*}
\centering
\epsscale{1.0}
\includegraphics[width=6in]{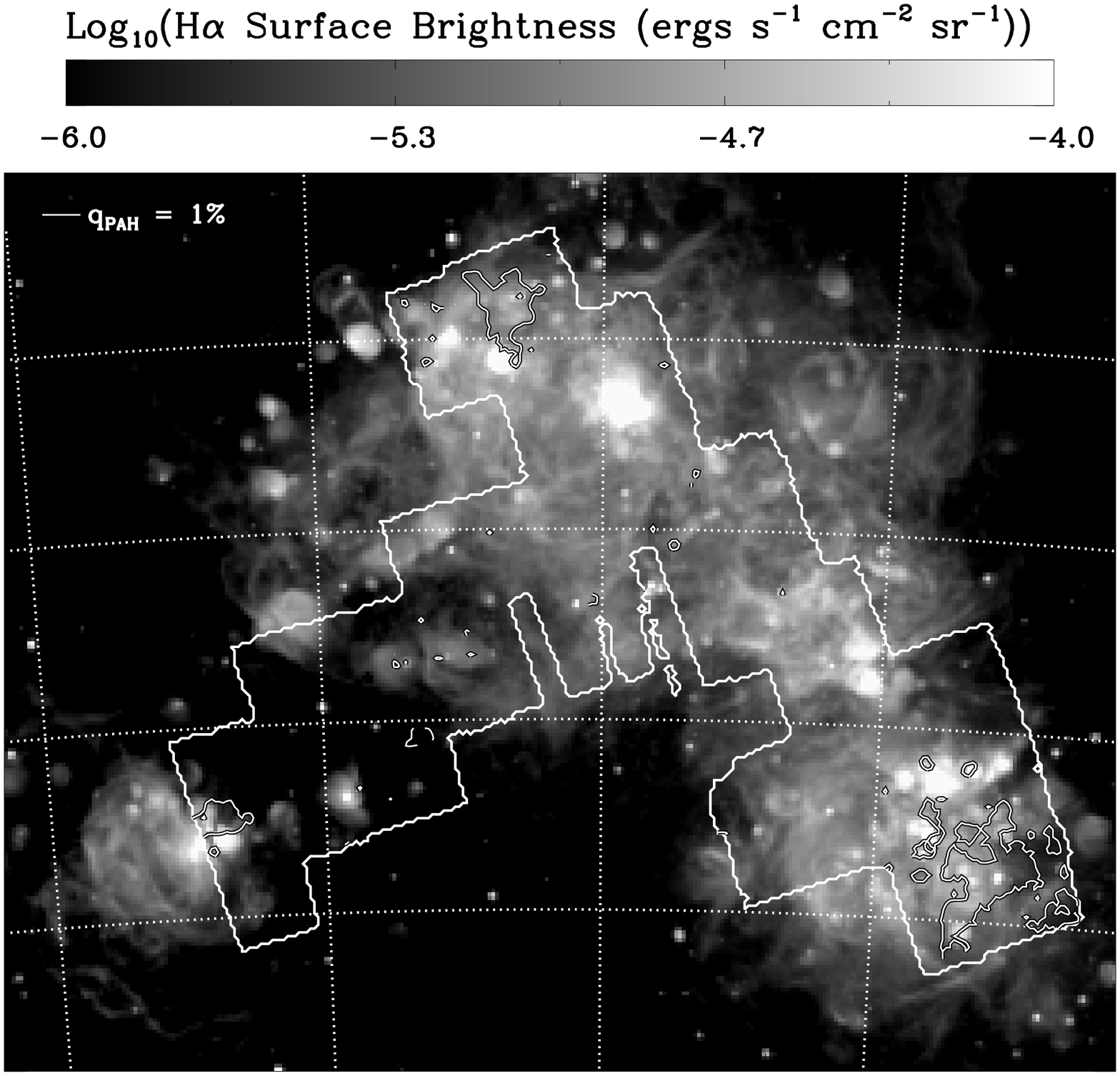}
\caption{MCELS H$\alpha$ image with the 1\% contour of \qpah\
overlayed. The regions of high PAH fraction are typically on the
outskirts of H II regions.}
\label{fig:hacomp}
\end{figure*}

We have so far shown that PAH fraction is high in regions of active
star-formation, associated with the presence of CO and molecular gas.
PAHs can also be destroyed in regions of active star-formation by the
intense UV fields produced by massive stars or in the H II regions
themselves by chemistry with H$^{+}$ \citep{giard94}.
Figure~\ref{fig:hacomp} shows the MCELS map of H$\alpha$ in the SMC
overlayed with the 1\% contour of \qpah.  From this
comparison it is clear that the regions of high PAH fraction are
typically on the outskirts of H II regions (i.e. the high \qpah\
regions and the H II regions are not co-spatial).  In particular, the
region around N 66, the largest H II region in the SMC, has a very low
PAH fraction.  We will discuss the influence of H II regions and
massive star-formation on the PAH fraction further in
Section~\ref{sec:uv}.

\subsection{The PAH Fraction in SMC B1 \# 1}

\begin{figure*}
\centering
\epsscale{1.0}
\includegraphics[width=5in]{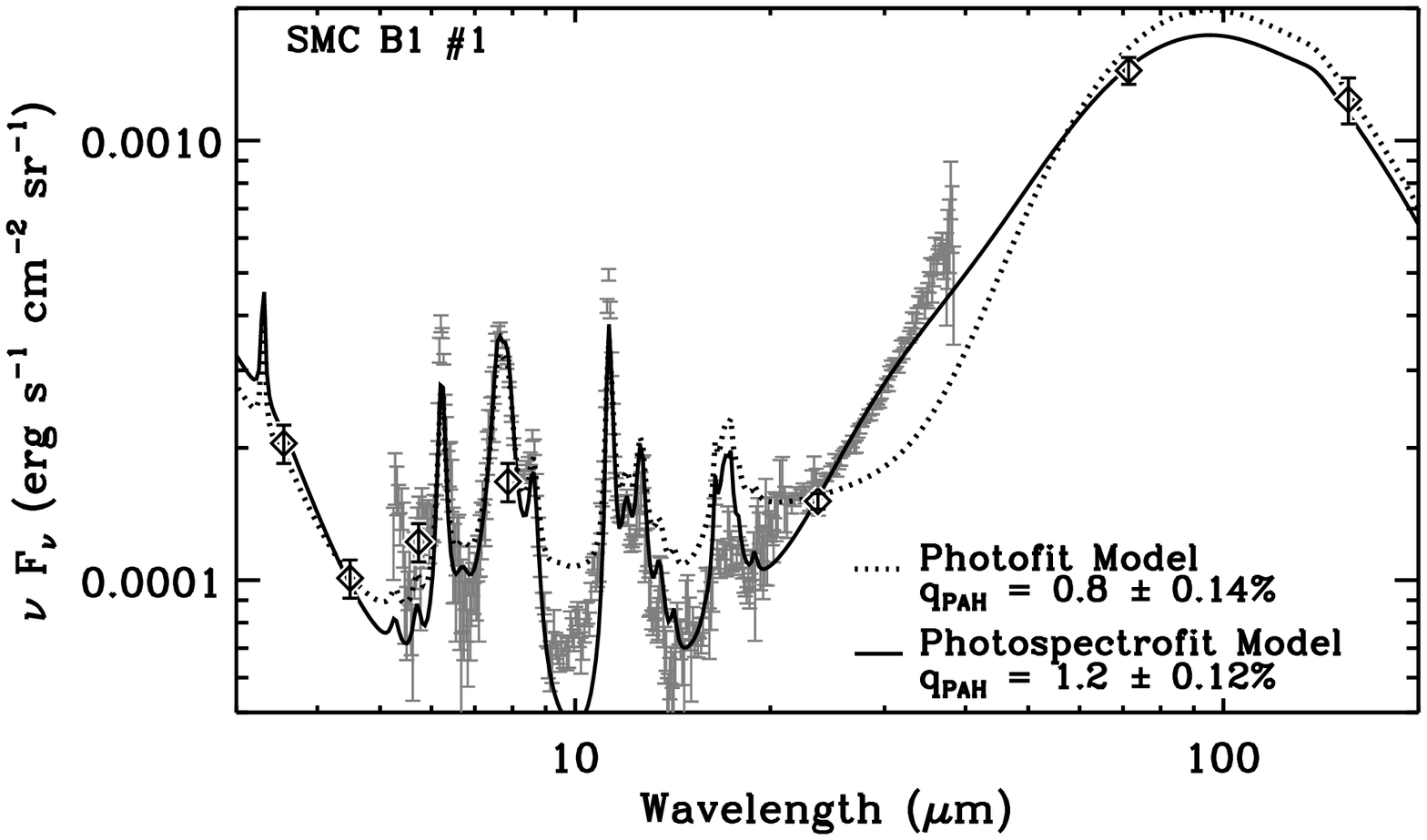}
\caption{Best fit models to the photometry and spectroscopy in SMC B1
\# 1.  The gray points show the IRS spectrum of SMC B1, while the
diamonds show the photometry from IRAC and MIPS.}
\label{fig:smcb1}
\end{figure*}

The molecular cloud SMC B1 \# 1 was the first location in the SMC
where the emission from PAHs was identified \citep{reach00}.  This
region has been studied extensively and the PAH emission spectrum has
been modeled by \citet{li02} who found that the PAH fraction in SMC B1
(\qpah $\sim 1.6$\%) was 8 times higher than the average fraction in
the Bar (\qpah $\sim 0.2$\%).  In addition, \citet{reach00} and
\citet{li02} noted unusual band ratios of the 6.2, 7.7, and 11.3
\micron\ features.  In Figure~\ref{fig:smcb1} we show the best fit
models for the spectrum and photometry of SMC B1.  We find \qpah $\sim
1.2\pm 0.1$\%, in relatively good agreement with \citet{li02}
considering the differences in resolution between our respective
datasets. We also reproduce the distinctive band ratios (11.3 and 6.2
features high compared to the 7.7 feature) seen with ISO.  SMC B1 does
not have a uniquely large PAH fraction compared to other locations in
the SW Bar.  

\subsection{PAH Fraction and the 2175\AA\ Bump}\label{sec:bump}

\begin{figure*}
\centering
\epsscale{1.0}
\includegraphics[width=6in]{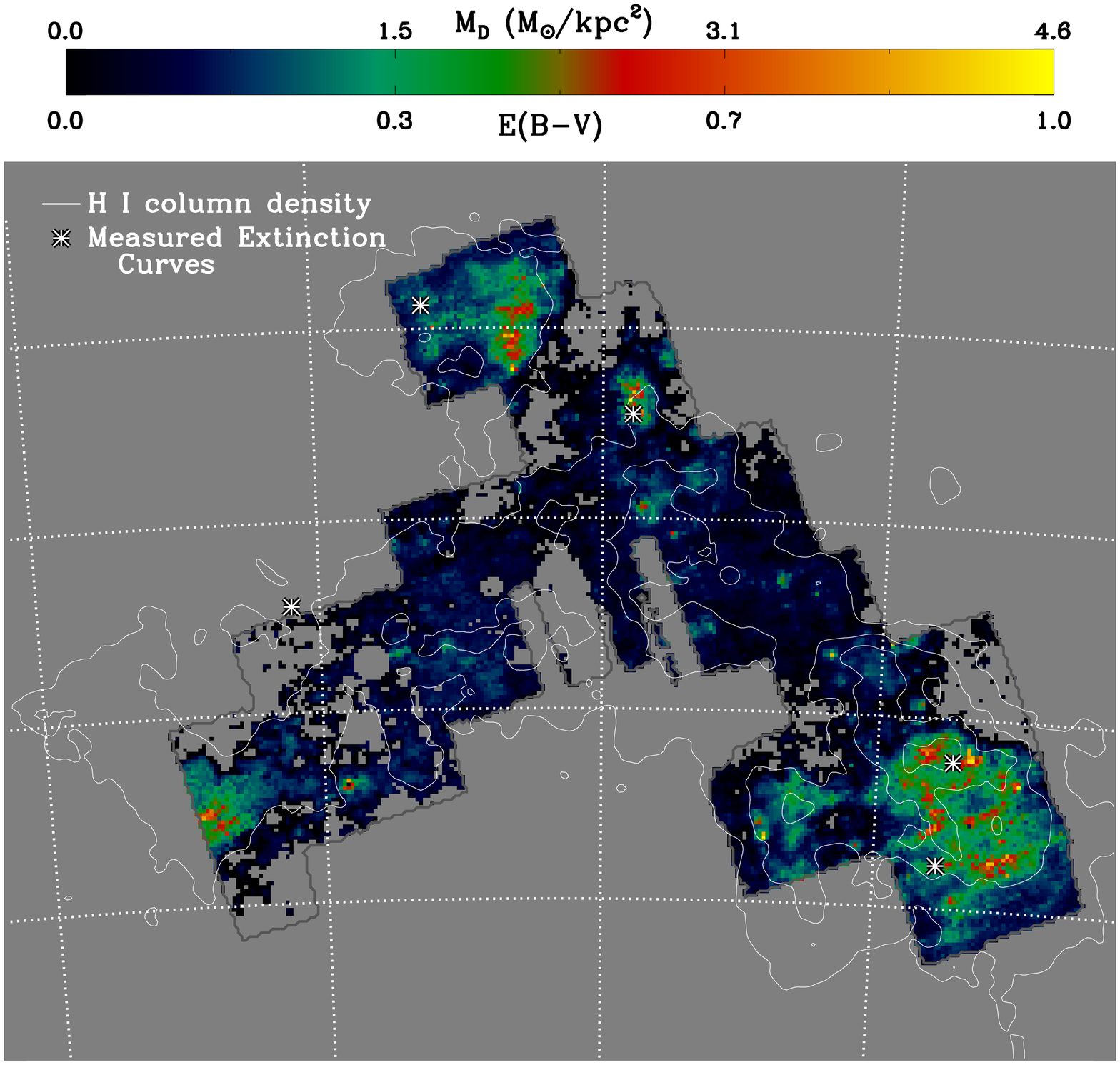}
\caption{The dust mass surface density and E(B$-$V) values derived from
our photofit models at each pixel overlayed with contours of neutral
hydrogen column density from \citet{stanimirovic99} at 6, 10, 14 and
$18\times 10^{21}$ cm$^{-2}$.  The locations of the five stars in the
SMC with extinction curves from \citet{gordon03} are shown with
asterisks.  Table~\ref{tab:ebv} shows a comparison of the E(B$-$V)
measured for those stars with the total line-of-sight E(B$-$V) we
calculate from the photofit model results.}
\label{fig:ebv}
\end{figure*}

On Figures~\ref{fig:qpahmap} and~\ref{fig:ebv} we show the locations
of the five stars with measurements of their UV extinction curves with
asterisks.  Of these stars, only one shows a 2175 \AA\ bump in its
extinction curve: AzV 456, which unfortunately lies just outside the
boundaries of the map.  The lack of a bump in the remaining stars has
been interpreted as evidence for a low PAH fraction along those
lines-of-sight, assuming that PAHs are the carrier of the bump
\citep{li02}.  To test this assertion, we list in Table~\ref{tab:ebv}
the measured values of E(B$-$V) for the five stars with extinction
curves from \citet{gordon03} along with the E(B$-$V) and \qpah\ we
calculate from our photofit model results at those positions.  For the
\citet{draine07a} models, the dust mass surface density M$_{D}$
and E(B$-$V) are related by a constant value of $2.16\times 10^{-6}$
mag (\msun\ kpc$^{-2}$)$^{-1}$.  

The comparison of the E(B$-$V) values shows that the stars are indeed
behind the majority of the dust along those lines of sight, and that
the \qpah\ values are slightly above the SMC average of 0.6\% (see
Section~\ref{sec:mapresults}).  A more detailed analysis of these
lines-of-sight will be presented in a follow-up paper with targeted
IRS spectroscopy to study the PAH emission in these regions.  Since we
do find that regions of high \qpah\ tend to be associated with
molecular gas, it may be the case that assuming the dust and PAHs are
uniformly mixed along each line of sight does not hold.  In that case,
the comparison of E(B$-$V) values may not be a good indicator of
whether these stars should show the 2175 \AA\ bump in their extinction
curves. In addition, our angular resolution is not high enough in
these maps to directly observe the structure of the dust emission in
the vicinity of these stars, so we can not definitively test the
PAH-2175 bump connection.  We note that one of the stars lies near N
66 in a region with very low ambient PAH fraction.  Some of the other
stars may fall in voids in the PAH distribution, but higher angular
resolution is necessary to understand the line of sight towards these
stars.  

\begin{deluxetable*}{lccccc}
\tablewidth{0pt}
\tabletypesize{\scriptsize}
\tablecolumns{6}
\tablecaption{Comparison with Extinction Curve Measurements}
\tablehead{ 
\multicolumn{1}{l}{Star} &
\multicolumn{1}{c}{R.A.} &
\multicolumn{1}{c}{Dec.} &
\multicolumn{1}{c}{E(B$-$V) (Gordon et al. 2003)} &
\multicolumn{1}{c}{E(B$-$V) (This Study)} &
\multicolumn{1}{c}{\qpah} \\
\multicolumn{1}{l}{} &
\multicolumn{1}{c}{(J2000)} &
\multicolumn{1}{c}{(J2000)} &
\multicolumn{1}{c}{(mag)} &
\multicolumn{1}{c}{(mag)} &
\multicolumn{1}{c}{(\%)}}
\startdata
AzV 18  & $0^{\rm h} 47^{\rm m} 12^\mathrm{s}$ & $-73^\circ 06\arcmin 33\arcsec$ & 0.167$\pm$0.013 & 0.25$\pm$0.07 & 0.4$\pm$ 0.1 \\
AzV 23  & $0^{\rm h} 47^{\rm m} 39^\mathrm{s}$ & $-73^\circ 22\arcmin 53\arcsec$ & 0.182$\pm$0.006 & 0.30$\pm$0.10 & 0.8$\pm$ 0.1 \\
AzV 214 & $0^{\rm h} 58^{\rm m} 55^\mathrm{s}$ & $-72^\circ 13\arcmin 17\arcsec$ & 0.147$\pm$0.012 & 0.34$\pm$0.13 & 0.7$\pm$ 0.2 \\
AzV 398 & $1^{\rm h} 06^{\rm m} 10^\mathrm{s}$ & $-71^\circ 56\arcmin 01\arcsec$ & 0.218$\pm$0.024 & 0.33$\pm$0.09 & 0.8$\pm$ 0.1 \\
AzV 456 & $1^{\rm h} 10^{\rm m} 56^\mathrm{s}$ & $-72^\circ 42\arcmin 56\arcsec$ & 0.263$\pm$0.016 & \nodata & \nodata
\enddata
\tablecomments{The E(B$-$V) calculations are described in
Section~\ref{sec:bump}.}
\label{tab:ebv}
\end{deluxetable*}

\section{What Governs the PAH Fraction in the SMC?}\label{sec:discussion}

\citet{draine07b} determined the PAH fraction in the SINGs galaxy
sample using identical techniques to what we have done here.  They
observed a effect very similar to what was seen by
\citet{engelbracht05} in that at a metallicity of $12 +$ log(O/H)
$\sim 8.0$ there is an abrupt change in the typical PAH fraction or
8/24 \micron\ ratio.  The SMC lies right at this transition
metallicity, so we hope to gain some insight into the processes at
work by studying its PAH fraction.  There is still a great
deal of uncertainty as to how PAHs form and how they are destroyed,
let alone how those processes balance in the ISM.  In the following
sections we address some of the proposed aspects of the PAH life-cycle
and elucidate what we can learn about them from the SMC.

\subsection{Formation by Carbon-Rich Evolved Stars} 

The formation of PAHs in the atmospheres of evolved stars is a well
established hypothesis for the source of interstellar PAHs.  PAH
emission bands have been observed in the spectrum of the carbon-rich
post-AGB stars where the radiation field increases in hardness and
intensity, more effectively exciting the PAH emission features
\citep{justtanont96}.  Despite these observations of PAH formation in
carbon-rich stars, it remains to be shown that they inject PAHs into
the ISM at the level needed to explain the abundance observed.  This,
of course, is a general problem in the ``stardust'' scenario
\citep{draine09}.  Recently, in the Large Magellanic Cloud,
\citet{matsuura09} have performed a detailed accounting of the dust
enrichment of the ISM by AGB stars and find a significant deficit
compared to the observed ISM dust mass.  

Assuming that the ``stardust'' picture is correct, \citet{galliano08}
and \citet{dwek98} hypothesize that the low fraction of PAHs in low
metallicity galaxies reflects the delay ($\sim 500$ Myr) in the
production of carbon dust from AGB stars relative to silicate dust
from core-collapse supernova. For the SMC in particular, this picture
has a number of issues.  Most importantly, there is evidence that the
SMC formed a large fraction of its stars more than 8 Gyr ago followed
by a subsequent burst of star formation 3 Gyr ago \citep{harris04}.
This long time scale makes the delayed PAH injection into the ISM by
AGB stars an unlikely explanation for the current observed PAH
deficiency.  Recent work by \citet{sloan09}, for example, argues that
carbon-rich AGB stars in low metallicity galaxies can start
contributing dust to the ISM in $\sim 300$ Myr.  

\begin{figure*}
\centering
\epsscale{1.0}
\includegraphics[width=6in]{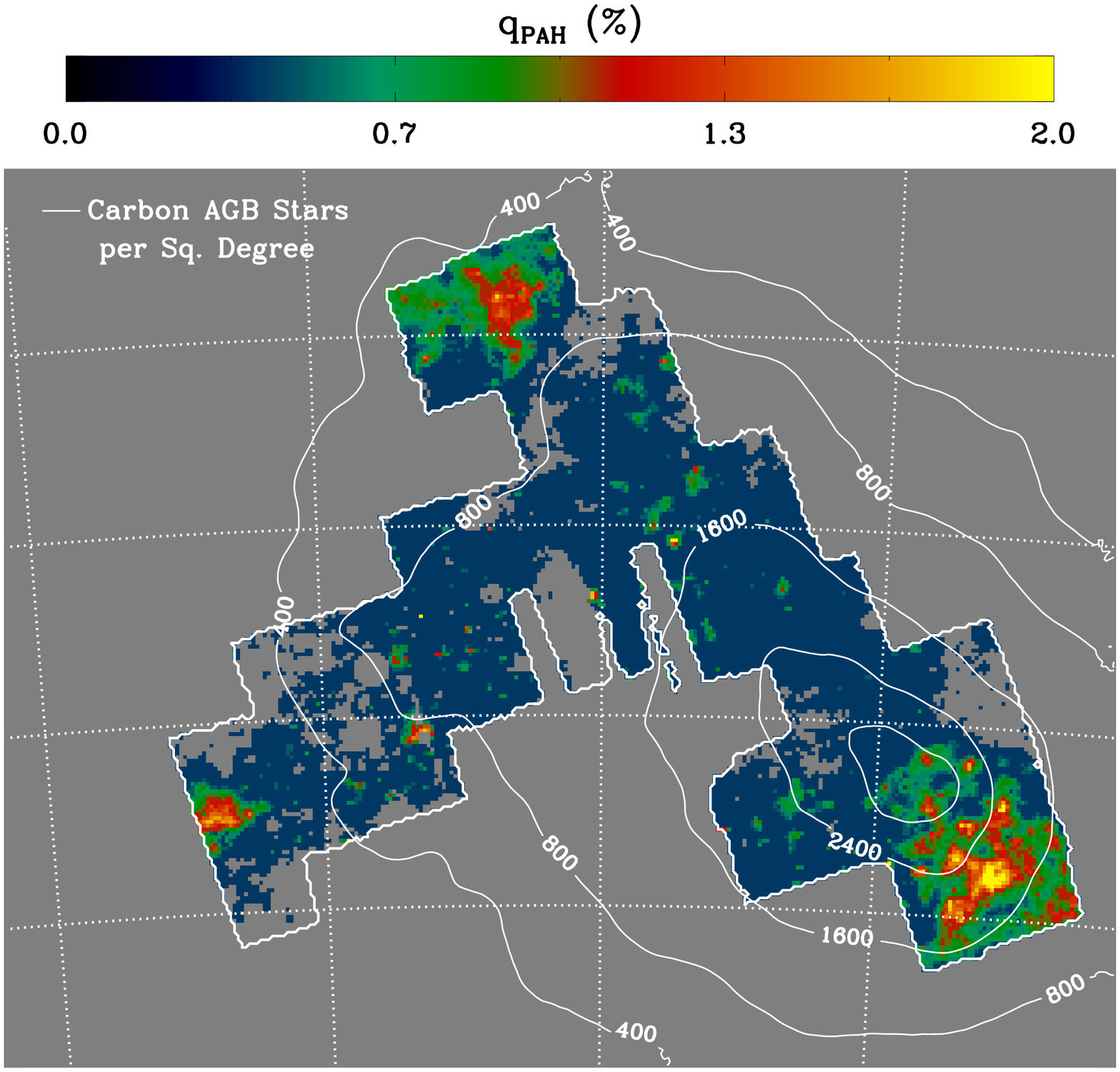}
\caption{Map of \qpah\ overlayed with contours of carbon AGB star
density determined from the 2MASS 6X point source catalog using the
selection criteria of \citet{cioni06}.  The contours are labeled with
the density of carbon stars per square degree.  There is no obvious
correspondence between the distribution of carbon stars and the
fraction of PAHs.}
\label{fig:cstar}
\end{figure*}

To evaluate the relationship between the current distribution of PAHs
and the input of PAHs from AGB stars, we show in
Figure~\ref{fig:cstar} the PAH fracion overlayed with contours of
the density of carbon AGB stars.  These contours are created using the
2MASS 6X point source catalog towards the SMC, selecting carbon stars
using the technique described in \citet{cioni06}.  The distribution of
carbon stars follows the observed ``spheroidal'' population of older
stars in the SMC very well \citep{zaritsky00,cioni00}.  The \qpah\
map, however, has no clear relationship to the distribution of carbon
stars. This is perhaps to be expected since the distribution of ISM
mass does not follow the stellar component either.  However, we note
that a different conclusion regarding the PAH fracion compared to
the distribution of AGB stars was recently reached by
\citet{paradis09}, who modeled dust SEDs across the Large Magellanic
Cloud using the SED models of \citet{desert90}.  They find evidence
that the fracion of PAHs in the LMC is highest in the region of the
stellar bar, where the concentration of AGB stars peaks.  This could
be an accidental coincidence, or it may reflect methodological
differences or it could indicate a different dominant mechanism of PAH
formation in the SMC and LMC or more efficient destruction of PAHs in
the SMC.

Although the distribution of PAHs may not resemble that of AGB stars,
we would expect the PAHs injected by those stars to be present in the
diffuse ISM and not preferentially in molecular clouds.  As we have
shown, however, the diffuse ISM PAH content is very low in the SMC.
As such, in order to reconcile the pathway of PAHs from AGB stars to
the diffuse ISM to molecular clouds, we must hypothesize a recent
event, occuring after the condensation of the current generation of
molecular clouds, which essentially cleared the diffuse ISM of
AGB-produced PAHs while leaving the shielded regions of molecular gas
with high PAH fractions.  Observations of giant molecular clouds
(GMCs) in the Magellanic Clouds and other nearby galaxies suggest that
the GMC lifetime is $\sim 25$ Myr \citep{fukui99,blitz07}, so the
event in question would have had to occur within the last 25 Myr or
so. In the absence of such an event (which we will investigate further
in a Section~\ref{sec:uv}), our observation of low PAH fraction in the
diffuse ISM, higher PAH fraction in molecular clouds, and the lack of
relation between the PAH fraction and the distribution of AGB stars
is strong evidence against AGB stars being the dominant force behind
the fraction of PAHs in the SMC.

\subsection{Formation and Destruction of PAHs in Shocks and Turbulence}

The shocks created by supernova explosions have a dramatic effect on
the content and size distribution of dust in the interstellar medium.
Upon encountering a shock wave, grains can be shattered via collisions
or sputtered by hot gas behind the shock or by motion of the
grain through the post-shock medium.  Because of their small mass-to-area
ratio, PAHs are well coupled to the gas and do not acquire large
relative velocities after the passage of a shock, so they will
primarily be sputtered only in hot post-shock gas behind fast ($v >
200$ km s$^{-1}$) shocks.  Calculations by \citet{jones96} suggest
that grain shattering could in fact be a net source of PAH material
for shocks between 50 and 200 km s$^{-1}$, converting $\sim 10$\% of
the initial grain mass into small PAH sized fragments.  Thus, it is
not immediately obvious what the net effect of interstellar shocks on
the fraction of PAHs will be.

Some studies attribute the low PAH fraction in low metallicity
galaxies to efficient destruction of PAHs by supernova shocks.
\citet{ohalloran06} found a correlation between decreasing PAH
emission and increased supernova activity as traced by the ratio of
mid-IR lines of [Fe II] and [Ne II].  Two issues with this
interpretation are that the mid-IR lines are tracing current supernova
activity, which only affects the PAH fraction in the immediate
vicinity of those remnants, and an increased supernova rate will have
recently been related to a higher UV field produced by the massive
stars, so it is difficult to disentangle the effects of shocks versus
intense UV fields. 

The distinctive nature of the SMC extinction curve might point toward
the influence of supernova shocks on the dust grain size distribution.
Four of the five extinction curves show similar characteristics: a
lack of the 2175 \AA\ bump \citep[the carrier of which is most likely
PAHs][]{li01} and a steeper far-UV rise indicating more small dust
grains relative to the Milky Way extinction curve
\citep{gordon03,cartledge05}.  Magellanic-type extinction has been
seen to arise in the Milky Way as well along the sightline towards HD
204827, which may be embedded in dust associated with a supernova
shock \citep{valencic03}.  However, there are many other viable
interpretations to explain the proportion of small dust grains,
including inhibition of grain growth in dense clouds. 

\begin{figure*}
\centering
\epsscale{1.0}
\includegraphics[width=6in]{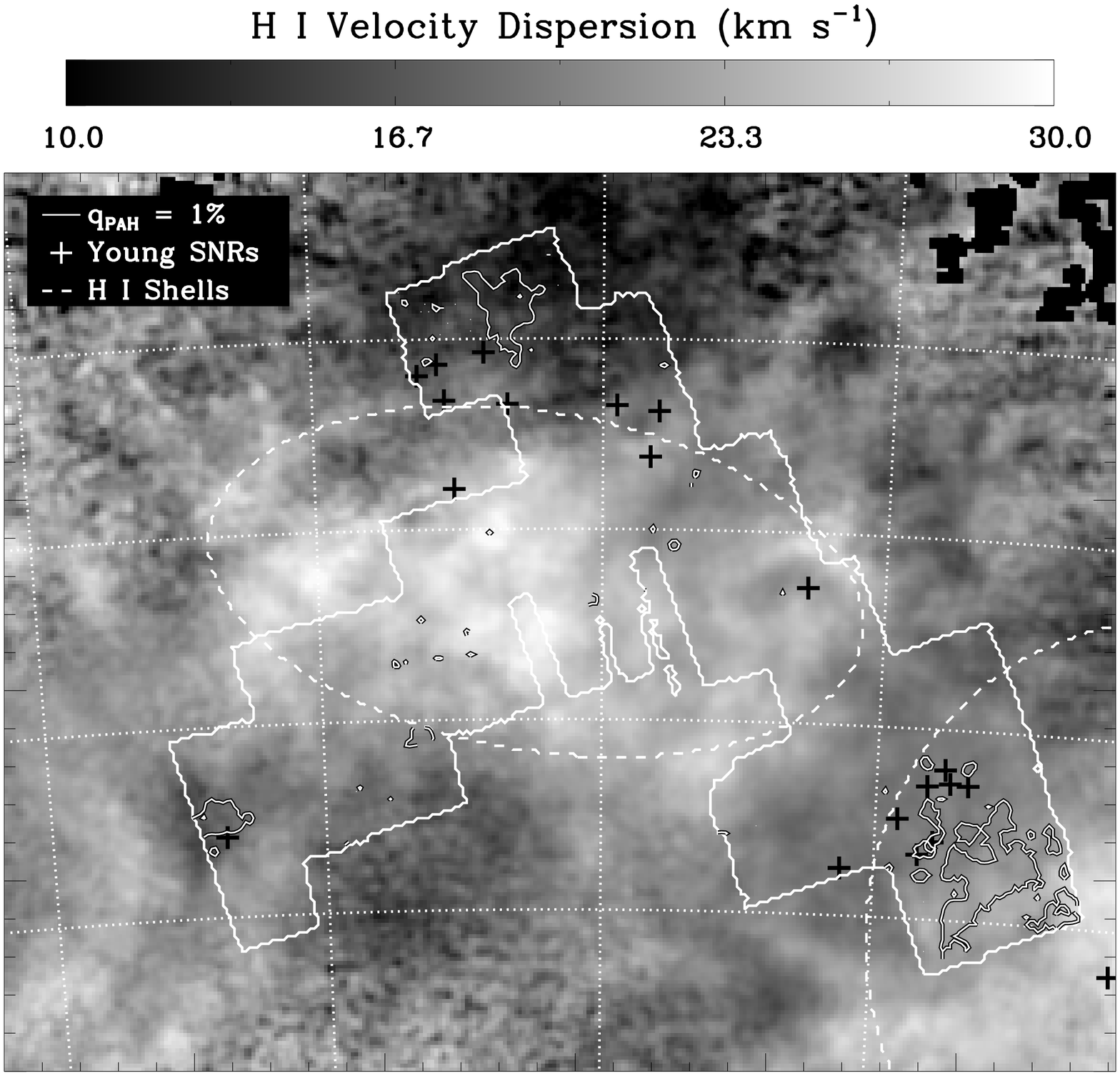}
\caption{SMC H I velocity dispersion from \citet{stanimirovic99}. The
approximate locations of the supergiant H I shells identified by
\citet{stanimirovic99} are shown with a dashed white line.  The
velocity dispersion mainly traces bulk motions of gas
along the line of sight, particularly highlighting the regions of
these shells.  We show the locations of young supernova remnants
identified in the ATCA survey of the SMC with black crosses.}
\label{fig:shocks}
\end{figure*}

In Figure~\ref{fig:shocks} we show a map of the H I velocity
dispersion in the SMC from \citet{stanimirovic99}.  The velocity
dispersion here mostly traces the regions where there are more than
one velocity component along the line of sight, particularly
highlighting the regions where \citet{stanimirovic99} find evidence
for supergiant shells in the H I distribution.  We show the
approximate locations of two of their shells that overlap our map.  If
supernovae are the source of these shells, \citet{stanimirovic99} find
that $\sim 1000$ supernovae are required to account for the kinetic
energy and the shells have dynamical ages of $\sim 20$ Myr.  The
supergiant shells provide indirect evidence for the effects of
supernovae on the ISM.  On Figure~\ref{fig:shocks} we also mark with
green crosses the locations of young ($\sim 1000-10000$ yr) supernova
remnants identified in the Australia Telescope Compact Array survey of
the SMC \citep{payne04}.

There is no clear trend relating \qpah\ to the boundaries
of the supergiant shells, although this may be an effect of depth
along the the line of sight.  The middle region of the Bar and Wing,
which is essentially devoid of PAHs is covered by one of the shells,
but the SW Bar, which hosts the largest concentration of PAHs in the
SMC is covered as well. Although there is an anticorrelation between
the young SNRs and large PAH fraction, the distribution of remnants
closely follows that of the massive star-forming regions (compare
Figures~\ref{fig:hacomp} and~\ref{fig:shocks}).  For this reason, it
is difficult to draw strong conclusions as to whether the supernova
shocks or the UV fields and H II regions created by their progenitor
stars were responsible for destroying PAHs in these areas.  

\begin{figure*}
\centering
\epsscale{1.0}
\includegraphics[width=6in]{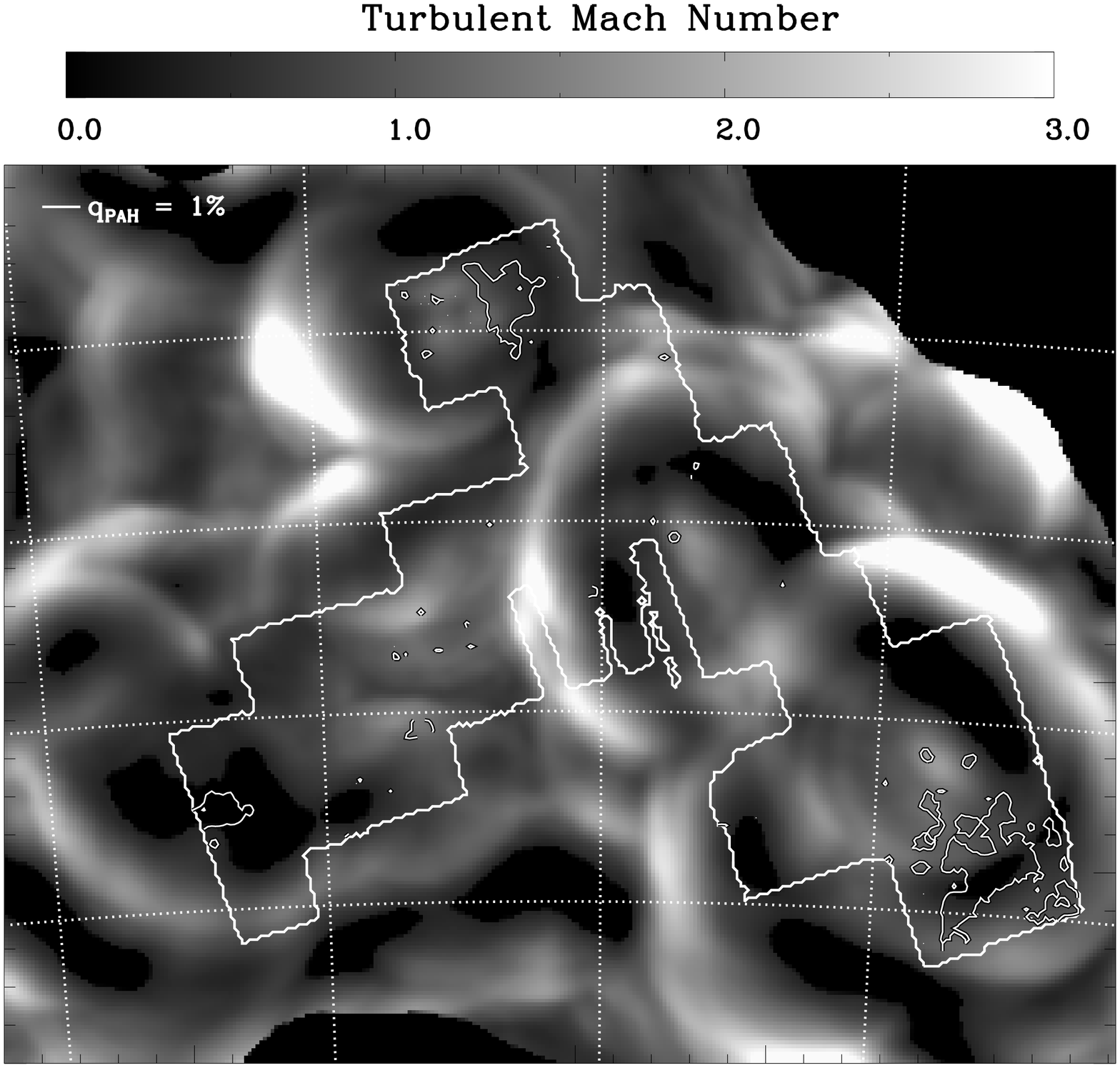}
\caption{Map of estimated turbulent Mach number from Burkhart et al (2009) in
prep, overlayed with the 1\% contour of \qpah. The Mach number map has an
angular resolution of $30\arcmin$.}
\label{fig:turb}
\end{figure*}

Similar to shocks, regions of high turbulent velocity may alter the
size distribution of dust grains via shattering.
\citet{miville-deschenes02} studied a region of high latitude cirrus
and argued that variations in small dust grain and PAH fractions 
could be related to the turbulent velocity field in the region.
\citet{burkhart10} have studied turbulence in the ISM of the SMC using
the H I observations of \citet{stanimirovic99}.  They present a map
showing an estimate of the sonic Mach number of turbulence in the SMC
based on higher order moments of H I column density and the results of
numerical simulations.  This map is quite distinct from the velocity
dispersion map shown in Figure~\ref{fig:shocks} which mostly
highlights the presence of bulk velocity motions along the line of
sight.  In Figure~\ref{fig:turb} we show the \citet{burkhart10} Mach
number map overlayed with the contours of \qpah.  We find that
there is no strong correlation between tubulent Mach number and \qpah,
although in general the regions of high \qpah\ are near
depressions in the Mach number distribution, opposite what we would
expect if grain shattering in turbulence was a major source of PAHs.
However, this simple comparison does not provide information about the
amount of dust affected by different levels of turbulence (e.g. some
of the low Mach number regions may represent a very small fraction of
the total ISM).  To account for this fact, we also calculate the dust
mass surface density weighted Mach number in regions with \qpah$>1$\%
and $<1$\% to be $0.8\pm0.3$ and $1.0\pm0.4$, respectively.  Although
there is no clear association between high sonic Mach number and
\qpah\ from this comparison, it is worth noting that if PAHs are
primarily associated with molecular gas, the Mach number derived from
H I observations may have little relevance to the creation or
destruction of PAHs in the SMC.

In general, we do not see clear cut evidence that turbulence or shocks
are the major drivers of the PAH fraction in the SMC.  Young SNRs are
found around H II regions, making it difficult to separate the effects
of radiation field and shocks in the destruction of PAHs in those
regions.  The supergiant H I shells observed in the SMC may be a
tracer of where supernova shocks have seriously affected the ISM, but
we see peaks of \qpah\ within their boundaries barring a
line-of-sight depth effect.  

\subsection{Destruction of PAHs by UV Fields}\label{sec:uv}

In low metallicity galaxies, the dust to gas ratio is decreased
\citep{lisenfeld98} and the effects of the UV field from regions of
massive star formation can be spread over a much larger area.  In
addition, the decreased metallicity may lead to harder UV fields
because of the lower line blanketing in stellar atmospheres.  These
changes can be traced by the ratios of mid-IR emission lines and work
by \citet{madden00} and others have shown that radiation fields are
harder in low metallicity environments.  \citet{gordon08} studied
star-forming regions in M 101, over a range of excitation conditions
and metallicities.  They found that the strength of PAH emission
decreases with increasing ionization parameter, suggesting that
processing by the UV field from these regions was the major force
behind the changing PAH fraction.

The exact mechanism of PAH destruction by UV fields is not entirely
clear.  Small PAHs can be destroyed by the ejection of an acetylene
group upon absorbing a UV photon \citep{allain96a}, but PAHs larger
than $\sim 50$ carbon atoms are relatively stable to these effects
under a range of UV field conditions.  If the PAHs are partially
dehydrogentated or highly ionized, they are more susceptible to
destruction \citep{allain96b}, so one possibility is that PAHs in low
metallicity galaxies tend to be more highly ionized, dehydrogenated or
smaller than their counterparts in higher metallicity galaxies.
However, there is little evidence from the spectra of low
metallicity galaxies \citep{engelbracht08,smith07} that PAHs are
different at low metallicity.

There is also a distinction to be made between the destruction of PAHs
by intense radiation fields in the immediate vicinity of H II regions
and a global decrease of the PAH fraction in the galaxy.  We see
evidence for PAH destruction near H II regions in our map (see
Figure~\ref{fig:hacomp}).  But can UV fields from massive star forming
regions be responsible for the low \qpah\ over the entire
galaxy?  There is evidence that star-forming regions in irregular
galaxies can ``leak'' a large fraction of their ionizing photons.
Observations of emission line ratios from the diffuse ionized gas
(DIG) in star-forming dwarfs indicate that dilution of radiation from
a central massive star-forming region which leaks a significant
fraction of its ionizing photons is a likely source for the DIG
\citep{martin97}.  In the SMC, models of N 66 suggest that $\sim$ 45\%
of the ionizing photons escape the H II region and go on to ionize the
diffuse ISM \citep{relano02}.  Thus, it is at least plausible that UV
fields may be an important driver of PAH destruction over the entire
galaxy.  

\begin{figure*}
\centering
\epsscale{1.0}
\plottwo{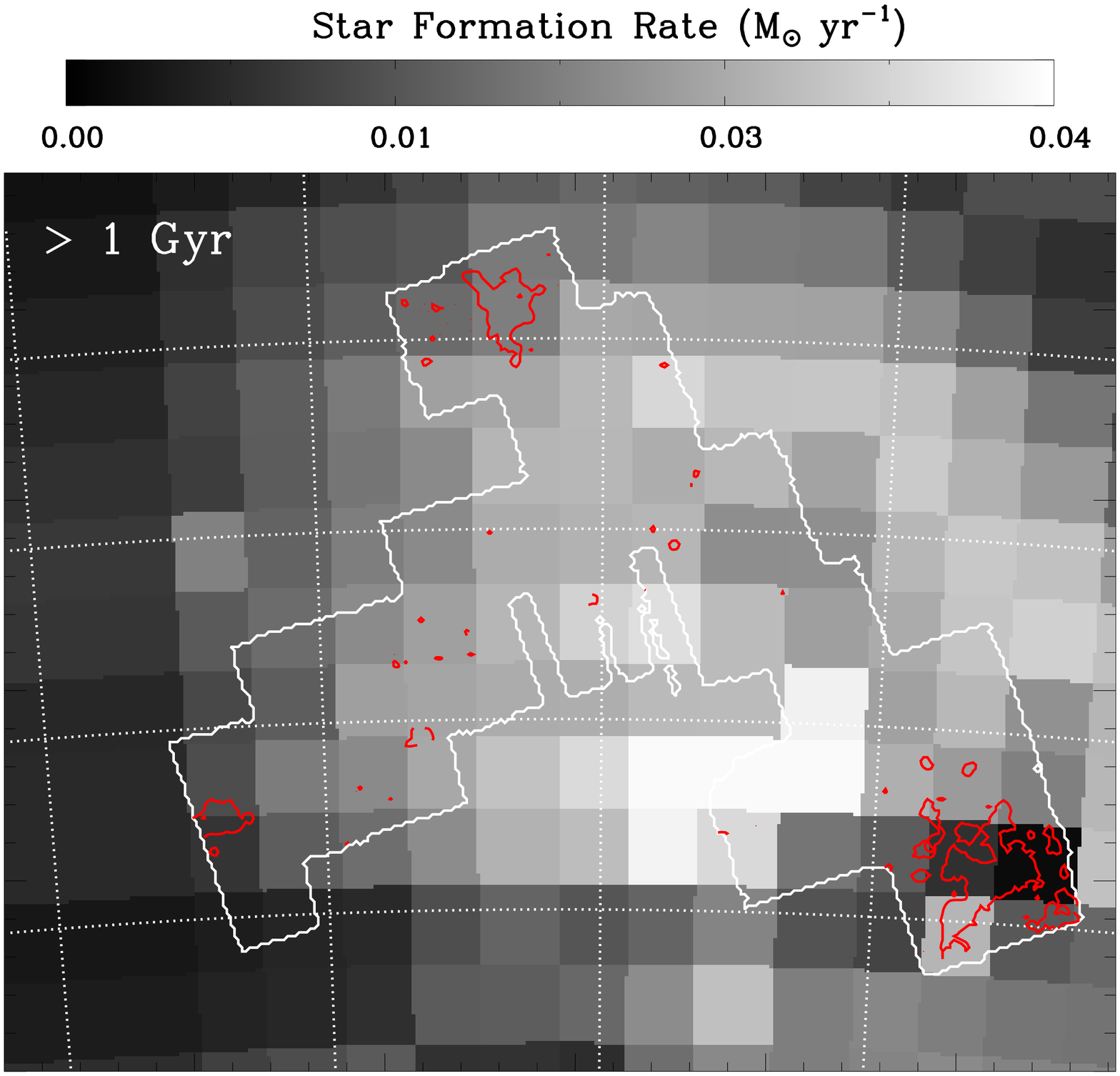}{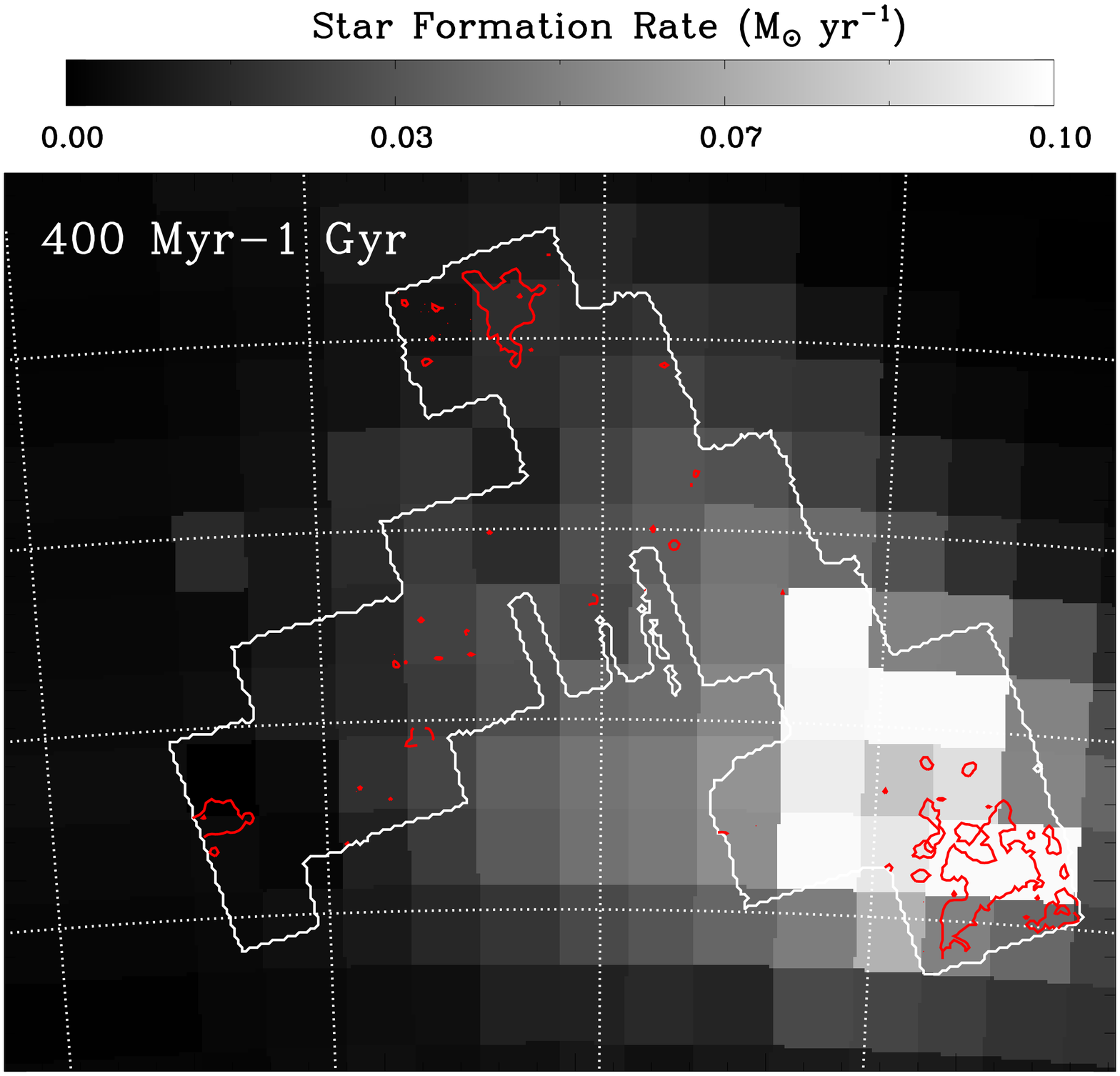}
\plottwo{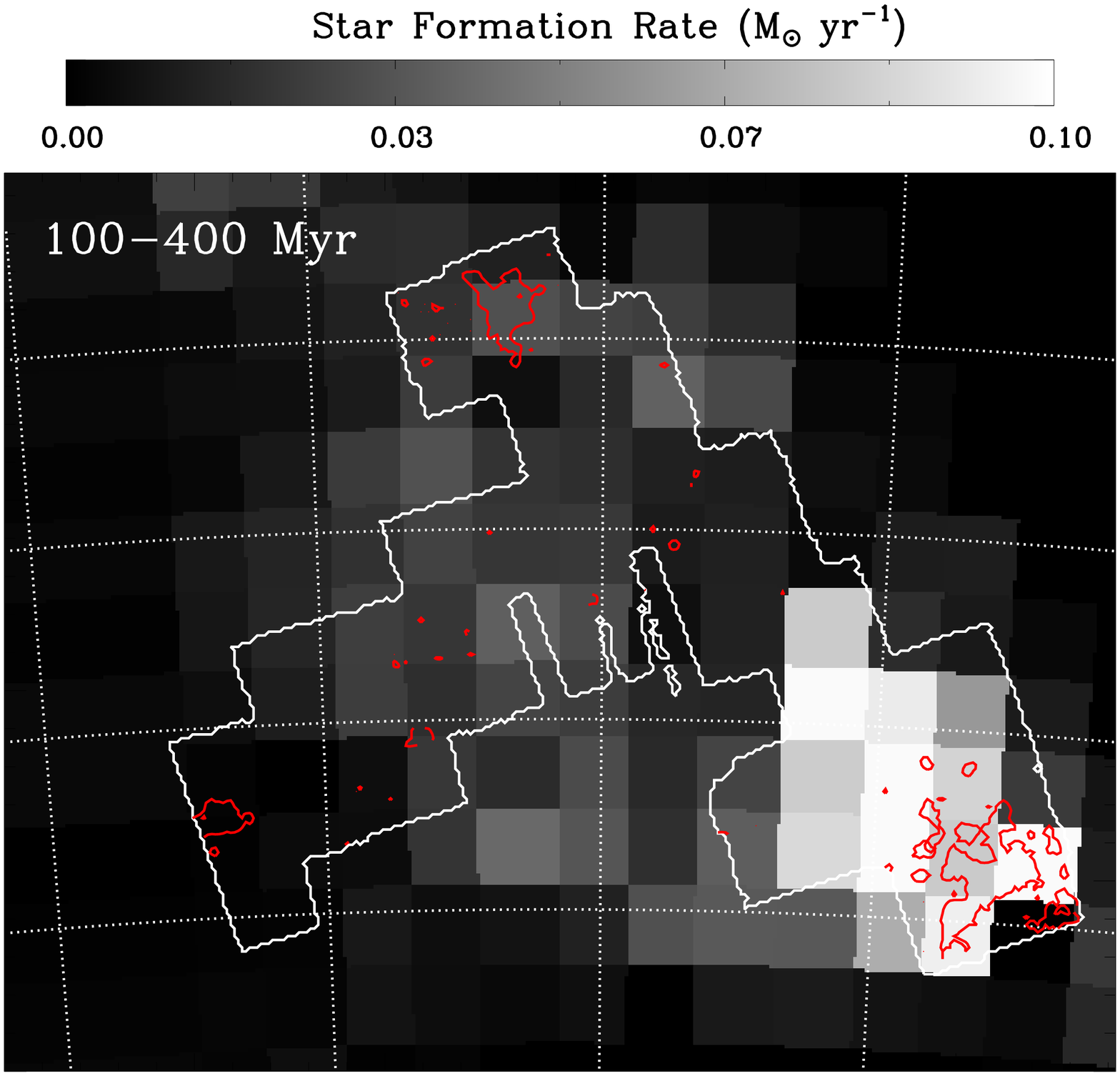}{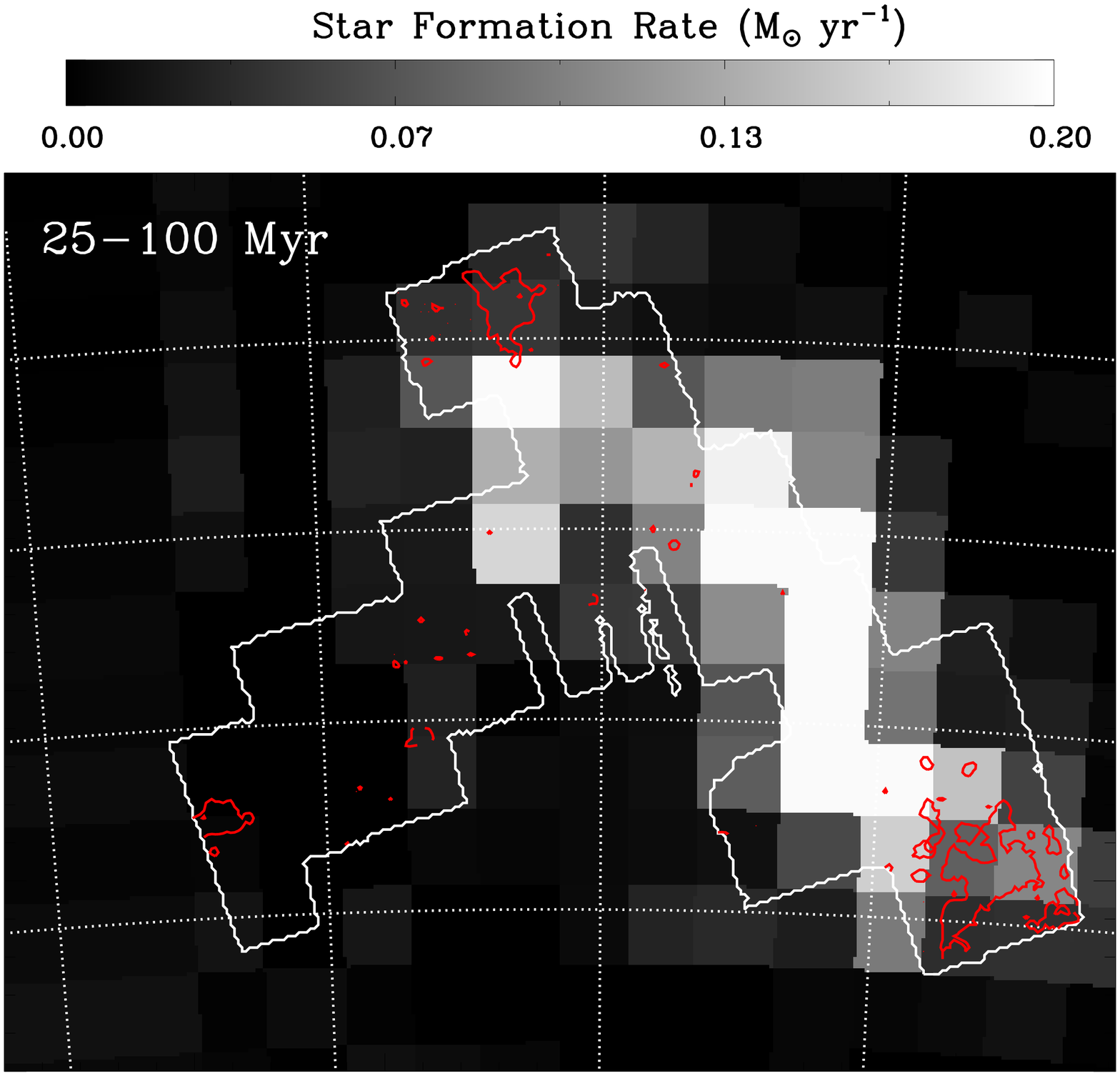}
\includegraphics[width=3.0in]{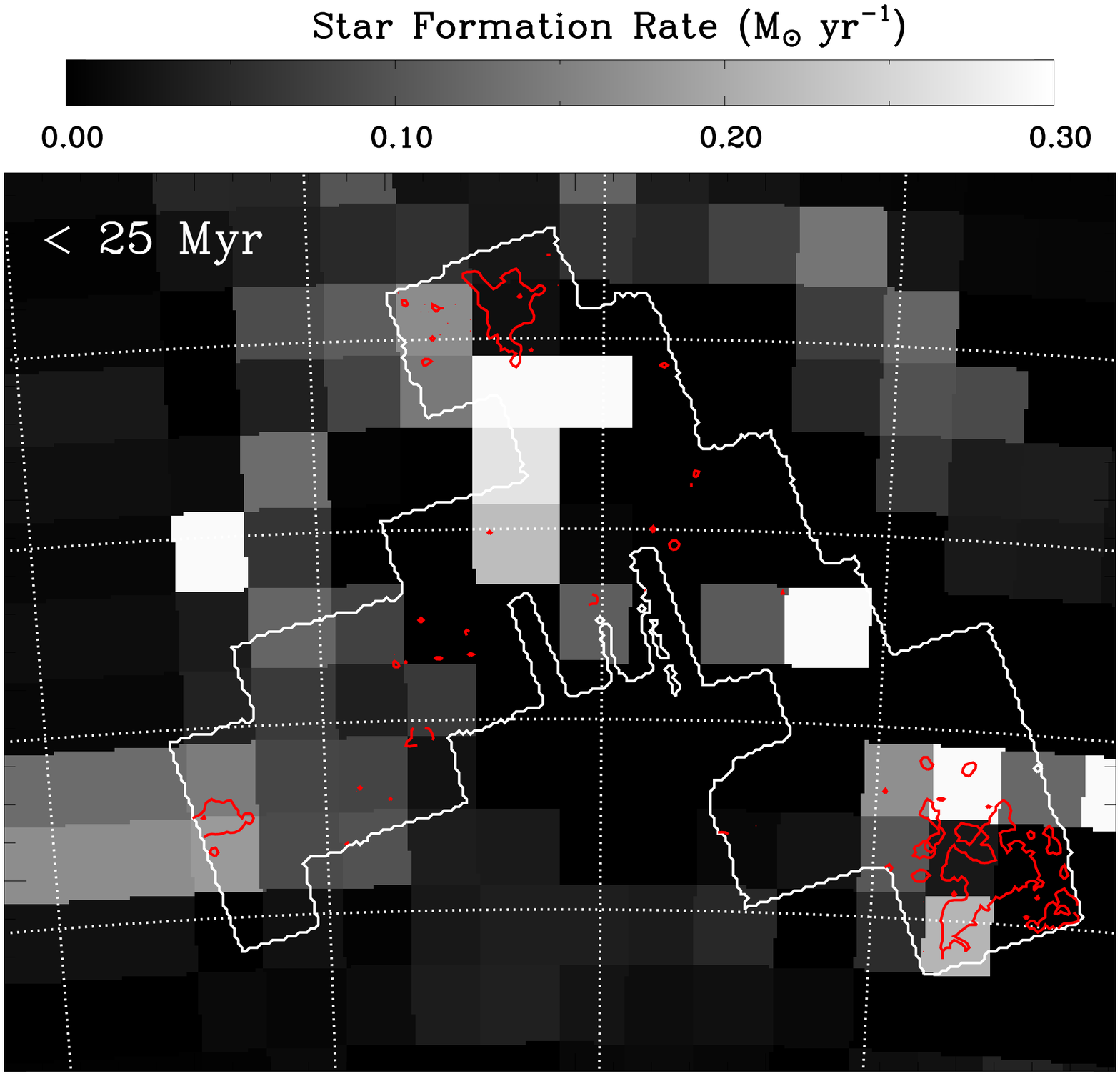}
\caption{Resolved star-formation history of the SMC from
\citet{harris04}. The time-ranges are listed in the upper left corners
of each image.}
\label{fig:sfh}
\end{figure*}

To evaluate the possibility that the low \qpah\ in the SMC is
due to destruction by UV fields, we examine the resolved
star-formation history of \citet{harris04} to search for recent
star-forming events that could have affected the PAH fraction in the
diffuse ISM through their UV fields.  In Figure~\ref{fig:sfh} we show
five panels illustrating the \citet{harris04} results regridded to
match our map of \qpah.  These panels show the total star
formation rate for bins older than 1 Gyr, from 400 Myr to 1 Gyr ago,
from 100 to 400 Myr ago, from 25 to 100 Myr ago and younger than 25
Myr. 

In general, panels which show the star-formation rate in bins older
than 100 Myr tend to follow the spheroidal distribution of older
stars, as traced by the carbon stars shown in Figure~\ref{fig:cstar}.
The $25-100$ Myr panel shows that star formation occurred along the
Bar, coinciding with the region in our \qpah\ map that is devoid of
PAHs.  The most recent star formation, which would overlap the time
when the current generation of molecular clouds condensed, seems to be
mainly associated with the outermost edges of the Bar, near where we
see PAHs.

PAH destruction as a side effect of massive star formation is
difficult to disentangle since the UV fields and subsequent supernovae
only affect the dust over a short period of time surrounding the star
formation event.  Between 25 and 100 Myr ago in the SMC, the
star-formation along the Bar, which most likely relates to the
supergiant bubble seen in the H I overlapping that location, may have
cleared the region of PAHs by some combination of UV fields and
shocks.  However, most of this activity likely occured before the
condensation of the molecular clouds \citep[$\sim 25$ Myr $>$
ago;][]{fukui99,blitz07,kawamura09} that currently have a high PAH
fraction, so there is still difficulty in reconciling the high \qpah\
in molecular clouds and the low \qpah\ in the diffuse ISM by
destruction alone.

\subsection{The Formation of PAHs in Molecular Clouds}

One of our primary observations in the SMC is that high PAH fractions 
occur along lines of sight through molecular gas.  We argue that such
a situation could arise in two scenarios: 1) AGB stars enrich the
diffuse ISM with PAHs, part of which is then incorporated into
molecular clouds. A subsequent event (e.g. SN) clears the PAHs in the
diffuse ISM.  Or, 2) PAH formation occurs in the molecular clouds.
These two scenarios are not mutually exclusive.  \citet{paradis09}
found enhanced PAH fraction in the LMC both in the stellar bar, which
hosts the highest concentration of AGB stars, and in molecular clouds.
It is possible that in the diffuse ISM of the SMC, AGB produced PAHs
are rapidly destroyed and all we observe are PAHs that formed in
molecular clouds.

\citet{greenberg00} propose a scenario by which PAHs could be formed
in dense clouds.  They argue that a layer of ices and organic material
forms on grains in dense clouds and the photoprocessing which occurs
in the transition from the dense cloud back to the diffuse ISM forms
PAHs, a phenomenon they explore through laboratory experimentation.
There is also observational evidence that hints at some of the
processing PAHs undergo in dense clouds.  \citet{rapacioli05} and
\citet{berne07} show variations in the PAH spectrum in
photo-dissociation regions consistent with PAHs being in clusters or
embedded in a carbonaceous matrix in the surrounding molecular gas and
emerging as free-flying PAHs closer to the exciting star.  

While it would not be out of question that a burst of star-formation
$\sim 25$ Myr ago could have a dramatic effect on the global diffuse
ISM of the SMC, a more likely scenario for the concentration of PAHs
in molecular clouds is that they form there.  Assuming that is the
case there are a number of interesting observations that we can make
about the formation of PAHs in molecular clouds in the SMC.  First,
the correspondence between the CO and \qpah\ is very
good, better even then the correspondence between the PAHs and the
Bolatto et al 2009 (in prep) map of molecular hydrogen inferred from
excess dust emission.  This suggests that the conditions for the
formation of CO and the formation of PAHs are similar.  It may be that
the UV field in the outskirts of the clouds, where H$_{2}$ can
self-shield but CO is dissociated, can disrupt the growth of mantles
on the grains, preventing a process that creates PAHs.
Alternatively, because the growth of mantles requires relatively high
densities, the presence of CO may allow the gas to cool and become
denser, increasing the rate of gas-grain reactions. This scenario could
be tested by comparing the distribution of CO and PAHs in a sample of
irregular galaxies.  \citet{leroy09} found that CO in the N 83 region
of the SMC is found where A$_V > 2$ through the cloud (A$_V > 1$ at
cloud center).  If CO survives and PAHs form at a similar extinction,
the decrease in the PAH fraction and the deficit of CO emission at
low metallicity may be related to the scarcity of regions where this
condition is met.  

Second, we observe that the typical \qpah\ in SMC molecular clouds is
$\sim 1-2$\%.  This is lower than the average Milky Way value (which
is for the diffuse ISM) and may represent some limit on the efficiency
of creating PAHs in molecular gas in the SMC.  In the low metallicity
galaxies studied by \citet{draine07b}, those with metallicities below
$12 +$ log(O/H) $\sim 8$ had a median PAH \qpah\ of 1\% with large
scatter.  This scatter could be due to the filling factor of molecular
gas. If PAHs are forming primarily in molecular clouds, the
differences observed in the PAH fraction in low versus high
metallicity galaxies may be related to the ability of PAHs to survive
and accumulate in the diffuse ISM and/or to the efficiency of PAH
formation in molecular gas, which can be decreased by the lower carbon
abundance and the more pervasive UV fields in low metallicity
environments.

Finally, in the scenario where PAHs are destroyed in the diffuse ISM
and formed in molecular clouds and AGB stars, the abrupt transition at
a metallicity of $12 +$ log(O/H) $\sim 8$ would represent mostly a
change in the efficiency of PAH destruction in the diffuse ISM.  If
PAHs are destroyed efficiently in the diffuse ISM of these systems,
all that is left are the PAHs created in molecular clouds.  A
resolved study of the PAH abundance in the SINGs galaxies using higher
resolution far-IR and millimeter-wave CO observations from Herschel
may be able to show whether the PAH fraction in the diffuse ISM
versus molecular clouds is changing across the transition metallicity.

\section{Summary \& Conclusions}\label{sec:conclusions}

We present results of fitting the \citet{draine07a} dust models to SEDs
and spectra obtained from S$^3$MC and S$^4$MC.  Our major results are
as follows:

\begin{enumerate}

\item Comparisons of best-fit dust models in regions with overlapping
      photometry and spectroscopy demonstrate that the IRAC and MIPS
      SED estimator for \qpah\ does not appear to be appreciably
      biased despite the absence of information between $8-24$ and
      $24-70$ \micron. When $5-38$ \micron\ IRS spectroscopy is added
      to the fitting constraints, the resulting ``photospectrofit''
      models yield \qpah\ values that are only slightly larger than
      the ``photofit'' model estimates.  The photospectrofit models
      have more dust at intermediate (T $\approx 60$K) temperatures,
      and less dust with T $\approx 200$ K.  The estimate for the
      average radiation intensity scale factor $\bar{U}$ and the dust
      surface density M$_{D}$ are nearly unaffected.

\item The PAH fraction in the SMC is low and variable.  As a fraction
      of the total dust mass, the highest PAH fractions we observe are
      about half of the Milky Way value, but most of the galaxy has
      PAH fractions at the lower limit of our models (\qpah= $0.4$\%),
      an order of magnitude lower than the Milky Way value. The
      average $\langle$\qpah$\rangle \sim 0.6$\% in the SMC.  This is
      consistent with the earlier estimate of a very low PAH fraction
      in the SMC \citep{li02} based on IRAS and DIRBE photometry.

\item The 8 \micron\ emission alone does not trace the \qpah\ well in
      the SMC.  We find that the 8/24 ratio is correlated with \qpah\
      and agrees with what one would predict from the metallicity of
      the SMC and the observed trends in \citet{engelbracht05}, but
      for a given \qpah\ there is a wide range of 8/24 ratios,
      depending on the intensity of the local starlight heating the
      grains.

\item The metallicity of the SMC places it at the transition where
      lower metallicity galaxies show a deficiency of PAHs and higher
      metallicity galaxies show approximately Milky Way level PAH
      fractions.  If the SMC is typical of galaxies in this region,
      the transition seems to represent a decrease in the filling
      factor of PAH-rich regions rather than a uniform decrease in the
      PAH fraction throughout the galaxy, although even the highest
      PAH fractions are still well below MW levels.

\item Lines-of-sight with molecular gas have an average \qpah\ of
      $\sim 1$\%, while lines of sight through the diffuse ISM have
      \qpah\ at least a factor of two less. 

\item We evaluate the various proposed drivers of the deficit of PAHs
      at low metallicity.  The distribution of PAHs in the
      SMC does not follow the carbon AGB star distribution, the
      regions of high turbulent Mach number, or the location of
      supergiant shells in the ISM.  
      
\item The low PAH fraction in the diffuse ISM versus the high PAH
      fraction in molecular regions leads us to propose that PAHs may
      be forming in molecular clouds and/or a recent event (perhaps
      related to star formation events in the last $\sim$ 25 Myr)
      destroyed a large fraction of the PAHs in the diffuse ISM.

\item We surmise that the global PAH fraction at low metallicities is
      a reflection of the amount of gas in these systems that is found
      at high extinction (A$_V > 1$).

\end{enumerate}

\acknowledgements

We would like to thank F. Winkler and C. Smith for providing the MCELS
data, B. Burkhart for providing the map of turbulent Mach number and
E. Muller for providing the ATCA/Parkes map of the Galactic
foreground.  We would also like to thank K. Gordon for the
non-linearity correction at 70 \micron. KS would like to thank Joshua
E. G. Peek and Kathryn Peek for helpful discussions and Carl Heiles
for helpful comments on the draft of this paper.  This work is
based on observations made with the \emph{Spitzer} Space Telescope,
which is operated by the Jet Propulsion Laboratory, California
Institute of Technology under a contract with NASA. This research was
supported in part by by NASA through awards issued by JPL/Caltech
(NASA-JPL Spitzer grant 1264151 awarded to Cycle 1 project 3316, and
grants 1287693 and 1289519 awarded to Cycle 3 project 30491).

{\it Facilities:} \facility{Spitzer ()}

\appendix
\subsection{Cross-Calibration of MIPS, IRAC and
IRS}\label{sec:crosscal}

Because we combine photometry and spectroscopy in our model fitting,
we investigated whether there were systematic differences between the
calibrations of the IRS and MIPS/IRAC instruments.  The SSC finds the
cross-calibration agrees to within 10\% for the three instruments.
However, most of these tests have been carried out on stellar sources,
whereas our observations are primarily concerned with extended
emission.  We have used the extended source corrections for IRAC
determined by \citet{reach05} and the slit loss correction function for IRS
distributed with Cubism.    

We have constructed 8 and 24 \micron\ maps from the spectral cubes
using the most recent versions of the spectral response curves
available from the SSC \citep[for IRAC;][]{hora08}.  We then compare
the IRS 8 and 24 \micron\ maps to the same regions in the IRAC and
MIPS mosaic after smoothing to eliminate any small PSF differences
between the two instruments.  Calibration differences between the
instruments present themselves as a non-unity slope in the comparison
of IRS 24 \micron\ to MIPS 24 \micron\ and IRS 8 \micron\ to IRAC 8
\micron.  For the 8 \micron\ comparison, we find that the IRS
photometry is higher than the IRAC photometry by factor of 1.1, on
average, although the slope in individual regions varies between 1.04
to 1.14.  At 24 \micron\, the IRS photometry is lower than the MIPS
photometry by a factor of 0.94 on average, varying between 0.91 and
0.96 region to region.  Individually, these differences are within the
quoted cross-calibration limits stated by the SSC, however without
correction they introduce scatter in the comparison of
spectroscopically and photometrically determined \qpah.

In addition to the slope, there is an additive offset between the IRS
and MIPS/IRAC photometry that is too large to be an issue with the
MIPS/IRAC mosaic foreground subtraction described previously (additive
offsets are $\sim 0.1-1$ MJy sr$^{-1}$ at 24 \micron\ and $\sim
0.01-0.1$ MJy sr$^{-1}$ at 8 \micron).  The source of this additive
offset is not definitively known, but we consider it very likely that
it results from slight under or over subtraction of the zodiacal light
foreground as determined by SPOT for the IRS cubes.  The uncertainties
of the DIRBE model are on the order of a few percent at the DIRBE
wavelengths, and particularly at 24 \micron\ where the zodiacal
foreground is very bright (on the order of tens of MJy sr$^{-1}$) and
a few percent error easily translates into offsets on the order of a
few tenths of a MJy sr$^{-1}$.  In order to eliminate these offsets,
we must make some assumption about the spectral shape of the
correction.  To do this we fix the ``off'' position zodiacal spectrum
and determine the percentage by which the ``map'' position spectrum
must be off to produce the offsets we observe.  We do this for each
AOR in the dataset and then add in the resulting correction spectrum.
The correction is less than 5\% in all cases.

The multiplicative differences in the calibration between IRS and MIPS
most likely result from the extended source calibration in IRS, since
MIPS does not require such a correction.  For the IRAC/IRS match-up,
the source is not obvious, since both instruments require an extended
source correction factor.  A further consideration in our work is that
we must apply a correction factor consistently across either all of
the IRS or all of the IRAC data, since we only have overlapping
information in a few regions, but we aim to determine the PAH
fractions over the whole S$^3$MC map.  There is no obvious way to
decide which to correct, the IRS or IRAC calibration, so we choose the
simplest route: we apply corrections to the IRS spectra based on the
factors necessary to match IRAC 8 \micron\ and MIPS 24 \micron.  We
determine a correction factor that depends linearly on wavelength
which matches the photometric to spectroscopic points in each region.
This correction factor is very small in the region of the SL1/LL2
overlap, so it does not affect our stitching of the orders.



\begin{thebibliography}{95}
\expandafter\ifx\csname natexlab\endcsname\relax\def\natexlab#1{#1}\fi

\bibitem[{{Allain} {et~al.}(1996{\natexlab{a}}){Allain}, {Leach}, \&
  {Sedlmayr}}]{allain96a}
{Allain}, T., {Leach}, S., \& {Sedlmayr}, E. 1996{\natexlab{a}}, \aap, 305, 602

\bibitem[{{Allain} {et~al.}(1996{\natexlab{b}}){Allain}, {Leach}, \&
  {Sedlmayr}}]{allain96b}
---. 1996{\natexlab{b}}, \aap, 305, 616

\bibitem[{{Allamandola} {et~al.}(1989){Allamandola}, {Tielens}, \&
  {Barker}}]{allamandola89}
{Allamandola}, L.~J., {Tielens}, A.~G.~G.~M., \& {Barker}, J.~R. 1989, \apjs,
  71, 733

\bibitem[{{Arendt} {et~al.}(1998){Arendt}, {Odegard}, {Weiland}, {Sodroski},
  {Hauser}, {Dwek}, {Kelsall}, {Moseley}, {Silverberg}, {Leisawitz},
  {Mitchell}, {Reach}, \& {Wright}}]{arendt98}
{Arendt}, R.~G., {Odegard}, N., {Weiland}, J.~L., {Sodroski}, T.~J., {Hauser},
  M.~G., {Dwek}, E., {Kelsall}, T., {Moseley}, S.~H., {Silverberg}, R.~F.,
  {Leisawitz}, D., {Mitchell}, K., {Reach}, W.~T., \& {Wright}, E.~L. 1998,
  \apj, 508, 74

\bibitem[{{Bakes} \& {Tielens}(1994)}]{bakes94}
{Bakes}, E.~L.~O., \& {Tielens}, A.~G.~G.~M. 1994, \apj, 427, 822

\bibitem[{{Bakes} \& {Tielens}(1998)}]{bakes98}
---. 1998, \apj, 499, 258

\bibitem[{{Bendo} {et~al.}(2008){Bendo}, {Draine}, {Engelbracht}, {Helou},
  {Thornley}, {Bot}, {Buckalew}, {Calzetti}, {Dale}, {Hollenbach}, {Li}, \&
  {Moustakas}}]{bendo08}
{Bendo}, G.~J., {Draine}, B.~T., {Engelbracht}, C.~W., {Helou}, G., {Thornley},
  M.~D., {Bot}, C., {Buckalew}, B.~A., {Calzetti}, D., {Dale}, D.~A.,
  {Hollenbach}, D.~J., {Li}, A., \& {Moustakas}, J. 2008, \mnras, 389, 629

\bibitem[{{Bernard} {et~al.}(2008){Bernard}, {Reach}, {Paradis}, {Meixner},
  {Paladini}, {Kawamura}, {Onishi}, {Vijh}, {Gordon}, {Indebetouw}, {Hora},
  {Whitney}, {Blum}, {Meade}, {Babler}, {Churchwell}, {Engelbracht}, {For},
  {Misselt}, {Leitherer}, {Cohen}, {Boulanger}, {Frogel}, {Fukui}, {Gallagher},
  {Gorjian}, {Harris}, {Kelly}, {Latter}, {Madden}, {Markwick-Kemper},
  {Mizuno}, {Mizuno}, {Mould}, {Nota}, {Oey}, {Olsen}, {Panagia},
  {Perez-Gonzalez}, {Shibai}, {Sato}, {Smith}, {Staveley-Smith}, {Tielens},
  {Ueta}, {Van Dyk}, {Volk}, {Werner}, \& {Zaritsky}}]{bernard08}
{Bernard}, J.-P., {Reach}, W.~T., {Paradis}, D., {Meixner}, M., {Paladini}, R.,
  {Kawamura}, A., {Onishi}, T., {Vijh}, U., {Gordon}, K., {Indebetouw}, R.,
  {Hora}, J.~L., {Whitney}, B., {Blum}, R., {Meade}, M., {Babler}, B.,
  {Churchwell}, E.~B., {Engelbracht}, C.~W., {For}, B.-Q., {Misselt}, K.,
  {Leitherer}, C., {Cohen}, M., {Boulanger}, F., {Frogel}, J.~A., {Fukui}, Y.,
  {Gallagher}, J., {Gorjian}, V., {Harris}, J., {Kelly}, D., {Latter}, W.~B.,
  {Madden}, S., {Markwick-Kemper}, C., {Mizuno}, A., {Mizuno}, N., {Mould}, J.,
  {Nota}, A., {Oey}, M.~S., {Olsen}, K., {Panagia}, N., {Perez-Gonzalez}, P.,
  {Shibai}, H., {Sato}, S., {Smith}, L., {Staveley-Smith}, L., {Tielens},
  A.~G.~G.~M., {Ueta}, T., {Van Dyk}, S., {Volk}, K., {Werner}, M., \&
  {Zaritsky}, D. 2008, \aj, 136, 919

\bibitem[{{Bern{\'e}} {et~al.}(2007){Bern{\'e}}, {Joblin}, {Deville}, {Smith},
  {Rapacioli}, {Bernard}, {Thomas}, {Reach}, \& {Abergel}}]{berne07}
{Bern{\'e}}, O., {Joblin}, C., {Deville}, Y., {Smith}, J.~D., {Rapacioli}, M.,
  {Bernard}, J.~P., {Thomas}, J., {Reach}, W., \& {Abergel}, A. 2007, \aap,
  469, 575

\bibitem[{{Blitz} {et~al.}(2007){Blitz}, {Fukui}, {Kawamura}, {Leroy},
  {Mizuno}, \& {Rosolowsky}}]{blitz07}
{Blitz}, L., {Fukui}, Y., {Kawamura}, A., {Leroy}, A., {Mizuno}, N., \&
  {Rosolowsky}, E. 2007, in Protostars and Planets V, ed. {B.~Reipurth,
  D.~Jewitt, \& K.~Keil}, 81--96

\bibitem[{{Bolatto} {et~al.}(2007){Bolatto}, {Simon}, {Stanimirovi{\'c}}, {van
  Loon}, {Shah}, {Venn}, {Leroy}, {Sandstrom}, {Jackson}, {Israel}, {Li},
  {Staveley-Smith}, {Bot}, {Boulanger}, \& {Rubio}}]{bolatto07}
{Bolatto}, A.~D., {Simon}, J.~D., {Stanimirovi{\'c}}, S., {van Loon}, J.~T.,
  {Shah}, R.~Y., {Venn}, K., {Leroy}, A.~K., {Sandstrom}, K., {Jackson}, J.~M.,
  {Israel}, F.~P., {Li}, A., {Staveley-Smith}, L., {Bot}, C., {Boulanger}, F.,
  \& {Rubio}, M. 2007, \apj, 655, 212

\bibitem[{{Bot} {et~al.}(2004){Bot}, {Boulanger}, {Lagache}, {Cambr{\'e}sy}, \&
  {Egret}}]{bot04}
{Bot}, C., {Boulanger}, F., {Lagache}, G., {Cambr{\'e}sy}, L., \& {Egret}, D.
  2004, \aap, 423, 567

\bibitem[{{Boulanger} {et~al.}(1996){Boulanger}, {Abergel}, {Bernard},
  {Burton}, {Desert}, {Hartmann}, {Lagache}, \& {Puget}}]{boulanger96}
{Boulanger}, F., {Abergel}, A., {Bernard}, J.-P., {Burton}, W.~B., {Desert},
  F.-X., {Hartmann}, D., {Lagache}, G., \& {Puget}, J.-L. 1996, \aap, 312, 256

\bibitem[{{Burkhart} {et~al.}(2010){Burkhart}, {Stanimirovi{\'c}},
  {Lazarian}, \& {Kowal}}]{burkhart10} 
{Burkhart}, B., {Stanimirovi{\'c}}, S., {Lazarian}, A., \& {Kowal}, G. 2010, \apj,
  708, 1204

\bibitem[{{Buss} {et~al.}(1993){Buss}, {Tielens}, {Cohen}, {Werner}, {Bregman},
  \& {Witteborn}}]{buss93}
{Buss}, Jr., R.~H., {Tielens}, A.~G.~G.~M., {Cohen}, M., {Werner}, M.~W.,
  {Bregman}, J.~D., \& {Witteborn}, F.~C. 1993, \apj, 415, 250

\bibitem[{{Calzetti} {et~al.}(2007){Calzetti}, {Kennicutt}, {Engelbracht},
  {Leitherer}, {Draine}, {Kewley}, {Moustakas}, {Sosey}, {Dale}, {Gordon},
  {Helou}, {Hollenbach}, {Armus}, {Bendo}, {Bot}, {Buckalew}, {Jarrett}, {Li},
  {Meyer}, {Murphy}, {Prescott}, {Regan}, {Rieke}, {Roussel}, {Sheth}, {Smith},
  {Thornley}, \& {Walter}}]{calzetti07}
{Calzetti}, D., {Kennicutt}, R.~C., {Engelbracht}, C.~W., {Leitherer}, C.,
  {Draine}, B.~T., {Kewley}, L., {Moustakas}, J., {Sosey}, M., {Dale}, D.~A.,
  {Gordon}, K.~D., {Helou}, G.~X., {Hollenbach}, D.~J., {Armus}, L., {Bendo},
  G., {Bot}, C., {Buckalew}, B., {Jarrett}, T., {Li}, A., {Meyer}, M.,
  {Murphy}, E.~J., {Prescott}, M., {Regan}, M.~W., {Rieke}, G.~H., {Roussel},
  H., {Sheth}, K., {Smith}, J.~D.~T., {Thornley}, M.~D., \& {Walter}, F. 2007,
  \apj, 666, 870

\bibitem[{{Cannon} {et~al.}(2006){Cannon}, {Smith}, {Walter}, {Bendo},
  {Calzetti}, {Dale}, {Draine}, {Engelbracht}, {Gordon}, {Helou}, {Kennicutt},
  {Leitherer}, {Armus}, {Buckalew}, {Hollenbach}, {Jarrett}, {Li}, {Meyer},
  {Murphy}, {Regan}, {Rieke}, {Rieke}, {Roussel}, {Sheth}, \&
  {Thornley}}]{cannon06a}
{Cannon}, J.~M., {Smith}, J.-D.~T., {Walter}, F., {Bendo}, G.~J., {Calzetti},
  D., {Dale}, D.~A., {Draine}, B.~T., {Engelbracht}, C.~W., {Gordon}, K.~D.,
  {Helou}, G., {Kennicutt}, Jr., R.~C., {Leitherer}, C., {Armus}, L.,
  {Buckalew}, B.~A., {Hollenbach}, D.~J., {Jarrett}, T.~H., {Li}, A., {Meyer},
  M.~J., {Murphy}, E.~J., {Regan}, M.~W., {Rieke}, G.~H., {Rieke}, M.~J.,
  {Roussel}, H., {Sheth}, K., \& {Thornley}, M.~D. 2006, \apj, 647, 293

\bibitem[{{Cartledge} {et~al.}(2005){Cartledge}, {Clayton}, {Gordon},
  {Rachford}, {Draine}, {Martin}, {Mathis}, {Misselt}, {Sofia}, {Whittet}, \&
  {Wolff}}]{cartledge05}
{Cartledge}, S.~I.~B., {Clayton}, G.~C., {Gordon}, K.~D., {Rachford}, B.~L.,
  {Draine}, B.~T., {Martin}, P.~G., {Mathis}, J.~S., {Misselt}, K.~A., {Sofia},
  U.~J., {Whittet}, D.~C.~B., \& {Wolff}, M.~J. 2005, \apj, 630, 355

\bibitem[{{Cherchneff} {et~al.}(1992){Cherchneff}, {Barker}, \&
  {Tielens}}]{cherchneff92}
{Cherchneff}, I., {Barker}, J.~R., \& {Tielens}, A.~G.~G.~M. 1992, \apj, 401,
  269

\bibitem[{{Cioni} {et~al.}(2006){Cioni}, {Girardi}, {Marigo}, \&
  {Habing}}]{cioni06}
{Cioni}, M.-R.~L., {Girardi}, L., {Marigo}, P., \& {Habing}, H.~J. 2006, \aap,
  452, 195

\bibitem[{{Cioni} {et~al.}(2000){Cioni}, {Habing}, \& {Israel}}]{cioni00}
{Cioni}, M.-R.~L., {Habing}, H.~J., \& {Israel}, F.~P. 2000, \aap, 358, L9

\bibitem[{{Cohen}(2009)}]{cohen09}
{Cohen}, M. 2009, \aj, 137, 3449

\bibitem[{{Cohen} {et~al.}(2007){Cohen}, {Green}, {Meade}, {Babler},
  {Indebetouw}, {Whitney}, {Watson}, {Wolfire}, {Wolff}, {Mathis}, \&
  {Churchwell}}]{cohen07}
{Cohen}, M., {Green}, A.~J., {Meade}, M.~R., {Babler}, B., {Indebetouw}, R.,
  {Whitney}, B.~A., {Watson}, C., {Wolfire}, M., {Wolff}, M.~J., {Mathis},
  J.~S., \& {Churchwell}, E.~B. 2007, \mnras, 374, 979

\bibitem[{{Dale} {et~al.}(2007){Dale}, {Gil de Paz}, {Gordon}, {Hanson},
  {Armus}, {Bendo}, {Bianchi}, {Block}, {Boissier}, {Boselli}, {Buckalew},
  {Buat}, {Burgarella}, {Calzetti}, {Cannon}, {Engelbracht}, {Helou},
  {Hollenbach}, {Jarrett}, {Kennicutt}, {Leitherer}, {Li}, {Madore}, {Martin},
  {Meyer}, {Murphy}, {Regan}, {Roussel}, {Smith}, {Sosey}, {Thilker}, \&
  {Walter}}]{dale07}
{Dale}, D.~A., {Gil de Paz}, A., {Gordon}, K.~D., {Hanson}, H.~M., {Armus}, L.,
  {Bendo}, G.~J., {Bianchi}, L., {Block}, M., {Boissier}, S., {Boselli}, A.,
  {Buckalew}, B.~A., {Buat}, V., {Burgarella}, D., {Calzetti}, D., {Cannon},
  J.~M., {Engelbracht}, C.~W., {Helou}, G., {Hollenbach}, D.~J., {Jarrett},
  T.~H., {Kennicutt}, R.~C., {Leitherer}, C., {Li}, A., {Madore}, B.~F.,
  {Martin}, D.~C., {Meyer}, M.~J., {Murphy}, E.~J., {Regan}, M.~W., {Roussel},
  H., {Smith}, J.~D.~T., {Sosey}, M.~L., {Thilker}, D.~A., \& {Walter}, F.
  2007, \apj, 655, 863

\bibitem[{{Desert} {et~al.}(1990){Desert}, {Boulanger}, \& {Puget}}]{desert90}
{Desert}, F.-X., {Boulanger}, F., \& {Puget}, J.~L. 1990, \aap, 237, 215

\bibitem[{{Draine}(2009)}]{draine09}
{Draine}, B.~T. 2009, ArXiv e-prints

\bibitem[{{Draine} {et~al.}(2007){Draine}, {Dale}, {Bendo}, {Gordon}, {Smith},
  {Armus}, {Engelbracht}, {Helou}, {Kennicutt}, {Li}, {Roussel}, {Walter},
  {Calzetti}, {Moustakas}, {Murphy}, {Rieke}, {Bot}, {Hollenbach}, {Sheth}, \&
  {Teplitz}}]{draine07b}
{Draine}, B.~T., {Dale}, D.~A., {Bendo}, G., {Gordon}, K.~D., {Smith},
  J.~D.~T., {Armus}, L., {Engelbracht}, C.~W., {Helou}, G., {Kennicutt}, Jr.,
  R.~C., {Li}, A., {Roussel}, H., {Walter}, F., {Calzetti}, D., {Moustakas},
  J., {Murphy}, E.~J., {Rieke}, G.~H., {Bot}, C., {Hollenbach}, D.~J., {Sheth},
  K., \& {Teplitz}, H.~I. 2007, \apj, 663, 866

\bibitem[{{Draine} \& {Li}(2007)}]{draine07a}
{Draine}, B.~T., \& {Li}, A. 2007, \apj, 657, 810

\bibitem[{{Dunne} {et~al.}(2003){Dunne}, {Eales}, {Ivison}, {Morgan}, \&
  {Edmunds}}]{dunne03}
{Dunne}, L., {Eales}, S., {Ivison}, R., {Morgan}, H., \& {Edmunds}, M. 2003,
  \nat, 424, 285

\bibitem[{{Dwek}(1998)}]{dwek98}
{Dwek}, E. 1998, \apj, 501, 643

\bibitem[{{Engelbracht} {et~al.}(2005){Engelbracht}, {Gordon}, {Rieke},
  {Werner}, {Dale}, \& {Latter}}]{engelbracht05}
{Engelbracht}, C.~W., {Gordon}, K.~D., {Rieke}, G.~H., {Werner}, M.~W., {Dale},
  D.~A., \& {Latter}, W.~B. 2005, \apjl, 628, L29

\bibitem[{{Engelbracht} {et~al.}(2008){Engelbracht}, {Rieke}, {Gordon},
  {Smith}, {Werner}, {Moustakas}, {Willmer}, \& {Vanzi}}]{engelbracht08}
{Engelbracht}, C.~W., {Rieke}, G.~H., {Gordon}, K.~D., {Smith}, J.-D.~T.,
  {Werner}, M.~W., {Moustakas}, J., {Willmer}, C.~N.~A., \& {Vanzi}, L. 2008,
  \apj, 678, 804

\bibitem[{{Fixsen} \& {Dwek}(2002)}]{fixsen02}
{Fixsen}, D.~J., \& {Dwek}, E. 2002, \apj, 578, 1009

\bibitem[{{Fixsen} {et~al.}(1998){Fixsen}, {Dwek}, {Mather}, {Bennett}, \&
  {Shafer}}]{fixsen98}
{Fixsen}, D.~J., {Dwek}, E., {Mather}, J.~C., {Bennett}, C.~L., \& {Shafer},
  R.~A. 1998, \apj, 508, 123

\bibitem[{{Fukui} {et~al.}(1999){Fukui}, {Mizuno}, {Yamaguchi}, {Mizuno},
  {Onishi}, {Ogawa}, {Yonekura}, {Kawamura}, {Tachihara}, {Xiao}, {Yamaguchi},
  {Hara}, {Hayakawa}, {Kato}, {Abe}, {Saito}, {Mano}, {Matsunaga}, {Mine},
  {Moriguchi}, {Aoyama}, {Asayama}, {Yoshikawa}, \& {Rubio}}]{fukui99}
{Fukui}, Y., {Mizuno}, N., {Yamaguchi}, R., {Mizuno}, A., {Onishi}, T.,
  {Ogawa}, H., {Yonekura}, Y., {Kawamura}, A., {Tachihara}, K., {Xiao}, K.,
  {Yamaguchi}, N., {Hara}, A., {Hayakawa}, T., {Kato}, S., {Abe}, R., {Saito},
  H., {Mano}, S., {Matsunaga}, K., {Mine}, Y., {Moriguchi}, Y., {Aoyama}, H.,
  {Asayama}, S.-i., {Yoshikawa}, N., \& {Rubio}, M. 1999, \pasj, 51, 745

\bibitem[{{Galliano} {et~al.}(2008){Galliano}, {Dwek}, \&
  {Chanial}}]{galliano08}
{Galliano}, F., {Dwek}, E., \& {Chanial}, P. 2008, \apj, 672, 214

\bibitem[{{Galliano} {et~al.}(2005){Galliano}, {Madden}, {Jones}, {Wilson}, \&
  {Bernard}}]{galliano05}
{Galliano}, F., {Madden}, S.~C., {Jones}, A.~P., {Wilson}, C.~D., \& {Bernard},
  J.-P. 2005, \aap, 434, 867

\bibitem[{{Giard} {et~al.}(1994){Giard}, {Bernard}, {Lacombe}, {Normand}, \&
  {Rouan}}]{giard94}
{Giard}, M., {Bernard}, J.~P., {Lacombe}, F., {Normand}, P., \& {Rouan}, D.
  1994, \aap, 291, 239

\bibitem[{{Gordon} {et~al.}(2003){Gordon}, {Clayton}, {Misselt}, {Landolt}, \&
  {Wolff}}]{gordon03}
{Gordon}, K.~D., {Clayton}, G.~C., {Misselt}, K.~A., {Landolt}, A.~U., \&
  {Wolff}, M.~J. 2003, \apj, 594, 279

\bibitem[{{Gordon} {et~al.}(2008){Gordon}, {Engelbracht}, {Rieke}, {Misselt},
  {Smith}, \& {Kennicutt}}]{gordon08}
{Gordon}, K.~D., {Engelbracht}, C.~W., {Rieke}, G.~H., {Misselt}, K.~A.,
  {Smith}, J.-D.~T., \& {Kennicutt}, Jr., R.~C. 2008, \apj, 682, 336

\bibitem[{{Greenberg} {et~al.}(2000){Greenberg}, {Gillette}, {Mu{\~n}oz Caro},
  {Mahajan}, {Zare}, {Li}, {Schutte}, {de Groot}, \&
  {Mendoza-G{\'o}mez}}]{greenberg00}
{Greenberg}, J.~M., {Gillette}, J.~S., {Mu{\~n}oz Caro}, G.~M., {Mahajan},
  T.~B., {Zare}, R.~N., {Li}, A., {Schutte}, W.~A., {de Groot}, M., \&
  {Mendoza-G{\'o}mez}, C. 2000, \apjl, 531, L71

\bibitem[{{Haas} {et~al.}(2002){Haas}, {Klaas}, \& {Bianchi}}]{haas02}
{Haas}, M., {Klaas}, U., \& {Bianchi}, S. 2002, \aap, 385, L23

\bibitem[{{Harris} \& {Zaritsky}(2004)}]{harris04}
{Harris}, J., \& {Zaritsky}, D. 2004, \aj, 127, 1531

\bibitem[{{Herbst}(1991)}]{herbst91}
{Herbst}, E. 1991, \apj, 366, 133

\bibitem[{{Heiles} \& {Troland} (2003)}]{heiles03}
{Heiles}, C. \& {Troland}, T.~H. 2003, \apj, 586, 1067

\bibitem[{{Hilditch} {et~al.}(2005){Hilditch}, {Howarth}, \&
  {Harries}}]{hilditch05}
{Hilditch}, R.~W., {Howarth}, I.~D., \& {Harries}, T.~J. 2005, \mnras, 357, 304

\bibitem[{{Hora} {et~al.}(2008){Hora}, {Carey}, {Surace}, {Marengo},
  {Lowrance}, {Glaccum}, {Lacy}, {Reach}, {Hoffmann}, {Barmby}, {Willner},
  {Fazio}, {Megeath}, {Allen}, {Bhattacharya}, \& {Quijada}}]{hora08}
{Hora}, J.~L., {Carey}, S., {Surace}, J., {Marengo}, M., {Lowrance}, P.,
  {Glaccum}, W.~J., {Lacy}, M., {Reach}, W.~T., {Hoffmann}, W.~F., {Barmby},
  P., {Willner}, S.~P., {Fazio}, G.~G., {Megeath}, S.~T., {Allen}, L.~E.,
  {Bhattacharya}, B., \& {Quijada}, M. 2008, \pasp, 120, 1233

\bibitem[{{Hunter} {et~al.}(2006){Hunter}, {Elmegreen}, \& {Martin}}]{hunter06}
{Hunter}, D.~A., {Elmegreen}, B.~G., \& {Martin}, E. 2006, \aj, 132, 801

\bibitem[{{Jackson} {et~al.}(2006){Jackson}, {Cannon}, {Skillman}, {Lee},
  {Gehrz}, {Woodward}, \& {Polomski}}]{jackson06}
{Jackson}, D.~C., {Cannon}, J.~M., {Skillman}, E.~D., {Lee}, H., {Gehrz},
  R.~D., {Woodward}, C.~E., \& {Polomski}, E. 2006, \apj, 646, 192

\bibitem[{{Jones} {et~al.}(1996){Jones}, {Tielens}, \& {Hollenbach}}]{jones96}
{Jones}, A.~P., {Tielens}, A.~G.~G.~M., \& {Hollenbach}, D.~J. 1996, \apj, 469,
  740

\bibitem[{{Jones} {et~al.}(1994){Jones}, {Tielens}, {Hollenbach}, \&
  {McKee}}]{jones94}
{Jones}, A.~P., {Tielens}, A.~G.~G.~M., {Hollenbach}, D.~J., \& {McKee}, C.~F.
  1994, \apj, 433, 797

\bibitem[{{Justtanont} {et~al.}(1996){Justtanont}, {Barlow}, {Skinner},
  {Roche}, {Aitken}, \& {Smith}}]{justtanont96}
{Justtanont}, K., {Barlow}, M.~J., {Skinner}, C.~J., {Roche}, P.~F., {Aitken},
  D.~K., \& {Smith}, C.~H. 1996, \aap, 309, 612

\bibitem[{{Kawamura} {et~al.}(2009){Kawamura}, {Mizuno}, {Minamidani},
  {Filipovi{\'c}}, {Staveley-Smith}, {Kim}, {Mizuno}, {Onishi}, {Mizuno}, \&
  {Fukui}}]{kawamura09}
{Kawamura}, A., {Mizuno}, Y., {Minamidani}, T., {Filipovi{\'c}}, M.~D.,
  {Staveley-Smith}, L., {Kim}, S., {Mizuno}, N., {Onishi}, T., {Mizuno}, A., \&
  {Fukui}, Y. 2009, \apjs, 184, 1

\bibitem[{{Kelsall} {et~al.}(1998){Kelsall}, {Weiland}, {Franz}, {Reach},
  {Arendt}, {Dwek}, {Freudenreich}, {Hauser}, {Moseley}, {Odegard},
  {Silverberg}, \& {Wright}}]{kelsall98}
{Kelsall}, T., {Weiland}, J.~L., {Franz}, B.~A., {Reach}, W.~T., {Arendt},
  R.~G., {Dwek}, E., {Freudenreich}, H.~T., {Hauser}, M.~G., {Moseley}, S.~H.,
  {Odegard}, N.~P., {Silverberg}, R.~F., \& {Wright}, E.~L. 1998, \apj, 508, 44

\bibitem[{{Krause} {et~al.}(2004){Krause}, {Birkmann}, {Rieke}, {Lemke},
  {Klaas}, {Hines}, \& {Gordon}}]{krause04}
{Krause}, O., {Birkmann}, S.~M., {Rieke}, G.~H., {Lemke}, D., {Klaas}, U.,
  {Hines}, D.~C., \& {Gordon}, K.~D. 2004, \nat, 432, 596

\bibitem[{{Kurt} \& {Dufour}(1998)}]{kurt98}
{Kurt}, C.~M., \& {Dufour}, R.~J. 1998, in Revista Mexicana de Astronomia y
  Astrofisica Conference Series, Vol.~7, Revista Mexicana de Astronomia y
  Astrofisica Conference Series, ed. R.~J. {Dufour} \& S.~{Torres-Peimbert},
  202--+

\bibitem[{{Latter}(1991)}]{latter91}
{Latter}, W.~B. 1991, \apj, 377, 187

\bibitem[{{Leroy} {et~al.}(2007){Leroy}, {Bolatto}, {Stanimirovic}, {Mizuno},
  {Israel}, \& {Bot}}]{leroy07}
{Leroy}, A., {Bolatto}, A., {Stanimirovic}, S., {Mizuno}, N., {Israel}, F., \&
  {Bot}, C. 2007, \apj, 658, 1027

\bibitem[{{Leroy} {et~al.}(2009){Leroy}, {Bolatto}, {Bot}, {Engelbracht},
  {Gordon}, {Israel}, {Rubio}, {Sandstrom}, \& {Stanimirovi{\'c}}}]{leroy09}
{Leroy}, A.~K., {Bolatto}, A., {Bot}, C., {Engelbracht}, C.~W., {Gordon}, K.,
  {Israel}, F.~P., {Rubio}, M., {Sandstrom}, K., \& {Stanimirovi{\'c}}, S.
  2009, ArXiv e-prints

\bibitem[{{Li} \& {Draine}(2001)}]{li01}
{Li}, A., \& {Draine}, B.~T. 2001, \apj, 554, 778

\bibitem[{{Li} \& {Draine}(2002)}]{li02}
---. 2002, \apj, 576, 762

\bibitem[{{Lisenfeld} \& {Ferrara}(1998)}]{lisenfeld98}
{Lisenfeld}, U. \& {Ferrara}, A. 1998, \apj, 496, 145

\bibitem[{{Madden}(2000)}]{madden00}
{Madden}, S.~C. 2000, New Astronomy Review, 44, 249

\bibitem[{{Madden} {et~al.}(2006){Madden}, {Galliano}, {Jones}, \&
  {Sauvage}}]{madden06}
{Madden}, S.~C., {Galliano}, F., {Jones}, A.~P., \& {Sauvage}, M. 2006, \aap,
  446, 877

\bibitem[{{Martin}(1997)}]{martin97}
{Martin}, C.~L. 1997, \apj, 491, 561

\bibitem[{{Mathis} {et~al.}(1983){Mathis}, {Mezger}, \& {Panagia}}]{mathis83}
{Mathis}, J.~S., {Mezger}, P.~G., \& {Panagia}, N. 1983, \aap, 128, 212

\bibitem[{{Matsuura} {et~al.}(2009){Matsuura}, {Barlow}, {Zijlstra},
  {Whitelock}, {Cioni}, {Groenewegen}, {Volk}, {Kemper}, {Kodama}, {Lagadec},
  {Meixner}, {Sloan}, \& {Srinivasan}}]{matsuura09}
{Matsuura}, M., {Barlow}, M.~J., {Zijlstra}, A.~A., {Whitelock}, P.~A.,
  {Cioni}, M.-R.~L., {Groenewegen}, M.~A.~T., {Volk}, K., {Kemper}, F.,
  {Kodama}, T., {Lagadec}, E., {Meixner}, M., {Sloan}, G.~C., \& {Srinivasan},
  S. 2009, \mnras, 396, 918

\bibitem[{{Meikle} {et~al.}(2007){Meikle}, {Mattila}, {Pastorello}, {Gerardy},
  {Kotak}, {Sollerman}, {Van Dyk}, {Farrah}, {Filippenko}, {H{\"o}flich},
  {Lundqvist}, {Pozzo}, \& {Wheeler}}]{meikle07}
{Meikle}, W.~P.~S., {Mattila}, S., {Pastorello}, A., {Gerardy}, C.~L., {Kotak},
  R., {Sollerman}, J., {Van Dyk}, S.~D., {Farrah}, D., {Filippenko}, A.~V.,
  {H{\"o}flich}, P., {Lundqvist}, P., {Pozzo}, M., \& {Wheeler}, J.~C. 2007,
  \apj, 665, 608

\bibitem[{{Miville-Desch{\^e}nes} {et~al.}(2002){Miville-Desch{\^e}nes},
  {Boulanger}, {Joncas}, \& {Falgarone}}]{miville-deschenes02}
{Miville-Desch{\^e}nes}, M.-A., {Boulanger}, F., {Joncas}, G., \& {Falgarone},
  E. 2002, \aap, 381, 209

\bibitem[{{Mizuno} {et~al.}(2001){Mizuno}, {Rubio}, {Mizuno}, {Yamaguchi},
  {Onishi}, \& {Fukui}}]{mizuno01}
{Mizuno}, N., {Rubio}, M., {Mizuno}, A., {Yamaguchi}, R., {Onishi}, T., \&
  {Fukui}, Y. 2001, \pasj, 53, L45

\bibitem[{{Moseley} {et~al.}(1989){Moseley}, {Dwek}, {Glaccum}, {Graham}, \&
  {Loewenstein}}]{moseley89}
{Moseley}, S.~H., {Dwek}, E., {Glaccum}, W., {Graham}, J.~R., \& {Loewenstein},
  R.~F. 1989, \nat, 340, 697

\bibitem[{{Mu{\~n}oz-Mateos} {et~al.}(2009){Mu{\~n}oz-Mateos}, {Gil de
  Paz}, {Boissier}, {Zamorano}, {Dale}, {P{\'e}rez-Gonz{\'a}lez},
  {Gallego}, {Madore}, {Bendo}, {Thornley}, {Draine}, {Boselli}, {Buat},
  {Calzetti}, {Moustakas}, \& {Kennicutt}}]{munoz-mateos09}
{Mu{\~n}oz-Mateos}, J.~C., {Gil de Paz}, A., {Boissier}, S.,
  {Zamorano}, J., {Dale}, D.~A., {P{\'e}rez-Gonz{\'a}lez}, P.~G.,
  {Gallego}, J., {Madore}, B.~F., {Bendo}, G., {Thornley}, M.~D.,
  {Draine}, B.~T., {Boselli}, A., {Buat}, V., {Calzetti}, D.,
  {Moustakas}, J., \& {Kennicutt}, R.~C. 2009, \apj, 701, 1965

\bibitem[{{O'Halloran} {et~al.}(2006){O'Halloran}, {Satyapal}, \&
  {Dudik}}]{ohalloran06}
{O'Halloran}, B., {Satyapal}, S., \& {Dudik}, R.~P. 2006, \apj, 641, 795

\bibitem[{{Osterbrock} \& {Ferland}(2006)}]{osterbrock06}
{Osterbrock}, D.~E., \& {Ferland}, G.~J. 2006, {Astrophysics of gaseous nebulae
  and active galactic nuclei}, ed. D.~E. {Osterbrock} \& G.~J. {Ferland}

\bibitem[{{Paradis} {et~al.}(2009){Paradis}, {Reach}, {Bernard}, {Block},
  {Engelbracht}, {Gordon}, {Hora}, {Indebetouw}, {Kawamura}, {Meade},
  {Meixner}, {Sewilo}, {Vijh}, \& {Volk}}]{paradis09}
{Paradis}, D., {Reach}, W.~T., {Bernard}, J.-P., {Block}, M., {Engelbracht},
  C.~W., {Gordon}, K., {Hora}, J.~L., {Indebetouw}, R., {Kawamura}, A.,
  {Meade}, M., {Meixner}, M., {Sewilo}, M., {Vijh}, U.~P., \& {Volk}, K. 2009,
  \aj, 138, 196

\bibitem[{{Payne} {et~al.}(2004){Payne}, {Filipovi{\'c}}, {Reid}, {Jones},
  {Staveley-Smith}, \& {White}}]{payne04}
{Payne}, J.~L., {Filipovi{\'c}}, M.~D., {Reid}, W., {Jones}, P.~A.,
  {Staveley-Smith}, L., \& {White}, G.~L. 2004, \mnras, 355, 44

\bibitem[{{Peeters} {et~al.}(2004){Peeters}, {Spoon}, \& {Tielens}}]{peeters04}
{Peeters}, E., {Spoon}, H.~W.~W., \& {Tielens}, A.~G.~G.~M. 2004, \apj, 613,
  986

\bibitem[{{Povich} {et~al.}(2007){Povich}, {Stone}, {Churchwell}, {Zweibel},
  {Wolfire}, {Babler}, {Indebetouw}, {Meade}, \& {Whitney}}]{povich07}
{Povich}, M.~S., {Stone}, J.~M., {Churchwell}, E., {Zweibel}, E.~G., {Wolfire},
  M.~G., {Babler}, B.~L., {Indebetouw}, R., {Meade}, M.~R., \& {Whitney}, B.~A.
  2007, \apj, 660, 346

\bibitem[{{Puget} \& {Leger}(1989)}]{puget89}
{Puget}, J.~L., \& {Leger}, A. 1989, \araa, 27, 161

\bibitem[{{Rapacioli} {et~al.}(2005){Rapacioli}, {Joblin}, \&
  {Boissel}}]{rapacioli05}
{Rapacioli}, M., {Joblin}, C., \& {Boissel}, P. 2005, \aap, 429, 193

\bibitem[{{Reach} {et~al.}(2000){Reach}, {Boulanger}, {Contursi}, \&
  {Lequeux}}]{reach00}
{Reach}, W.~T., {Boulanger}, F., {Contursi}, A., \& {Lequeux}, J. 2000, \aap,
  361, 895

\bibitem[{{Reach} {et~al.}(2005){Reach}, {Megeath}, {Cohen}, {Hora}, {Carey},
  {Surace}, {Willner}, {Barmby}, {Wilson}, {Glaccum}, {Lowrance}, {Marengo}, \&
  {Fazio}}]{reach05}
{Reach}, W.~T., {Megeath}, S.~T., {Cohen}, M., {Hora}, J., {Carey}, S.,
  {Surace}, J., {Willner}, S.~P., {Barmby}, P., {Wilson}, G., {Glaccum}, W.,
  {Lowrance}, P., {Marengo}, M., \& {Fazio}, G.~G. 2005, \pasp, 117, 978

\bibitem[{{Rela{\~n}o} {et~al.}(2002){Rela{\~n}o}, {Peimbert}, \&
  {Beckman}}]{relano02}
{Rela{\~n}o}, M., {Peimbert}, M., \& {Beckman}, J. 2002, \apj, 564, 704

\bibitem[{{Rodrigues} {et~al.}(1997){Rodrigues}, {Magalhaes}, {Coyne}, \&
  {Piirola}}]{rodrigues97}
{Rodrigues}, C.~V., {Magalhaes}, A.~M., {Coyne}, G.~V., \& {Piirola}, V. 1997,
  \apj, 485, 618

\bibitem[{{Rubin} {et~al.}(2009){Rubin}, {Hony}, {Madden}, {Tielens},
  {Meixner}, {Indebetouw}, {Reach}, {Ginsburg}, {Kim}, {Mochizuki}, {Babler},
  {Block}, {Bracker}, {Engelbracht}, {For}, {Gordon}, {Hora}, {Leitherer},
  {Meade}, {Misselt}, {Sewilo}, {Vijh}, \& {Whitney}}]{rubin09}
{Rubin}, D., {Hony}, S., {Madden}, S.~C., {Tielens}, A.~G.~G.~M., {Meixner},
  M., {Indebetouw}, R., {Reach}, W., {Ginsburg}, A., {Kim}, S., {Mochizuki},
  K., {Babler}, B., {Block}, M., {Bracker}, S.~B., {Engelbracht}, C.~W., {For},
  B.-Q., {Gordon}, K., {Hora}, J.~L., {Leitherer}, C., {Meade}, M., {Misselt},
  K., {Sewilo}, M., {Vijh}, U., \& {Whitney}, B. 2009, \aap, 494, 647

\bibitem[{{Sandstrom} {et~al.}(2009){Sandstrom}, {Bolatto}, {Stanimirovi{\'c}},
  {van Loon}, \& {Smith}}]{sandstrom09}
{Sandstrom}, K.~M., {Bolatto}, A.~D., {Stanimirovi{\'c}}, S., {van Loon},
  J.~T., \& {Smith}, J.~D.~T. 2009, \apj, 696, 2138

\bibitem[{{Sloan} {et~al.}(2007){Sloan}, {Jura}, {Duley}, {Kraemer},
  {Bernard-Salas}, {Forrest}, {Sargent}, {Li}, {Barry}, {Bohac}, {Watson}, \&
  {Houck}}]{sloan07}
{Sloan}, G.~C., {Jura}, M., {Duley}, W.~W., {Kraemer}, K.~E., {Bernard-Salas},
  J., {Forrest}, W.~J., {Sargent}, B., {Li}, A., {Barry}, D.~J., {Bohac},
  C.~J., {Watson}, D.~M., \& {Houck}, J.~R. 2007, \apj, 664, 1144

\bibitem[{{Sloan} {et~al.}(2009){Sloan}, {Matsuura}, {Zijlstra}, {Lagadec},
  {Groenewegen}, {Wood}, {Szyszka}, {Bernard-Salas}, \& {van Loon}}]{sloan09}
{Sloan}, G.~C., {Matsuura}, M., {Zijlstra}, A.~A., {Lagadec}, E.,
  {Groenewegen}, M.~A.~T., {Wood}, P.~R., {Szyszka}, C., {Bernard-Salas}, J.,
  \& {van Loon}, J.~T. 2009, Science, 323, 353

\bibitem[{{Smith} \& {Hancock}(2009)}]{smith09}
{Smith}, B.~J., \& {Hancock}, M. 2009, \aj, 138, 130

\bibitem[{{Smith} {et~al.}(2007){Smith}, {Draine}, {Dale}, {Moustakas},
  {Kennicutt}, {Helou}, {Armus}, {Roussel}, {Sheth}, {Bendo}, {Buckalew},
  {Calzetti}, {Engelbracht}, {Gordon}, {Hollenbach}, {Li}, {Malhotra},
  {Murphy}, \& {Walter}}]{smith07}
{Smith}, J.~D.~T., {Draine}, B.~T., {Dale}, D.~A., {Moustakas}, J.,
  {Kennicutt}, Jr., R.~C., {Helou}, G., {Armus}, L., {Roussel}, H., {Sheth},
  K., {Bendo}, G.~J., {Buckalew}, B.~A., {Calzetti}, D., {Engelbracht}, C.~W.,
  {Gordon}, K.~D., {Hollenbach}, D.~J., {Li}, A., {Malhotra}, S., {Murphy},
  E.~J., \& {Walter}, F. 2007, \apj, 656, 770

\bibitem[{{Smith} \& {The MCELS Team}(1999)}]{smith99}
{Smith}, R.~C., \& {The MCELS Team}. 1999, in IAU Symposium, Vol. 190, New
  Views of the Magellanic Clouds, ed. Y.-H. {Chu}, N.~{Suntzeff}, J.~{Hesser},
  \& D.~{Bohlender}, 28--+

\bibitem[{{Stanimirovic} {et~al.}(1999){Stanimirovic}, {Staveley-Smith},
  {Dickey}, {Sault}, \& {Snowden}}]{stanimirovic99}
{Stanimirovic}, S., {Staveley-Smith}, L., {Dickey}, J.~M., {Sault}, R.~J., \&
  {Snowden}, S.~L. 1999, \mnras, 302, 417

\bibitem[{{Sugerman} {et~al.}(2006){Sugerman}, {Ercolano}, {Barlow}, {Tielens},
  {Clayton}, {Zijlstra}, {Meixner}, {Speck}, {Gledhill}, {Panagia}, {Cohen},
  {Gordon}, {Meyer}, {Fabbri}, {Bowey}, {Welch}, {Regan}, \&
  {Kennicutt}}]{sugerman06}
{Sugerman}, B.~E.~K., {Ercolano}, B., {Barlow}, M.~J., {Tielens}, A.~G.~G.~M.,
  {Clayton}, G.~C., {Zijlstra}, A.~A., {Meixner}, M., {Speck}, A., {Gledhill},
  T.~M., {Panagia}, N., {Cohen}, M., {Gordon}, K.~D., {Meyer}, M., {Fabbri},
  J., {Bowey}, J.~E., {Welch}, D.~L., {Regan}, M.~W., \& {Kennicutt}, R.~C.
  2006, Science, 313, 196

\bibitem[{{Tielens} {et~al.}(1987){Tielens}, {Seab}, {Hollenbach}, \&
  {McKee}}]{tielens87}
{Tielens}, A.~G.~G.~M., {Seab}, C.~G., {Hollenbach}, D.~J., \& {McKee}, C.~F.
  1987, \apjl, 319, L109

\bibitem[{{Valencic} {et~al.}(2003){Valencic}, {Clayton}, {Gordon}, \&
  {Smith}}]{valencic03}
{Valencic}, L.~A., {Clayton}, G.~C., {Gordon}, K.~D., \& {Smith}, T.~L. 2003,
  \apj, 598, 369

\bibitem[{{Walter} {et~al.}(2007){Walter}, {Cannon}, {Roussel}, {Bendo},
  {Calzetti}, {Dale}, {Draine}, {Helou}, {Kennicutt}, {Moustakas},
  {Rieke},{Armus}, {Engelbracht}, {Gordon}, {Hollenbach}, {Lee}, {Li},
  {Meyer}, {Murphy}, {Regan}, {Smith}, {Brinks}, {de Blok}, {Bigiel}, \&
  {Thornley}}]{walter07}
{Walter}, F., {Cannon}, J.~M., {Roussel}, H., {Bendo}, G.~J.,
  {Calzetti}, D., {Dale}, D.~A., {Draine}, B.~T., {Helou}, G., {Kennicutt},
   Jr., R.~C., {Moustakas}, J., {Rieke}, G.~H., {Armus}, L.,
  {Engelbracht}, C.~W., {Gordon},K., {Hollenbach}, D.~J., {Lee}, J., {Li}, A., {Meyer},
   M.~J., {Murphy}, E.~J., {Regan}, M.~W., {Smith}, J., {Brinks}, E., 
   {de Blok}, W.~J.~G., {Bigiel}, F., \& {Thornley}, M.~D. 2007, \apj, 661, 102

\bibitem[{{Weingartner} \& {Draine}(2001)}]{weingartner01}
{Weingartner}, J.~C., \& {Draine}, B.~T. 2001, \apj, 548, 296

\bibitem[{{Wu} {et~al.}(2006){Wu}, {Charmandaris}, {Hao}, {Brandl},
  {Bernard-Salas}, {Spoon}, \& {Houck}}]{wu06}
{Wu}, Y., {Charmandaris}, V., {Hao}, L., {Brandl}, B.~R., {Bernard-Salas}, J.,
  {Spoon}, H.~W.~W., \& {Houck}, J.~R. 2006, \apj, 639, 157

\bibitem[{{Zaritsky} {et~al.}(2000){Zaritsky}, {Harris}, {Grebel}, \&
  {Thompson}}]{zaritsky00}
{Zaritsky}, D., {Harris}, J., {Grebel}, E.~K., \& {Thompson}, I.~B. 2000,
  \apjl, 534, L53

\end{thebibliography}

\end{document}